\documentclass[%
 reprint,
 amsmath,amssymb,
 aps,
pra,
]{revtex4-1}

\usepackage{subfigure}
\usepackage{graphicx}
\usepackage{dcolumn}
\usepackage{bm}
\usepackage{amsmath,amsfonts,amssymb}
\DeclareMathOperator{\sech}{sech}
\usepackage[ruled,vlined]{algorithm2e}

\begin{document}


\title{Performance of superadiabatic stimulated Raman adiabatic passage in the presence of dissipation and Ornstein-Uhlenbeck dephasing}

\author{Kostas Blekos}

\author{Dionisis Stefanatos}
\email{dionisis@post.harvard.edu}

\author{Emmanuel Paspalakis}

\affiliation{Materials Science Department, School of Natural Sciences, University of Patras, Patras 26504, Greece}

\date{\today}

\begin{abstract}

In this paper we evaluate the performance of two superadiabatic stimulated Raman adiabatic passage (STIRAP) protocols derived from Gaussian and sin-cos pulses, under dissipation and Ornstein-Uhlenbeck noise in the energy levels. We find that for small amplitudes of Stokes and pump pulses, the population transfer is mainly achieved directly through the counterdiabatic pulse, while for large amplitudes the conventional STIRAP path dominates. This kind of ``hedging" leads to a remarkable robustness against dissipation in the lossy intermediate state. For small pulse amplitudes and increasing noise correlation time the performance is decreased, since the dominant counterdiabatic pulse is affected more, while for large pulse amplitudes, where the STIRAP path dominates, the efficiency is degraded more for intermediate correlation times (compared to the pulse duration). For the Gaussian superadiabatic STIRAP protocol we also investigate the effect of delay between pump and Stokes pulses and find that under the presence of noise the performance is improved for increasing delay. We conclude that the Gaussian protocol with suitably chosen delay and the sin-cos protocol perform quite well even under severe noise conditions. The present work is expected to have a broad spectrum of applications, since STIRAP has a crucial role in modern quantum technology.

\end{abstract}

\maketitle

\section{Introduction}

One of the most successful quantum control methods for population transfer between the levels of a quantum system is Stimulated Raman adiabatic passage (STIRAP) \cite{Bergmann98,Kobrak98,Vitanov01,Vitanov17,Bergmann19}. The prototype STIRAP system consists of three energy levels in the $\Lambda$-configuration. In order to transfer population from state $|1\rangle$ to state $|3\rangle$, passing through the intermediate level $|2\rangle$, two laser pulses are applied in counterintuitive order, the Stokes pulse coupling states $|2\rangle-|3\rangle$ and the pump pulse coupling states $|1\rangle-|2\rangle$. A coherent superposition is formed by states $|1\rangle$ and $|3\rangle$, which adiabatically evolves from state $|1\rangle$ initially to state $|3\rangle$ finally, while the lossy intermediate state $|2\rangle$ is barely populated. STIRAP finds a wide range of applications in modern quantum science, from optical wavequides \cite{Dreisow09} and matter waves \cite{Menchon16} to nitrogen-vacancy centers in diamond \cite{Golter14} and superconducting quantum circuits \cite{Kumar16}, for more details see the recently published roadmap \cite{Bergmann19}. The most important advantage of STIRAP is its robustness against moderate variations of experimental parameters. Its major drawback is that adiabatic transfer requires long times, leading to reduced efficiency when undesirable interactions with the environment, for example decoherence and dissipation, are present.

During the last decade, a series of methods characterized as \emph{shortcuts to adiabaticity} have been developed with the aim to improve the performance of quantum adiabatic evolution by reducing the required duration \cite{Odelin19,Unayan97,Demirplak03,Berry09,Motzoi09,Chen10a,Masuda10,Deffner14,Claeys19}. The common basic idea of these techniques is to drive faster the quantum system at the same final state as the slow adiabatic evolution. This goal is achieved either bypassing the intermediate adiabatic states, or by introducing an extra term in the Hamiltonian to suppress the diabatic transitions and evolve the system along the adiabatic path of the original Hamiltonian. The latter approach is called \emph{superadiabatic, assisted adiabatic passage} or \emph{transitionless tracking algorithm}. Both methods are widely exploited in modern quantum technologies, as is thoroughly discussed in the recently published review \cite{Odelin19}, and have also been used to increase STIRAP efficiency, see for example Refs. \cite{Li16,Clerk16,Du16,Zhou17,Mortensen18} and \cite{Demirplak03,Demirplak05,Chen10b,Giannelli14,Torosov14,Masuda15,Masuda15b,Vepsalainen19,Impens19} for the first and second method, respectively. 

The influence of noise on the efficiency of STIRAP has been the subject of several works. The effect of dephasing caused by classical Ornstein-Ulhenbeck noise \cite{Fox88} in the level energies was studied in Ref. \cite{Demirplak02}, while the dephasing due to quantum baths was evaluated in Refs. \cite{Shi03,Vitanov04,Zeng19}. In the context of superconducting artificial atoms, the influence of broadband colored noise on population transfer was considered in Refs. \cite{Falci13,DiStefano16}. Along with these studies which focus on conventional STIRAP, there are also works which examine the performance of STIRAP shortcuts in the presence of noise. For the case where no additional counterdiabatic field is used, i.e. the classical STIRAP framework is preserved, there are some recent works in the physical context of nitrogen-vacancy centers in diamond \cite{Zhou17,Boyers19}. Specifically, in Ref. \cite{Zhou17} the efficiency of STIRAP shortcuts developed in Refs. \cite{Ibanez12,Clerk16,Li16} was investigated in the presence of dissipation and spectral diffusion, while in Ref. \cite{Boyers19} the performance of a Floquet-engineered shortcut under colored noise was evaluated. In our recent work \cite{Stefanatos20} we studied the efficiency of the shortcuts derived in \cite{Ibanez12,Li16}, in the presence of classical Ornstein-Ulhenbeck noise in the energy levels, as in Ref. \cite{Demirplak02}. There are also a few works \cite{Issoufa14,Masuda15} which study the influence of noise on \emph{superadiabatic} STIRAP (SA-STIRAP, in the spirit of Giannelli and Arimondo \cite{Giannelli14}), where an extra counterdiabatic term is exploited in the Hamiltonian. Specifically, the authors of Ref. \cite{Masuda15} consider the effect of non-Gaussian distribution of energy fluctuations, while those of Ref. \cite{Issoufa14} study the effect of ground state dephasing.

In this article we evaluate the performance, in the presence of noise and dissipation, of two SA-STIRAP protocols derived from Gaussian and sin-cos pulses \cite{Giannelli14,Laine96,Chen12}.  We use classical Ornstein-Ulhenbeck noise processes with exponential correlation functions \cite{Fox88} in the energy levels, as in Refs. \cite{Demirplak02,Stefanatos20}. This type of noise provides a relatively simple means to study the effect of colored noise, with non-zero correlation time. It has been used to model the fluctuations in the energy levels for molecules in a liquid \cite{Demirplak02}. It may also be present in the phases of the applied laser fields, appearing in the rotating wave approximation as a corresponding noise term in the energy levels \cite{Demirplak05}. Also note that the sensitivity of shortcuts to adiabaticity to Ornstein-Ulhenbeck noise has also been studied in the context of fast shuttling of an atom using a moving optical lattice \cite{Lu20}. From numerical simulations, we find that for small amplitudes of Stokes and pump pulses the population transfer is mainly achieved directly from level $|1\rangle$ to level $|3\rangle$ through the counterdiabatic pulse, while for large amplitudes the STIRAP path $|1\rangle\rightarrow|2\rangle\rightarrow|3\rangle$ dominates. The presence of two alternative paths leading to the target state results in a remarkable robustness against dissipation in the lossy intermediate level. For small pulse amplitudes and increasing noise correlation time the performance is decreased, since the dominant counterdiabatic pulse is affected more. For large pulse amplitudes the efficiency is degraded more for intermediate correlation times (compared to the pulse duration), while it is better for small or large noise correlation times. This behavior is similar to that observed for conventional STIRAP \cite{Demirplak02}, something expected since for large amplitudes STIRAP is the dominant mechanism for population transfer. For Gaussian SA-STIRAP we also investigate the effect of delay between pump and Stokes pulses and find that under the presence of noise the performance is improved as the delay increases (at least up to the considered values). We conclude that the Gaussian SA-STIRAP protocol with suitably chosen delay and the sin-cos SA-STIRAP protocol perform quite well even under severe noise conditions.

The structure of the article is as follows. In the next section we derive for completeness the Gaussian and sin-cos SA-STIRAP protocols in the absence of noise. In section \ref{sec:noise} we study their performance when dissipation and noise are present. Section \ref{sec:conclusion} concludes this paper.

\section{SA-STIRAP in the absence of dissipation and dephasing}

\label{sec:no_noise}

The reference Hamiltonian for STIRAP in both one-photon and two-photon resonance is
\begin{equation}
\label{H0}
H_0(t)=\frac{\hbar}{2}
\left(
\begin{array}{ccc}
0 & \Omega_p(t)  & 0\\
\Omega_p(t) & -i\Gamma & \Omega_s(t) \\
0 & \Omega_s(t)  & 0
\end{array}
\right),
\end{equation}
where $\Omega_p(t), \Omega_s(t)$ are the Rabi frequencies for the pump and Stokes lasers, respectively, and $\Gamma$ is the dissipation rate from level $|2\rangle$. In this section we derive SA-STIRAP for the ideal case $\Gamma=0$
while in the next section we study the effect of nonzero dissipation and dephasing.

If we define the time-dependent amplitude $\Omega(t)$ and mixing angle $\theta(t)$ through the relations
\begin{equation}
\Omega(t)=\sqrt{\Omega_p^2(t)+\Omega_s^2(t)},\quad \tan{\theta(t)}=\frac{\Omega_p(t)}{\Omega_s(t)},
\end{equation}
then the instantaneous eigenstates of $H_0(t)$ are
\begin{equation}
\label{eigenvectors}
|\phi_{0}(t)\rangle=
\left(\begin{array}{c}
    \cos{\theta}\\
    0\\
    -\sin{\theta}
\end{array}\right),
\quad
|\phi_{\pm}(t)\rangle=\frac{1}{\sqrt{2}}
\left(\begin{array}{c}
    \sin{\theta}\\
    \pm 1 \\
    \cos{\theta}
\end{array}\right),
\end{equation}
with corresponding eigenvalues
\begin{equation}\label{eigenvalues}
E_0(t)=0,\quad E_{\pm}(t)=\pm\hbar\frac{\Omega(t)}{2} \, .
\end{equation}
In conventional STIRAP the angle $\theta$ changes slowly (adiabatically) from the initial value $\theta(t_i)=0$ at $t=t_i$ to the final value $\theta(t_f)=\pi/2$ at $t=t_f$, while the the system follows the dark adiabatic state $|\phi_{0}(t)\rangle=\cos\theta(t)|1\rangle-\sin\theta(t)|3\rangle$, from the initial state $|1\rangle$ to the final state $|3\rangle$.

The adiabatic approximation fails for fast variations of angle $\theta$, in which case some population remains in levels $|1\rangle$ and $|2\rangle$ at the final time. A method to accomplish the desired population transfer to level $|3\rangle$ in arbitrarily short times is to add in Hamiltonian $H_0$ an extra counter-diabatic Hamiltonian $H_{cd}$, which cancels the diabatic terms arising when $H_0$ is transformed to the time-dependent adiabatic basis,
\begin{eqnarray}
\label{counter_diabatic}
H_{cd}(t)&=&i\hbar\sum_{n=0,\pm}\Big[|\dot{\phi}_n(t)\rangle\langle\phi_n(t)|\nonumber\\
         & &-\langle\phi_n(t)|\dot{\phi}_n(t)\rangle|\phi_n(t)\rangle\langle\phi_n(t)|\Big].\nonumber
\end{eqnarray}
Using Eq. (\ref{eigenvectors}) we find for the STIRAP system
\begin{equation}
\label{Hcd3}
H_{cd}(t)=\frac{\hbar}{2}
\left(
\begin{array}{ccc}
0 & 0  & i\Omega_d\\
0 & 0 & 0 \\
-i\Omega_d & 0  & 0
\end{array}
\right),
\end{equation}
where
\begin{equation}
\Omega_d(t)=2\dot{\theta}(t)
\end{equation}
is a $\pi$-pulse connecting directly states $|1\rangle$ and $|3\rangle$, since $\int_{t_i}^{t_f} \Omega_d(t)dt=\int_{t_i}^{t_f} 2\dot{\theta}(t)dt=\pi$.
Under the total Hamiltonian $H=H_0+H_{cd}$ the system can track with perfect fidelity and for arbitrarily short times the dark state $|\phi_{0}(t)\rangle$ of the reference Hamiltonian $H_0$, and for this reason the method is called transitionless tracking algorithm.

Note that usually the direct transition between levels $|1\rangle$ and $|3\rangle$ is electric dipole forbidden. One physical system where such transition is possible is described in Refs. \cite{Unayan97,Giannelli14}. It is a $\Lambda$-system between $J=1$ Zeeman sublevels and an excited $J=0$ state. The coupling between the $|1\rangle$ and $|3\rangle$ states may be accomplished by a magnetic dipole interaction between the atomic or molecular angular momentum $\mathbf{J}$ and an external magnetic field.  Another example, also discussed in Ref. \cite{Giannelli14}, contains a ladder system which includes both single-photon transitions for the two electric dipole allowed transitions and a two-photon transition for the electric dipole forbidden transition. Different systems where all transitions in the three-level system are electric dipole allowed exist in asymmetric quantum systems with broken inversion symmetry, e.g. asymmetric molecules \cite{Vitanov01,Kral03a}. We note that the applicability of the method is not restricted to atoms or molecules, but can also be exploited for population transfer in artificial atoms, like superconducting quantum structures and semiconductor nanostructures \cite{Vepsalainen19,Kral01a,Nori05a}. For example, in the recent experiment \cite{Vepsalainen19}, SA-STIRAP was used in the first three states of a superconducting transmon circuit, to transfer population between the ground state and the second excited state, by combining two single-photon transitions and a two-photon transition. The advantage of the SA-STIRAP method, compared to a direct $\pi$-pulse between states $|1\rangle$ and $|3\rangle$ or classical STIRAP with only pump and Stokes fields, is its increased fidelity and robustness, since it combines the useful characteristics of both methods. The increased fidelity and robustness is particularly significant for quantum technology applications.

In the next section we will evaluate the performance of two SA-STIRAP protocols under noise. For the first protocol the pump and Stokes pulses are Gaussian \cite{Giannelli14}
\begin{equation}
\label{gaussian}
\Omega_p(t)=\Omega_0 e^{-\left(\frac{t-\tau}{T}\right)^2},\quad \Omega_s(t)=\Omega_0 e^{-\left(\frac{t+\tau}{T}\right)^2},
\end{equation}
with corresponding counterdiabatic pulse
\begin{equation}
\label{cd_gaussian}
\Omega_d(t)=\frac{4\tau}{T^2}\sech{\left(\frac{4\tau t}{T^2}\right)}.
\end{equation}
Note that $2\tau$ is the delay between the Gaussian pulses and $T$ is the pulse width.
For the second protocol the pump and Stokes pulses have the sin-cos shape \cite{Giannelli14,Chen12,Laine96}
\begin{equation}
\label{sincos}
\Omega_p(t)=\Omega_0 \sin{\left (\frac{\pi t}{2T}\right )},\quad \Omega_s(t)=\Omega_0 \cos{\left (\frac{\pi t}{2T}\right )},
\end{equation}
while the corresponding counterdiabatic pulse is constant,
\begin{equation}
\label{cd_sincos}
\Omega_d(t)=\frac{\pi}{T}.
\end{equation}

\section{Performance of SA-STIRAP protocols under dissipation and Ornstein-Uhlenbeck dephasing}

\label{sec:noise}

We study the effect of noise by adding to the total Hamiltonian a noise term,
\begin{equation}
H(t)=H_0(t)+H_{cd}(t)+H_{\epsilon}(t),
\end{equation}
where
\begin{equation}
H_{\epsilon}(t)=\frac{\hbar}{2}
\left(
\begin{array}{ccc}
\epsilon_1(t) & 0  & 0\\
0 & \epsilon_2(t) & 0 \\
0 & 0  & \epsilon_3(t)
\end{array}
\right)
\end{equation}
and $\epsilon_i(t), i=1,2,3$, are independent Ornstein-Uhlenbeck noise process. The latter are defined by the stochastic differential equations \cite{Fox88}
\begin{equation}
\dot{\epsilon}_i=-\frac{1}{\tau_c}\epsilon_i+\frac{1}{\tau_c}g_i,
\end{equation}
where $g_i(t), i=1,2,3$, are independent Gaussian white noises with zero mean and correlations
\begin{equation}
\langle g_i(t)g_j(t+\tau)\rangle=2\sigma^2\tau_c\delta_{ij}\delta(\tau).
\end{equation}
The stochastic processes $\epsilon_i$ are also Gaussian with zero mean, correlations
\begin{equation}
R_{ij}(\tau)=\langle \epsilon_i(t)\epsilon_j(t+\tau)\rangle=\sigma^2\delta_{ij}e^{-|\tau|/\tau_c},
\end{equation}
and steady state probability distributions
\begin{equation}
P_i(\epsilon)=\frac{1}{\sqrt{2\pi\sigma^2}}e^{-\frac{\epsilon^2}{2\sigma^2}},
\end{equation}
where note that parameter $\tau_c$ corresponds to the noise correlation time. The power spectral density of each Ornstein-Uhlenbeck process is
\begin{equation}
\label{spectrum}
S_{ii}(\omega)=\int_{-\infty}^{\infty}R_{ii}(\tau)e^{-i\omega\tau}d\tau=\frac{2\sigma^2\tau_c}{1+(\omega\tau_c)^2}
\end{equation}
and the expectation value of the total power is
\begin{equation}
\langle \epsilon_i^2(t)\rangle=R_{ii}(0)=\frac{1}{2\pi}\int_{-\infty}^{\infty}S_{ii}(\omega)d\omega=\sigma^2.
\end{equation}

We implement the Ornstein-Uhlenbeck processes using the algorithm described in Ref.\ \cite{Fox88}, with specific standard deviation $\sigma=5/T$, corresponding to fixed noise power, and three correlation times $\tau_c/T=0.008, 0.08, 0.8$. More details about the numerical implementation can be found in the Appendix. 
In the following Figs. \ref{fig:gaussian25}-\ref{fig:sin} we present simulation results for both STIRAP and SA-STIRAP using various pulses. In all these figures we plot the transfer fidelity $F$ (average final population of level $|3\rangle$ over 200 stochastic runs) versus the peak amplitude $\Omega_0$ of pump and Stokes pulses, for various values of delay $\tau$ and dissipation $\Gamma$. We start with Fig. \ref{fig:gaussian25}, where we use Gaussian pulses with delay parameter $\tau/T=1/4$. In Figs. \ref{fig:gaussian250}, \ref{fig:gaussian251}, \ref{fig:gaussian254}, \ref{fig:gaussian2510} we display results for the conventional STIRAP and four dissipation values, $\Gamma=0, 1/T, 4/T, 10/T$, respectively. Solid blue line corresponds to the case where there is no dephasing noise, while the other three lines correspond to different noise correlation times: $\tau_c/T=0.008$, (orange dashed line), $\tau_c/T=0.08$ (green dotted line), and $\tau_c/T=0.8$ (red dashed-dotted line). For the cases where dephasing noise is present, we also show the 98\% confidence interval, i.e. the range of values where lies the fidelity of 98\% of the stochastic runs. In Figs. \ref{fig:sa_gaussian250}, \ref{fig:sa_gaussian251}, \ref{fig:sa_gaussian254}, \ref{fig:sa_gaussian2510} we display similar results but for the corresponding SA-STIRAP protocol (with the additional counterdiabatic pulse), also for $\Gamma=0, 1/T, 4/T, 10/T$, respectively. In Figs. \ref{fig:gaussian33}, \ref{fig:gaussian50}, \ref{fig:gaussian75} we show analogous results for Gaussian protocols with delay parameters $\tau/T=1/3, 1/2, 3/4$, respectively, while in Fig. \ref{fig:sin} for the sin-cos protocol.

Several interesting observations can be made from these figures. First note that in the absence of noise or for small noise correlation time the STIRAP fidelity shows an oscillatory behavior for smaller delay values $\tau/T$, Figs. \ref{fig:gaussian250} and \ref{fig:gaussian330}, which disappears as $\tau/T$ increases, Figs. \ref{fig:gaussian500} and \ref{fig:gaussian750}. As discussed in Ref. \cite{Giannelli14}, these fidelity maxima occur when the Rabi oscillations between the levels are synchronized with the interaction time. Also observe that, although the first maximum in Figs. \ref{fig:gaussian250} and \ref{fig:gaussian330} appears at values $\Omega_0$ lower than those needed in Figs. \ref{fig:gaussian500} and \ref{fig:gaussian750}, high levels of fidelity are maintained only for a narrow window around this maximum. An oscillatory fidelity is also observed for STIRAP with sin-cos pulses in the absence of dephasing noise or for noise with small $\tau_c$, see Fig. \ref{fig:sin500}. The STIRAP efficiency is reduced in the presence of dephasing noise, as well as for nonzero dissipation, requiring larger values of $\Omega_0$, see the first column of Figs. \ref{fig:gaussian25}-\ref{fig:sin}.

We now move to discuss the performance of SA-STIRAP protocols. The most important observation when comparing the first (STIRAP) and second (SA-STIRAP) columns of Figs. \ref{fig:gaussian25} to \ref{fig:sin} is that the presence of the counterdiabatic pulse leads to nozero fidelity even for small values of $\Omega_0$. In the absence of dephasing noise a perfect transfer is achieved for every value of $\Omega_0$, while the efficiency is reduced when noise is present. From these plots becomes obvious that for SA-STIRAP and in the case of small $\Omega_0$ the population is mainly transferred directly from state $|1\rangle$ to $|3\rangle$ using the counterdiabatic pulse, while for larger $\Omega_0$ dominates the STIRAP path passing through the intermediate state $|2\rangle$. This is also demonstrated in Fig. \ref{fig:sa_gaussian504_P}, where the final total population $P$ in all three levels is plotted versus $\Omega_0$, for Gaussian SA-STIRAP with delay $\tau/T=1/2$ and dissipation rate $\Gamma=4/T$. Observe that, as $\Omega_0$ increases, there appear population losses ($P<1$), indicating that part of the population passes through the lossy level $|2\rangle$. For lower $\Omega_0$ the transfer to level $|3\rangle$ cannot be completed and most of the  population in the intermediate state $|2\rangle$ is dissipated. As $\Omega_0$ increases the STIRAP path is established and the transfer to state $|3\rangle$ is completed, as indicated by the convergence of total population $P$ to the fidelity $F$ when comparing Figs. \ref{fig:sa_gaussian504_P} and \ref{fig:sa_gaussian504}.

We emphasize that SA-STIRAP is different than the shortcut method used in our recent work \cite{Stefanatos20} since, in addition to the pump and Stokes pulses, it exploits the extra field $\Omega_d$ connecting directly states $|1\rangle$ and $|3\rangle$. The advantage of using the additional field in the presence of dissipation is demonstrated in Fig. \ref{fig:SA_ST_vs_STA}, where note that for simplicity we ignore dephasing noise. In Fig. \ref{fig:STA_pulses} we display the shortcut pump and Stokes pulses, derived from Gaussian profiles, that we used in Ref. \cite{Stefanatos20} (see Fig. 2 there), while in Fig. \ref{fig:STA_F} we plot the corresponding time evolution of the third level population $p_3$, for different values of dissipation rate $\Gamma=0, 1/T, 4/T, 10/T$ (from top to bottom). Observe that, as the dissipation increases, the transfer fidelity at the final time drops considerably. The reason is that the transfer is accomplished through the lossy intermediate state $|2\rangle$. In Fig. \ref{fig:SA_ST_pulses} we plot the Gaussian SA-STIRAP pump and Stokes pulses, as well as the counterdiabatic pulse $\Omega_d$ (red solid line), for $\Omega_0=5/T$ and $\tau/T=3/4$, while in Fig. \ref{fig:SA_ST_F} the corresponding time evolution of population $p_3$, for the same dissipation values as before, $\Gamma=0, 1/T, 4/T, 10/T$. Observe that now the effect of dissipation can be hardly distinguished. The reason is that the transfer is mostly accomplished directly from $|1\rangle$ to $|3\rangle$ by the extra field $\Omega_d$. For comparison we mention that the common area of SA-STIRAP pump and Stokes pulses is 8.8558 units, while those of the shortcut pulses are 10.3563 and 12.3597 units. By inspecting the second columns of Figs. \ref{fig:gaussian25}-\ref{fig:sin}, it becomes evident that SA-STIRAP shows a remarkable robustness against dissipation $\Gamma$, even in the presence of dephasing noise.

Another interesting observation in the case of Gaussian SA-STIRAP is that for fixed delay $\tau$ between the Stokes and pump pulses, the fidelity for small $\Omega_0$ drops with increasing noise correlation time $\tau_c$, as is evident in the second columns of Figs. \ref{fig:gaussian25}-\ref{fig:gaussian75}. This behavior can be explained as follows. Recall that in the range of small $\Omega_0$ the desired transfer is achieved through the counterdiabatic pulse given in Eq. (\ref{cd_gaussian}) which, for fixed delay $\tau$, is specified. The Fourier transform of this pulse is
\begin{equation}
\label{Fourier_cd_gaussian}
F_d(\omega)=\int_{-\infty}^{\infty}\Omega_d(t)e^{-i\omega t}dt=\pi\sech{\left(\frac{\pi T^2}{8\tau}\omega\right)},
\end{equation}
from which it is obvious that the pulse spectrum is mainly concentrated between $0$ and $8\tau/(\pi T^2)$. Now observe from Eq. (\ref{spectrum}) that, as the correlation time $\tau_c$ increases, the noise power is concentrated in lower frequencies around zero, affecting thus more the baseband counterdiabatic pulse. This is also demonstrated in Fig. \ref{fig:changetau_c}. The sin-cos protocol also exhibits a similar behavior for increasing $\tau_c$, see the second column of Fig. \ref{fig:sin}. A closely related observation is that for fixed $\tau_c$, $\Gamma$ and increasing $\tau$, i.e. when examining the same type of curve in subfigures belonging to the same row of the second column across Figs. \ref{fig:gaussian25} to \ref{fig:gaussian75} (f.e. green dotted curve corresponding to $\tau_c/T=0.08$ across Figs. \ref{fig:sa_gaussian250} to \ref{fig:sa_gaussian750}, where $\Gamma=0$), the fidelity for small $\Omega_0$ increases. The explanation is that, as the delay $\tau$ increases, the spectrum of the counterdiabatic pulse broadens and thus is less affected by the baseband noise, as shown in Fig. \ref{fig:changetau}.

For large $\Omega_0$ it is evident from Figs. \ref{fig:gaussian25}-\ref{fig:sin} that the fidelity for dephasing noise with large correlation time $\tau_c/T=0.8$ (red dashed-dotted line) is similar or in most cases better than that for noise with intermediate correlation time $\tau_c/T=0.08$ (green dotted line). This behavior has been observed for STIRAP in Ref. \cite{Demirplak02} and is confirmed here also for SA-STIRAP since, as we pointed out above, for large $\Omega_0$ the STIRAP path dominates. We next try to understand this behavior using Fig. \ref{fig:large_omega}, where we employ Gaussian STIRAP with delay $\tau/T=1/2$, dissipation $\Gamma=0$ and no dephasing noise, thus the reference for fidelity is Fig. \ref{fig:gaussian500}. In Fig. \ref{fig:P2W60} we plot the time evolution of level $|2\rangle$ population $p_2$ and in Fig. \ref{fig:BlochW60} the trajectory on the Bloch sphere, for large $\Omega_0=60/T$.  The oscillations of $p_2$, most of which occur with a period of about $0.2T$, correspond to the revolution of Bloch vector (red line) around the instantaneous total field which, as the pump and Stokes pulses evolve, rotates from the north pole to the equator (on the plane of the meridian shown). Dephasing noise with small correlation time $\tau_c/T=0.008$ is averaged during an oscillation period and its net effect is small, as pointed out in \cite{Demirplak02}. For large $\tau_c/T=0.8$, the noise remains constant for large parts of the evolution, and its influence is reduced since $\Omega_0=60/T>\sigma=5/T$.  It is the noise with intermediate correlation time $\tau_c/T=0.08$, comparable with the period of oscillations, which affects the most the desired transfer, and this is clearly demonstrated in Fig. \ref{fig:gaussian500} for $\Omega_0=60/T$. In Figs. \ref{fig:P2W15}, \ref{fig:BlochW15}, we plot the population $p_2$ and the trajectory on the Bloch sphere, respectively, for lower $\Omega_0=15/T$. Observe that here the period of oscillations is comparable to $\tau_c/T=0.8$, consequently the noise with this correlation time is now more effective, as depicted in Fig. \ref{fig:gaussian500} for $\Omega_0=15/T$.

Closing, we would like to emphasize that, by inspecting fidelity in Figs. \ref{fig:gaussian25} to \ref{fig:sin}, we conclude that the best performance under severe noise conditions is observed for Gaussian SA-STIRAP with delay $\tau/T=3/4$ and the sin-cos SA-STIRAP protocol.

\begin{figure*}[t]
 \centering
		\begin{tabular}{cc}
     	\subfigure[$\ $]{
	            \label{fig:gaussian250}
	            \includegraphics[width=.37\linewidth]{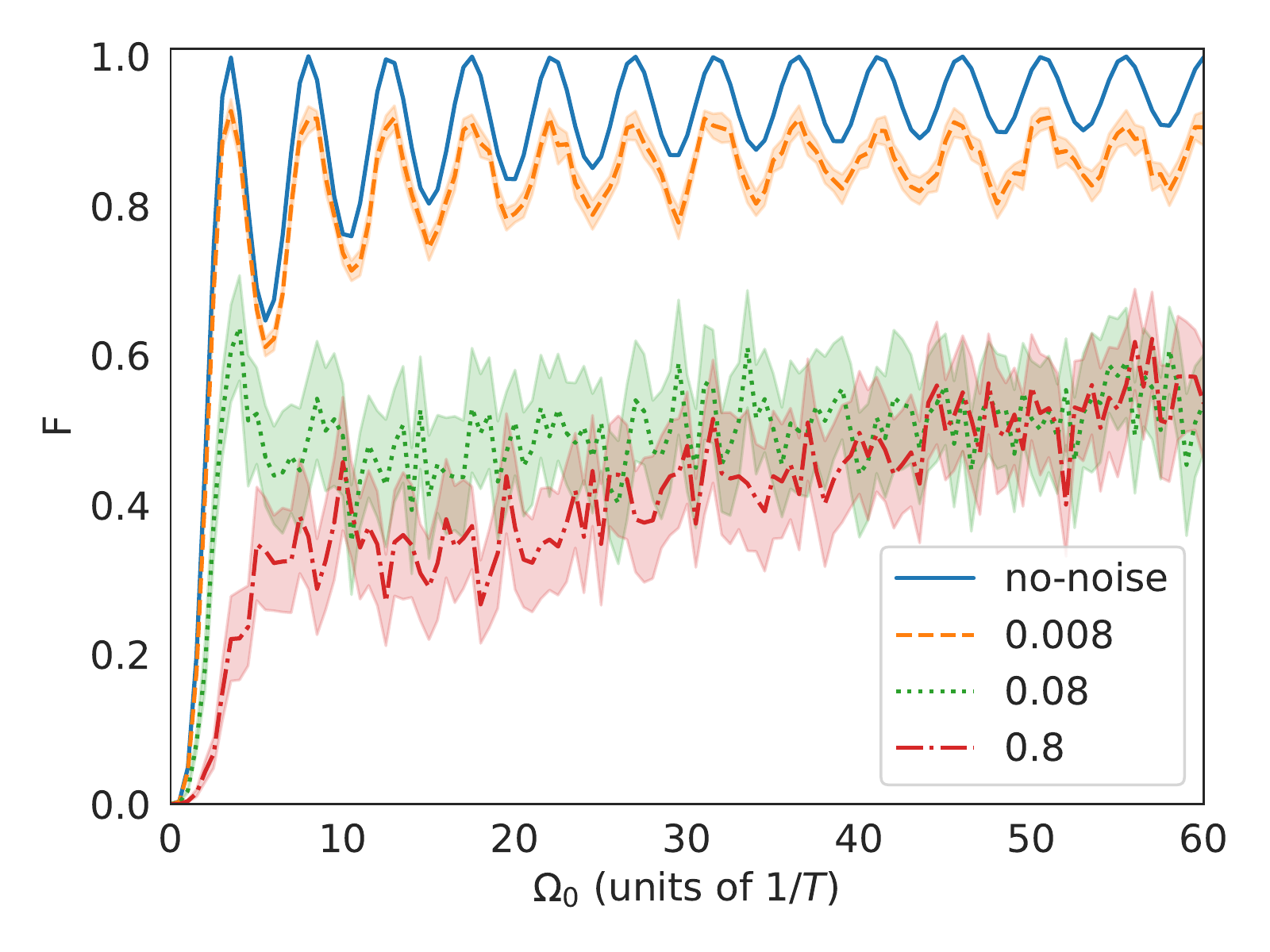}} &
        \subfigure[$\ $]{
	            \label{fig:sa_gaussian250}
	            \includegraphics[width=.37\linewidth]{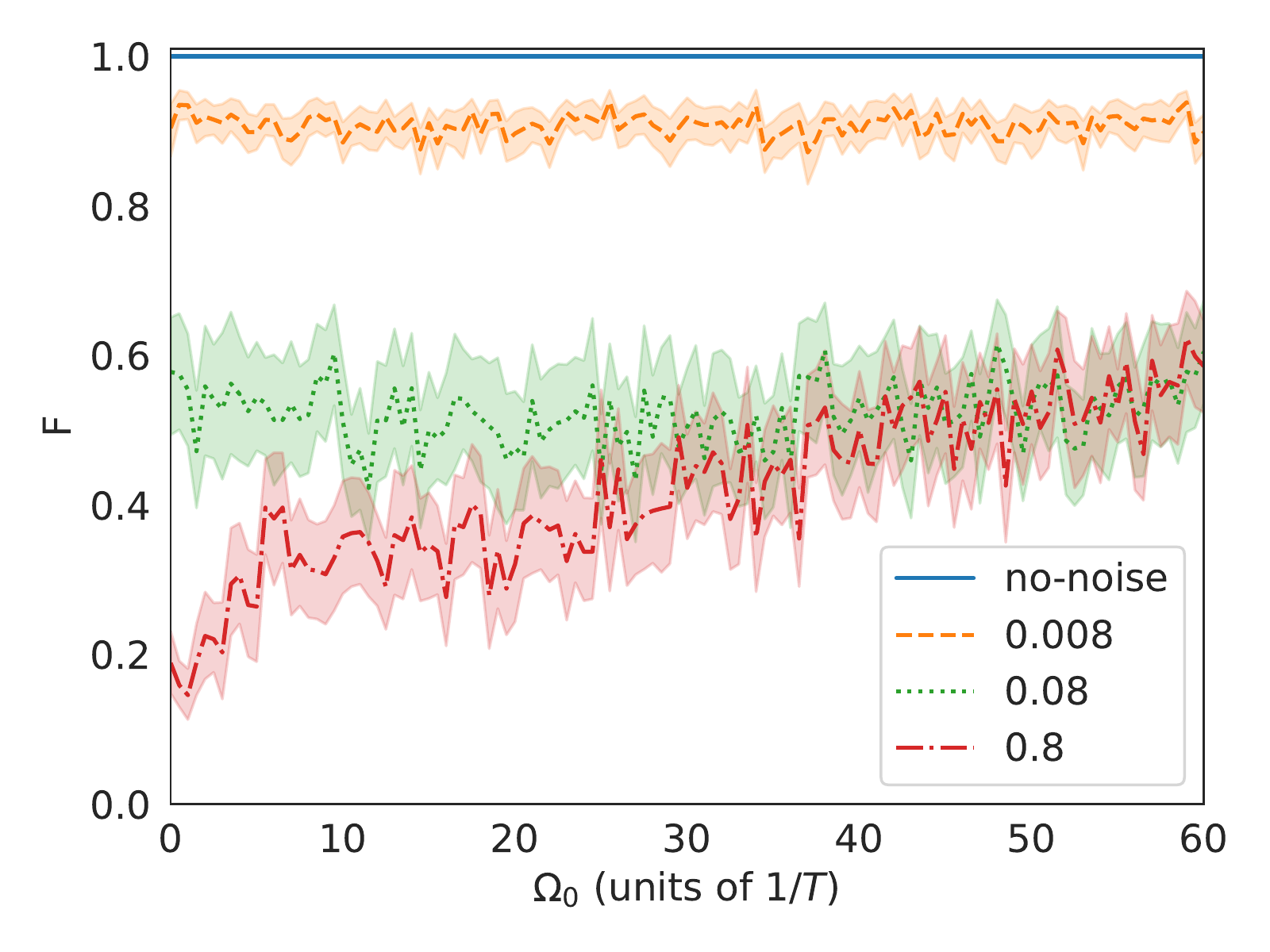}} \\
        \subfigure[$\ $]{
	            \label{fig:gaussian251}
	            \includegraphics[width=.37\linewidth]{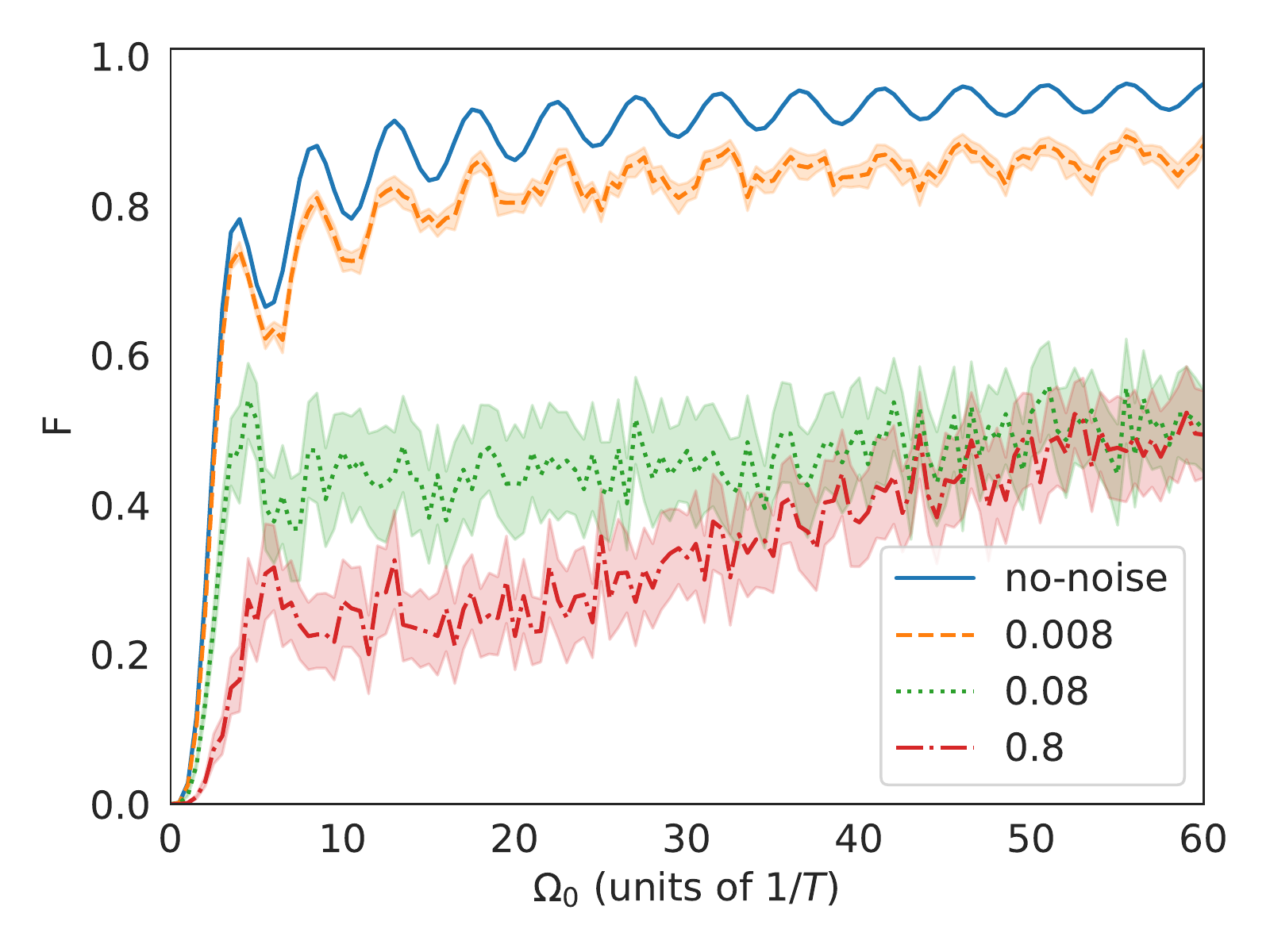}} &
        \subfigure[$\ $]{
	            \label{fig:sa_gaussian251}
	            \includegraphics[width=.37\linewidth]{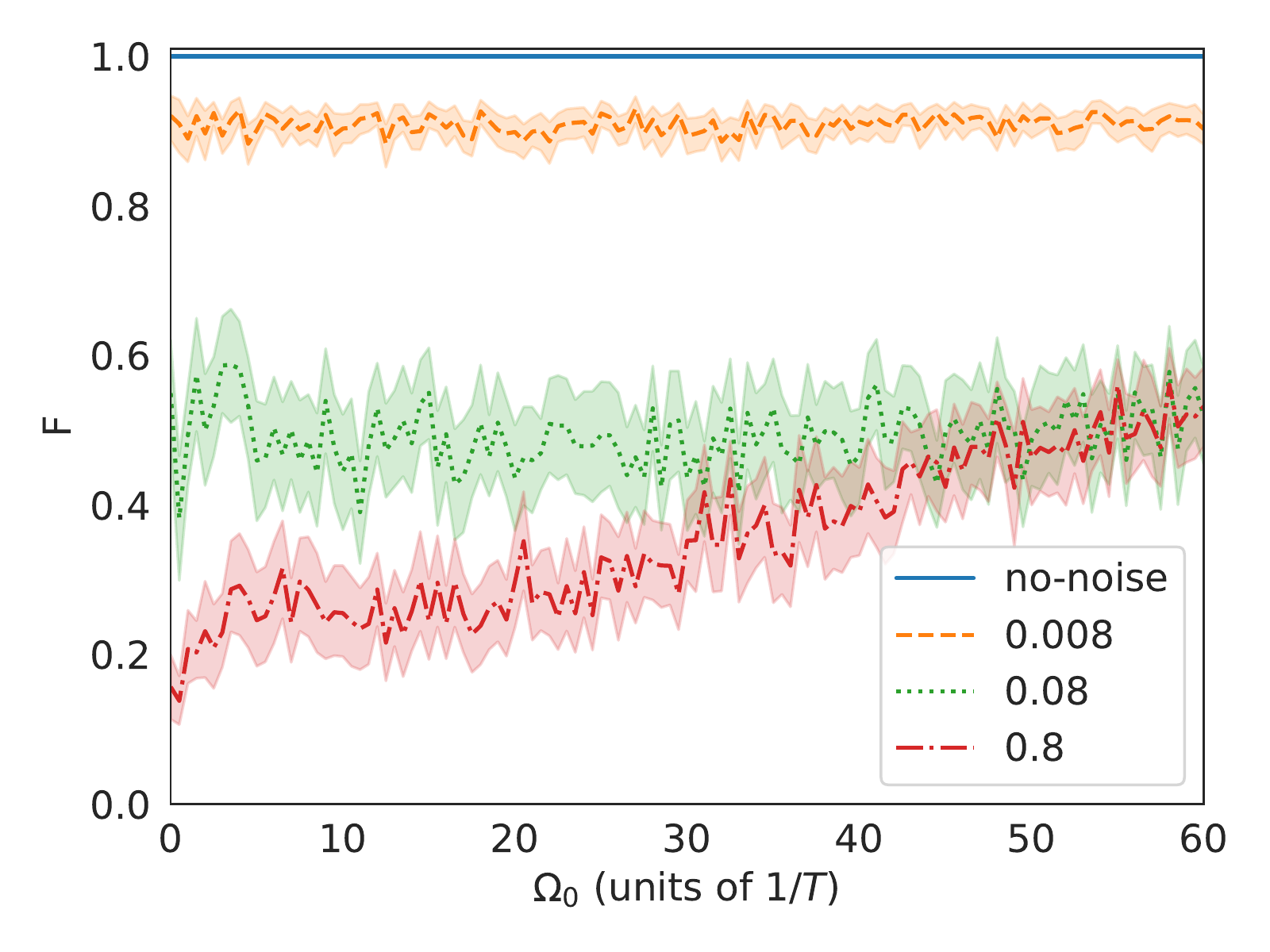}} \\
        \subfigure[$\ $]{
	            \label{fig:gaussian254}
	            \includegraphics[width=.37\linewidth]{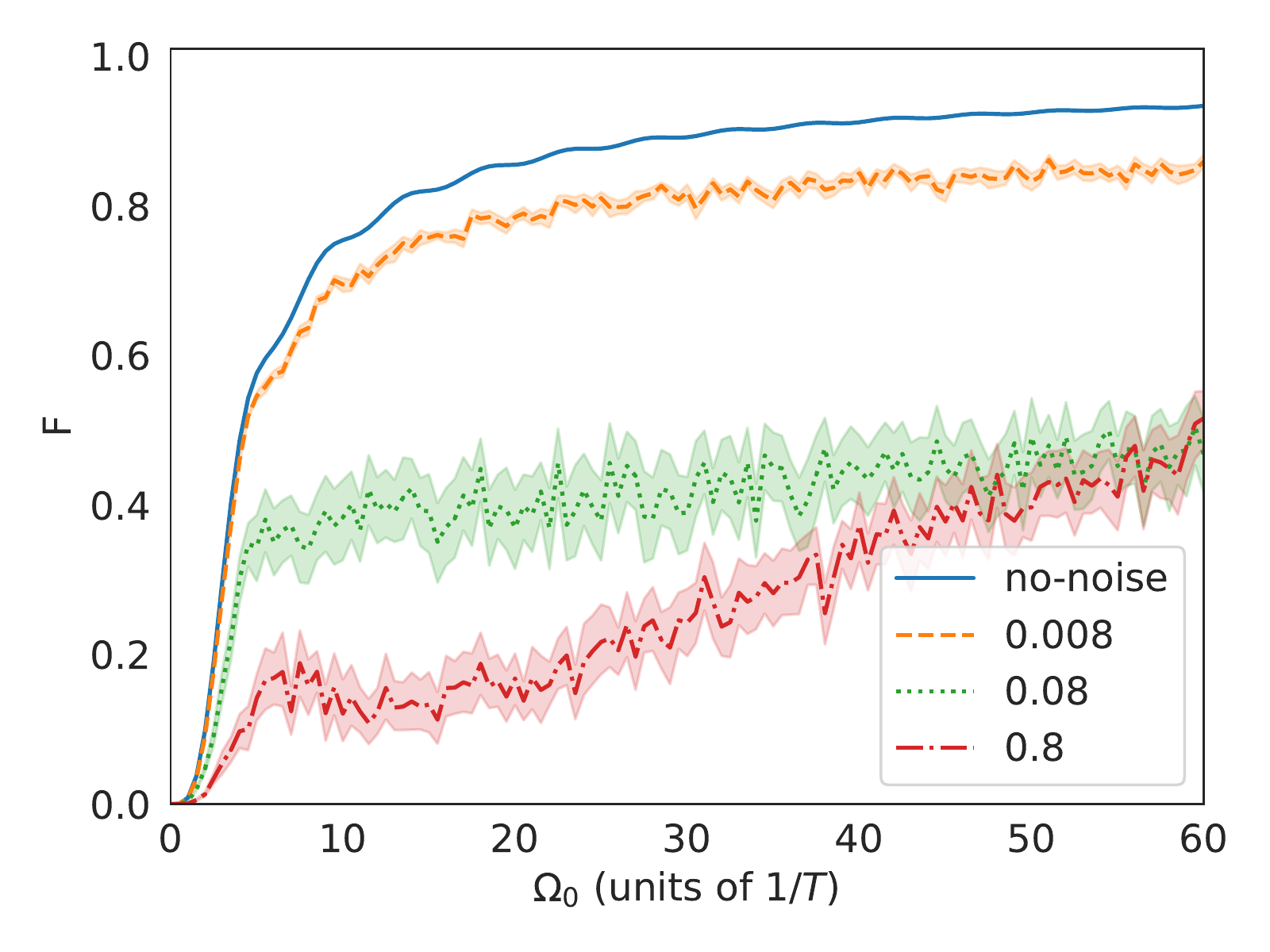}} &
        \subfigure[$\ $]{
	            \label{fig:sa_gaussian254}
	            \includegraphics[width=.37\linewidth]{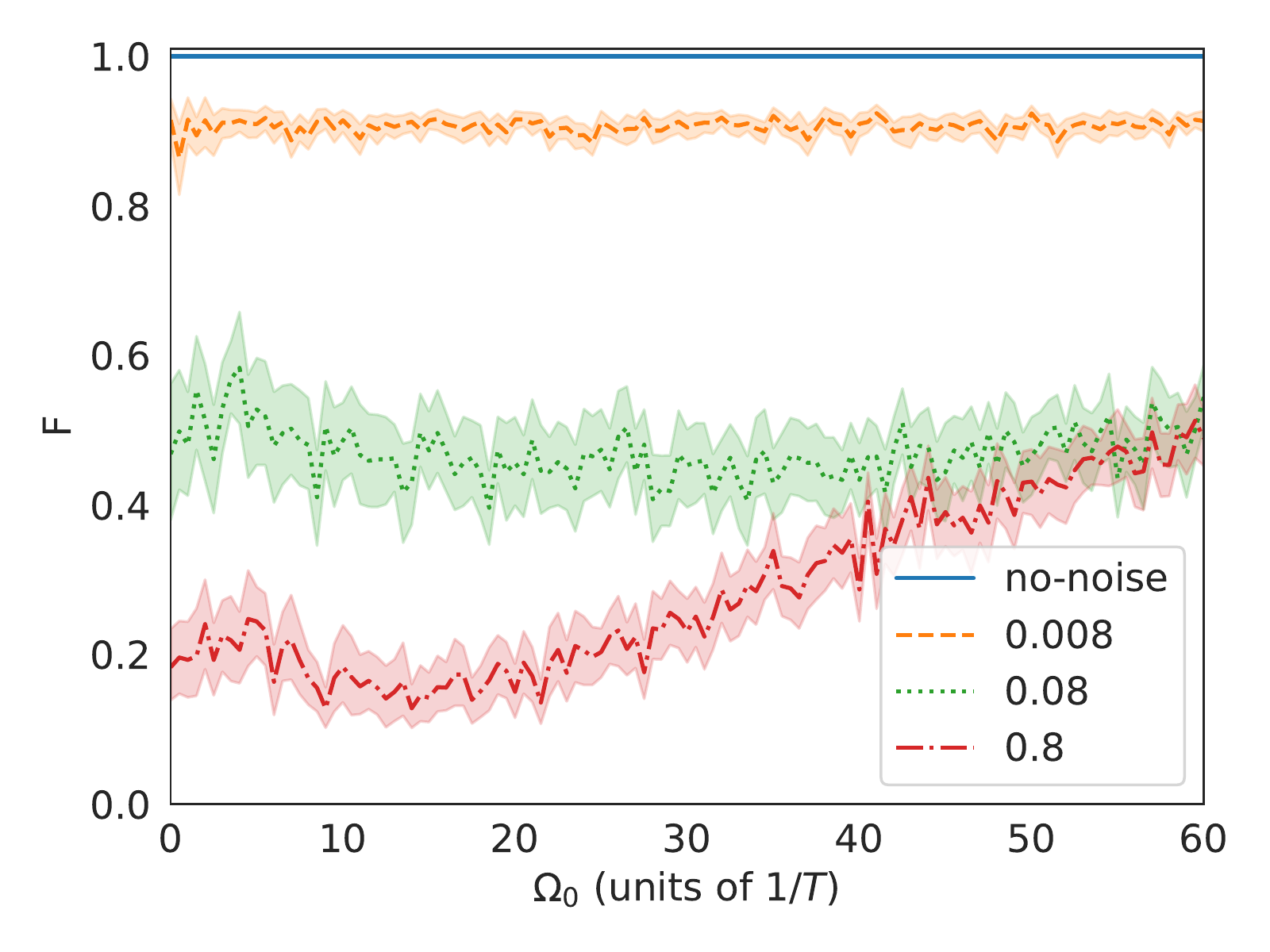}} \\
        \subfigure[$\ $]{
	            \label{fig:gaussian2510}
	            \includegraphics[width=.37\linewidth]{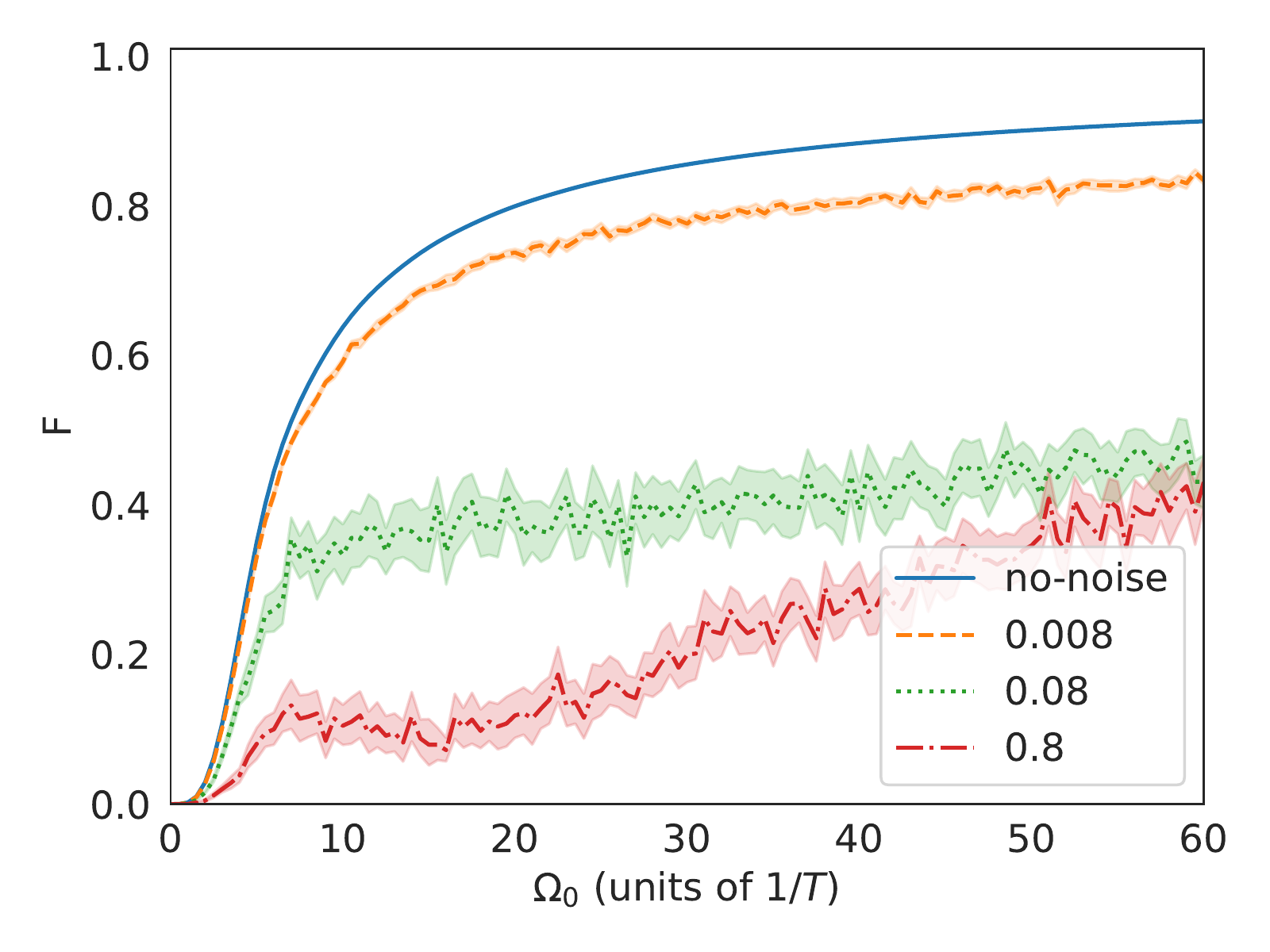}} &
        \subfigure[$\ $]{
	            \label{fig:sa_gaussian2510}
	            \includegraphics[width=.37\linewidth]{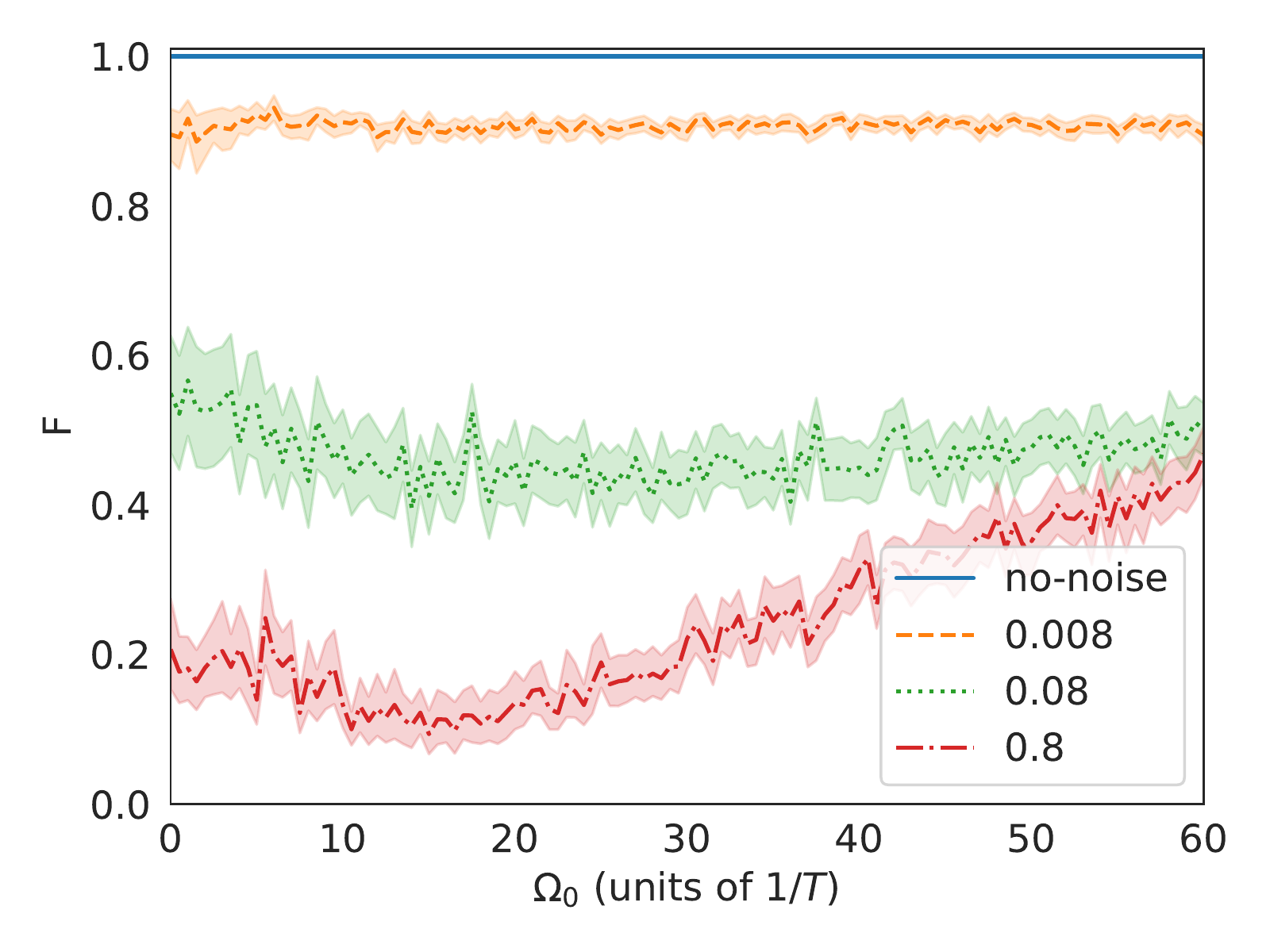}}
		\end{tabular}
\caption{(Color online) Fidelity versus $\Omega_0$ for Gaussian STIRAP and SA-STIRAP with delay $\tau/T=1/4$ between Stokes and pump pulses, for various values of the dephasing noise correlation time $\tau_c/T$, shown in the inset, and dissipation rates $\Gamma$: (a, c, e, g) STIRAP with $\Gamma=0, 1/T, 4/T, 10/T$, respectively, (b, d, f, h) SA-STIRAP with $\Gamma=0, 1/T, 4/T, 10/T$, respectively.}
\label{fig:gaussian25}
\end{figure*}

\begin{figure*}[t]
 \centering
		\begin{tabular}{cc}
     	\subfigure[$\ $]{
	            \label{fig:gaussian330}
	            \includegraphics[width=.37\linewidth]{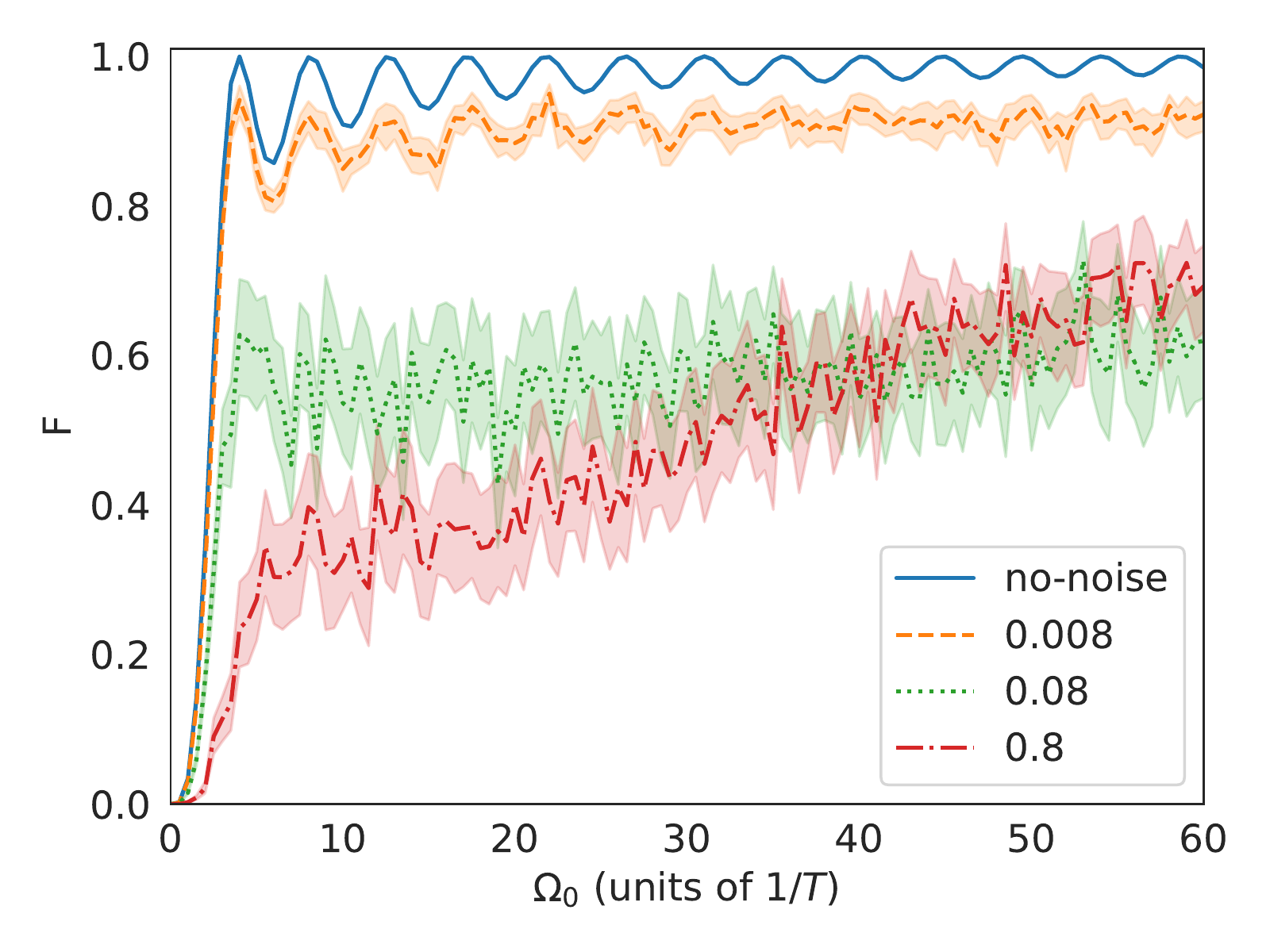}} &
        \subfigure[$\ $]{
	            \label{fig:sa_gaussian330}
	            \includegraphics[width=.37\linewidth]{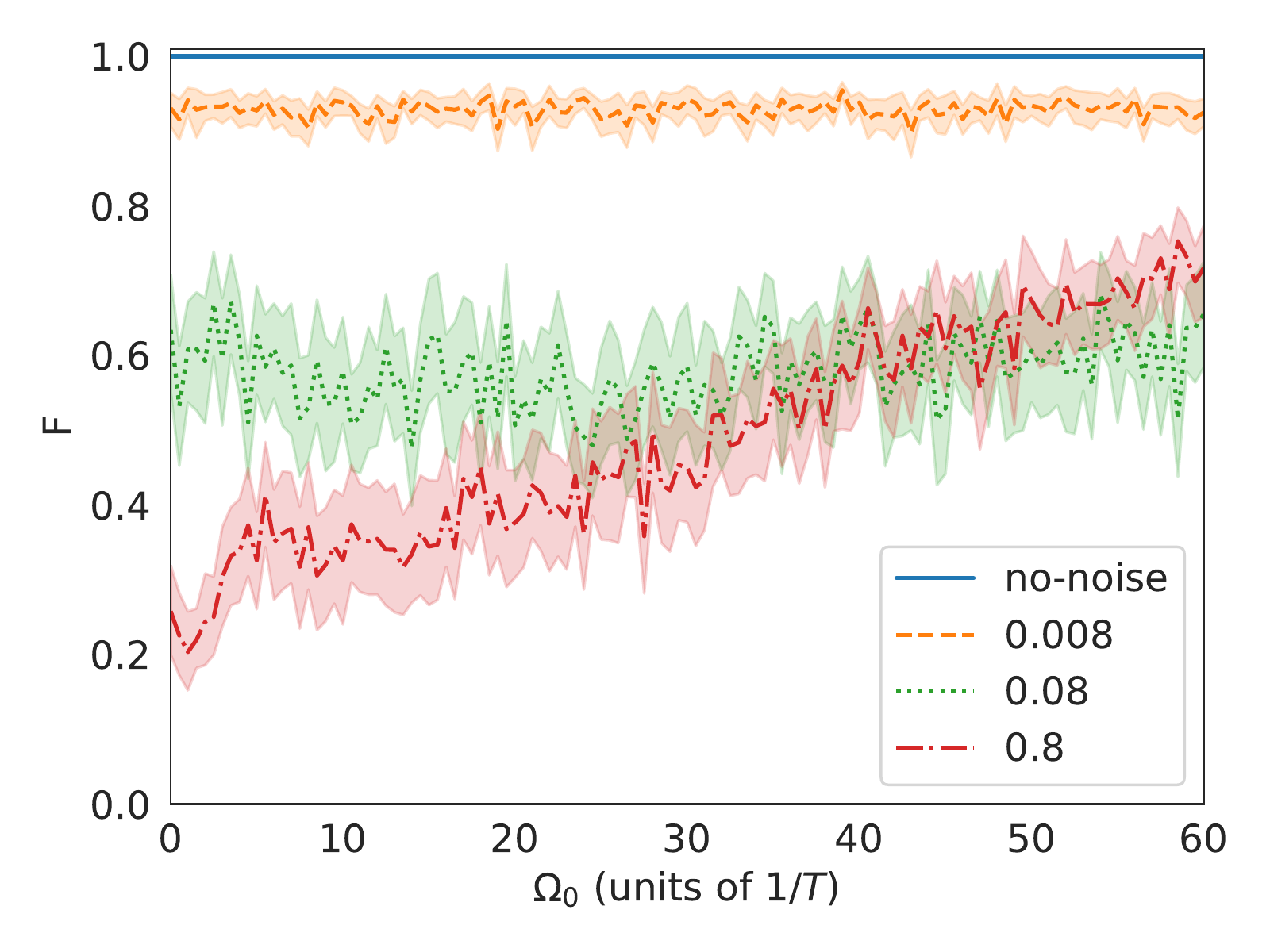}} \\
        \subfigure[$\ $]{
	            \label{fig:gaussian331}
	            \includegraphics[width=.37\linewidth]{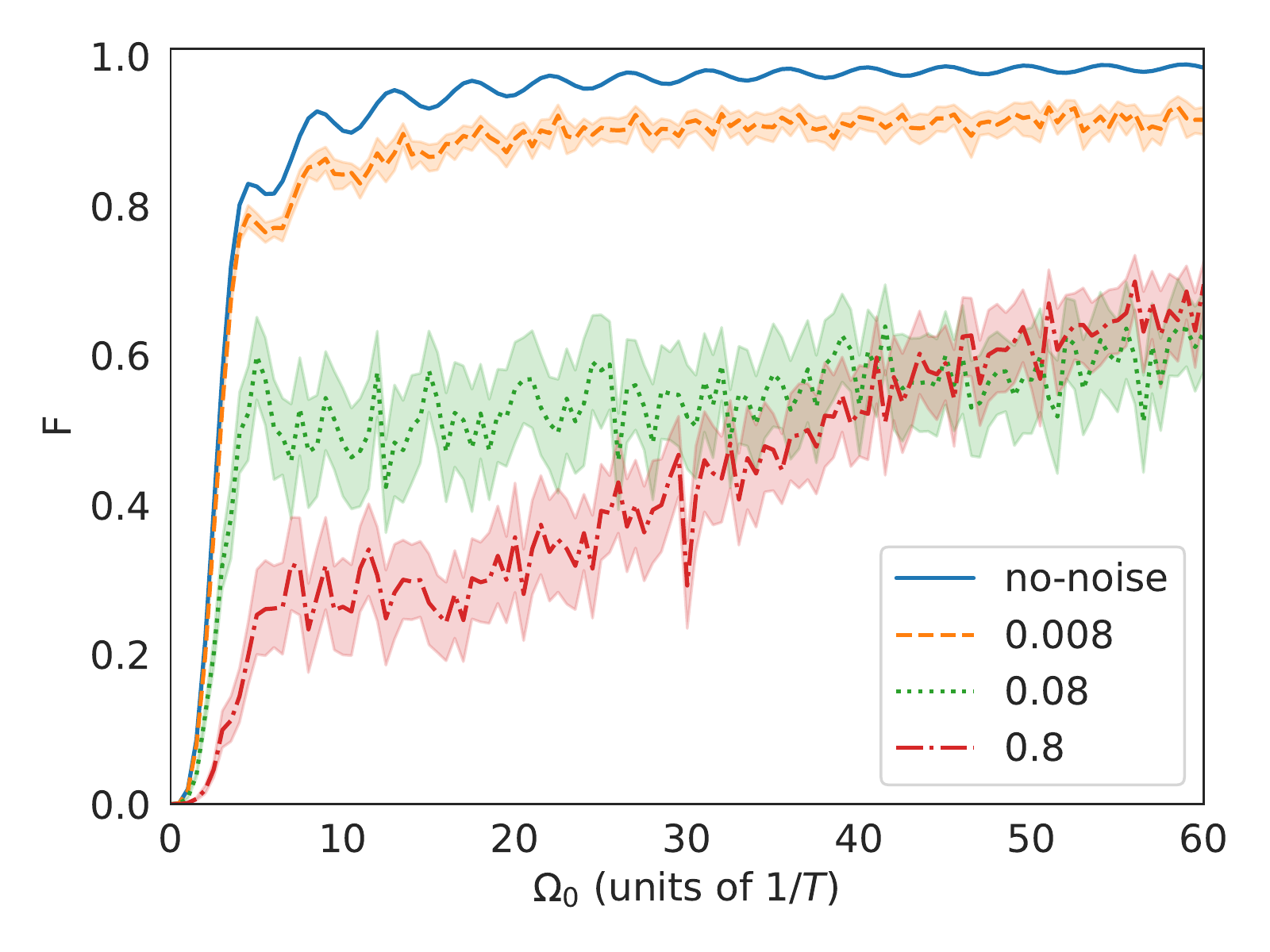}} &
        \subfigure[$\ $]{
	            \label{fig:sa_gaussian331}
	            \includegraphics[width=.37\linewidth]{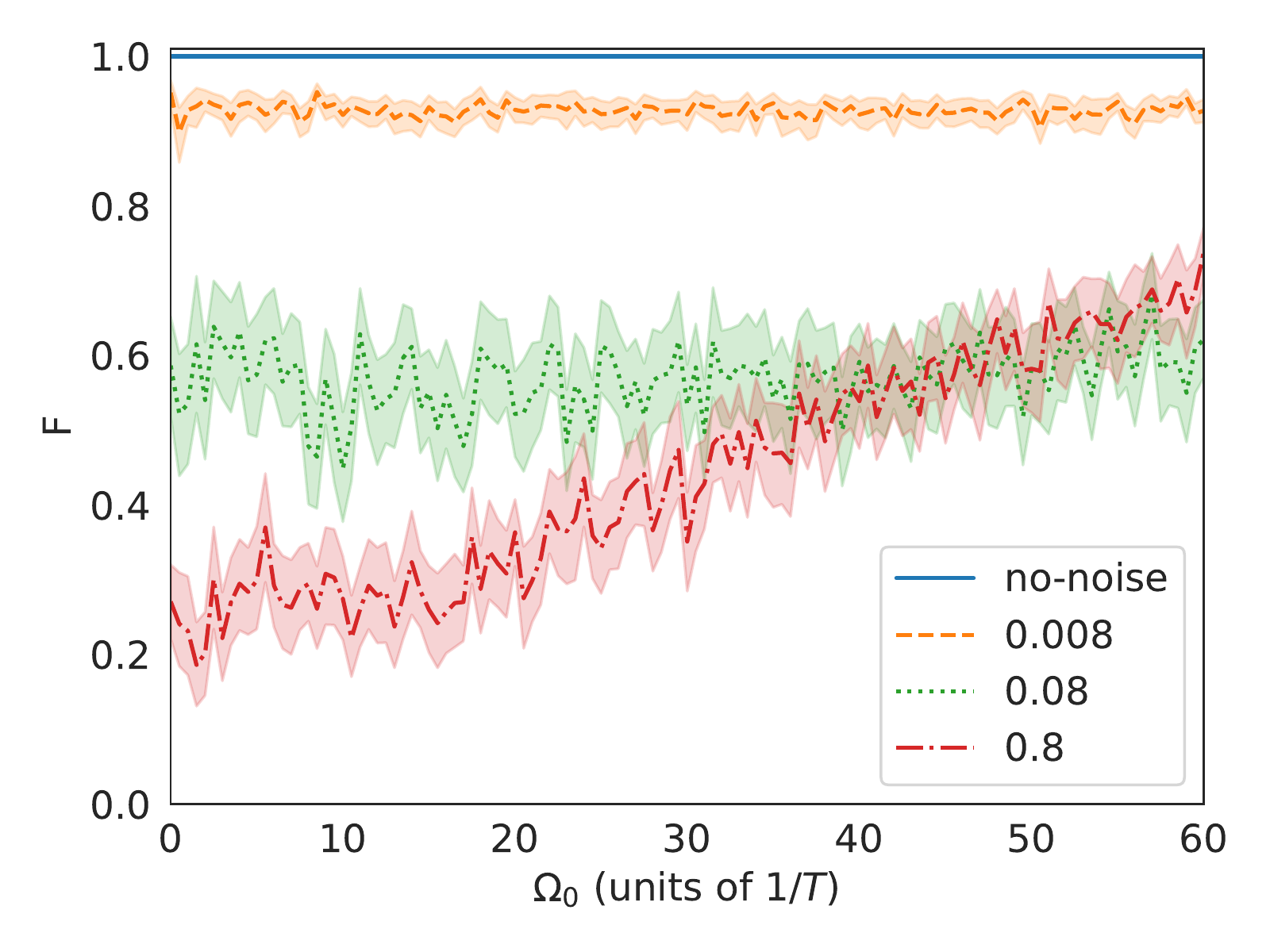}} \\
        \subfigure[$\ $]{
	            \label{fig:gaussian334}
	            \includegraphics[width=.37\linewidth]{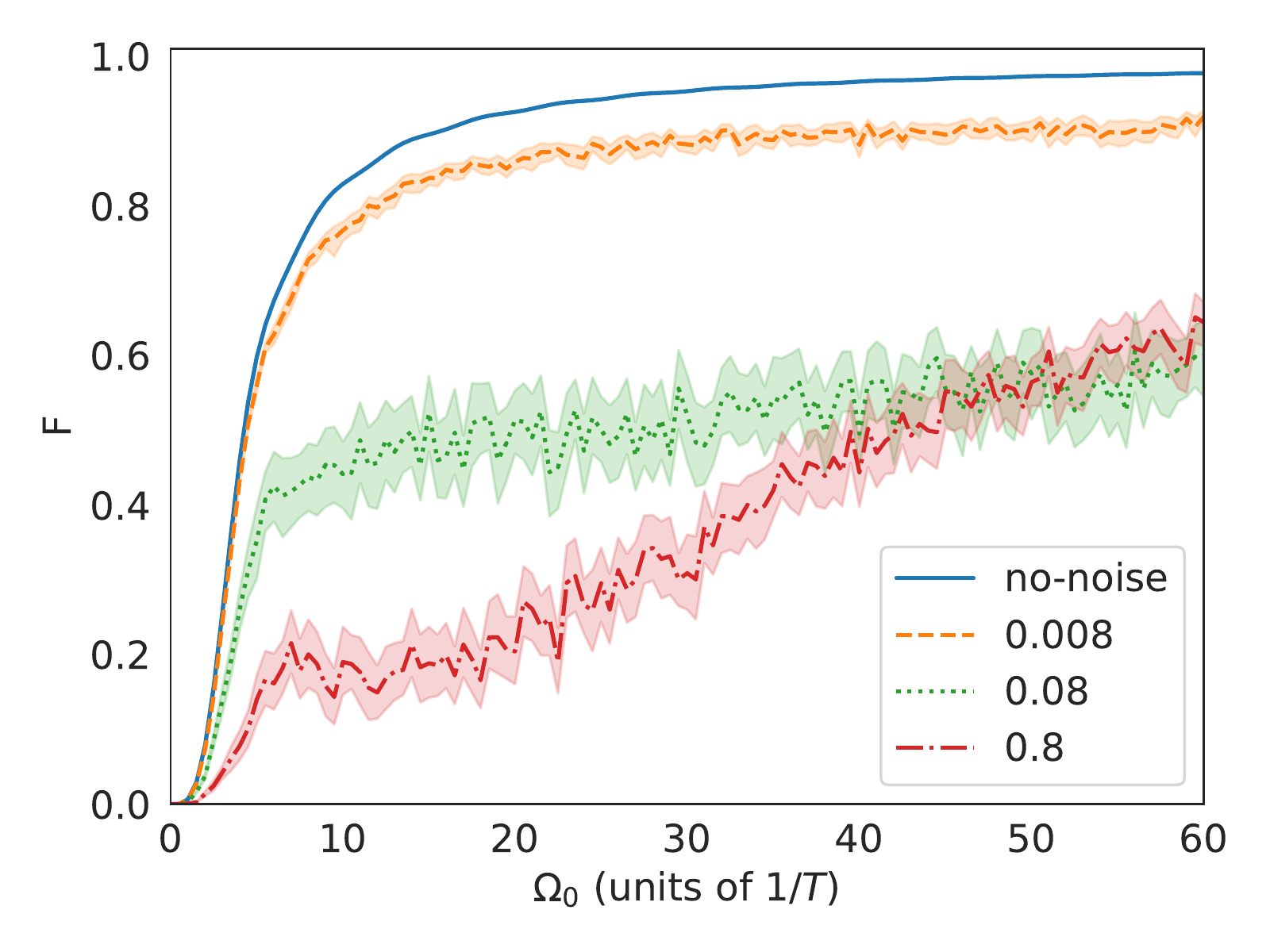}} &
        \subfigure[$\ $]{
	            \label{fig:sa_gaussian334}
	            \includegraphics[width=.37\linewidth]{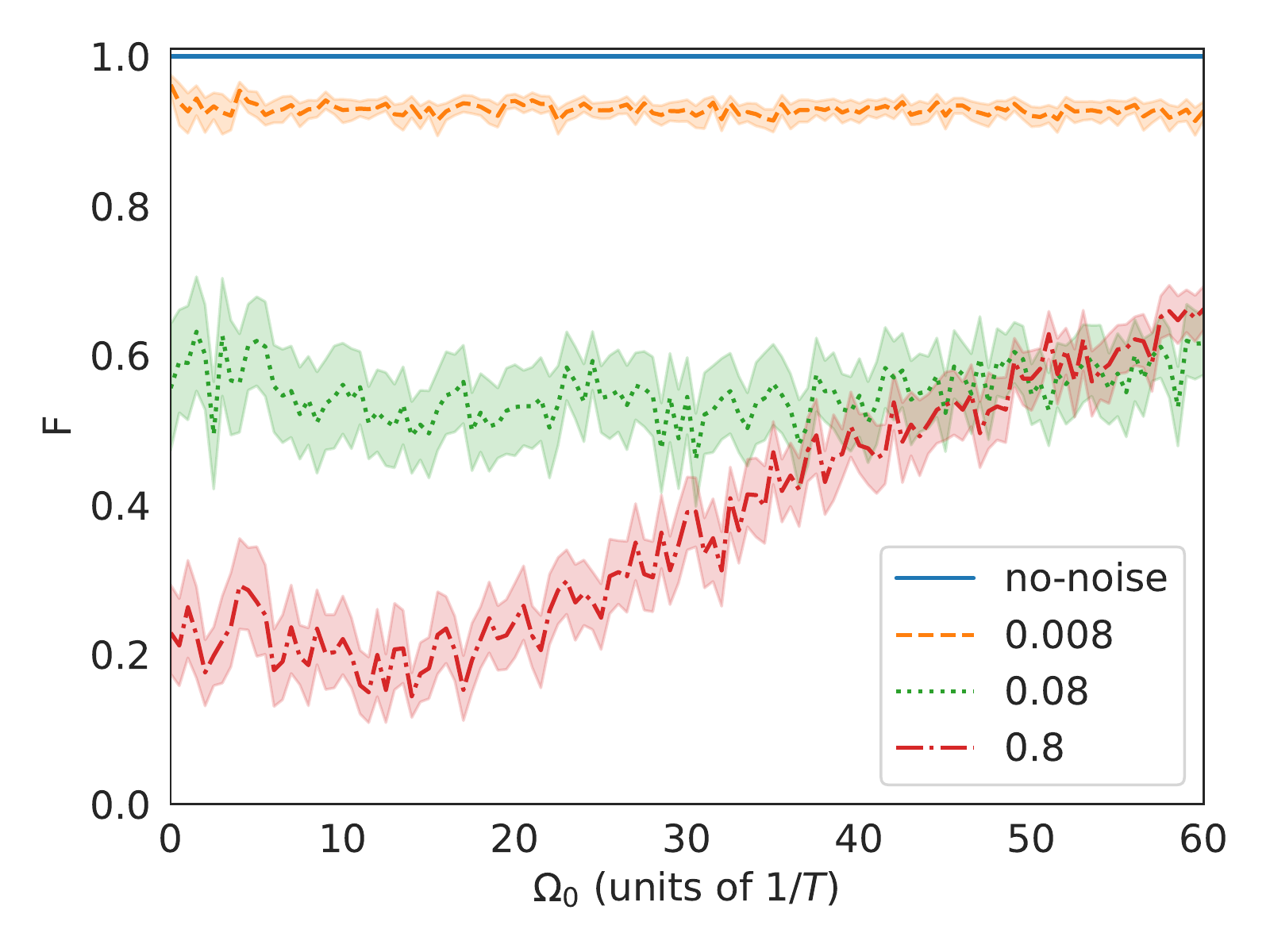}} \\
        \subfigure[$\ $]{
	            \label{fig:gaussian3310}
	            \includegraphics[width=.37\linewidth]{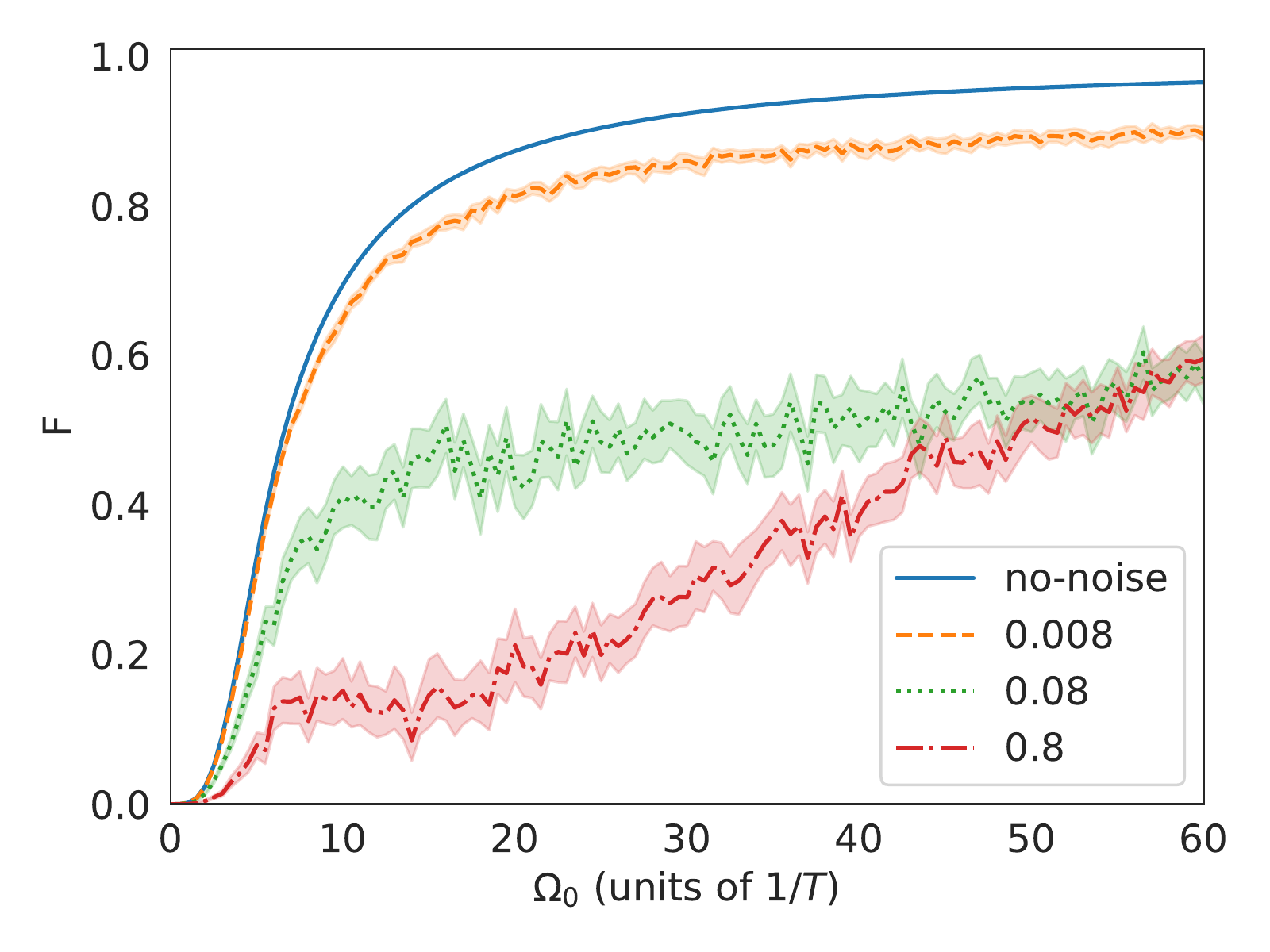}} &
        \subfigure[$\ $]{
	            \label{fig:sa_gaussian3310}
	            \includegraphics[width=.37\linewidth]{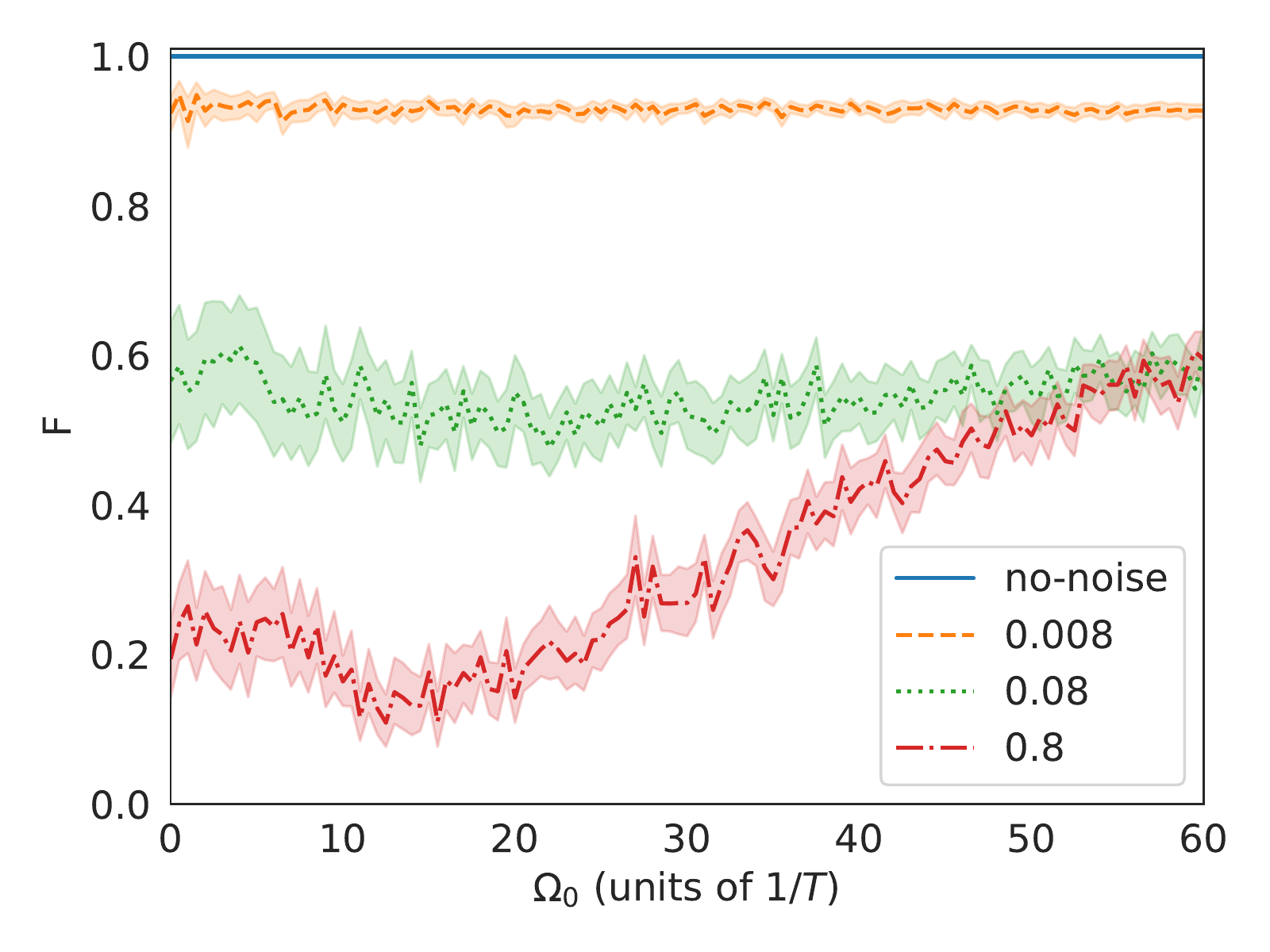}}
		\end{tabular}
\caption{(Color online) Fidelity versus $\Omega_0$ for Gaussian STIRAP and SA-STIRAP with delay $\tau/T=1/3$ between Stokes and pump pulses, for various values of the dephasing noise correlation time $\tau_c/T$, shown in the inset, and dissipation rates $\Gamma$: (a, c, e, g) STIRAP with $\Gamma=0, 1/T, 4/T, 10/T$, respectively, (b, d, f, h) SA-STIRAP with $\Gamma=0, 1/T, 4/T, 10/T$, respectively.}
\label{fig:gaussian33}
\end{figure*}

\begin{figure*}[t]
 \centering
		\begin{tabular}{cc}
     	\subfigure[$\ $]{
	            \label{fig:gaussian500}
	            \includegraphics[width=.37\linewidth]{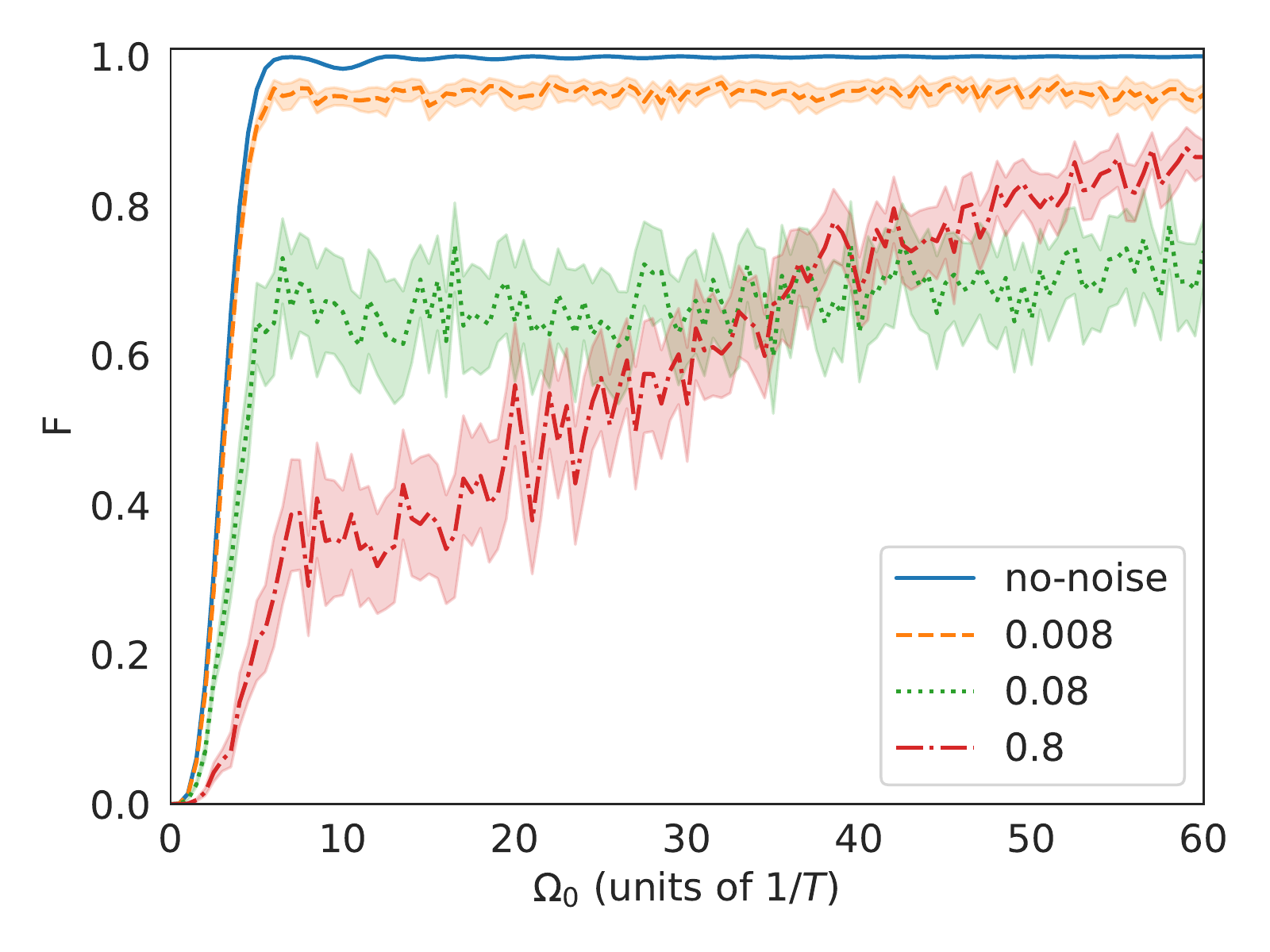}} &
        \subfigure[$\ $]{
	            \label{fig:sa_gaussian500}
	            \includegraphics[width=.37\linewidth]{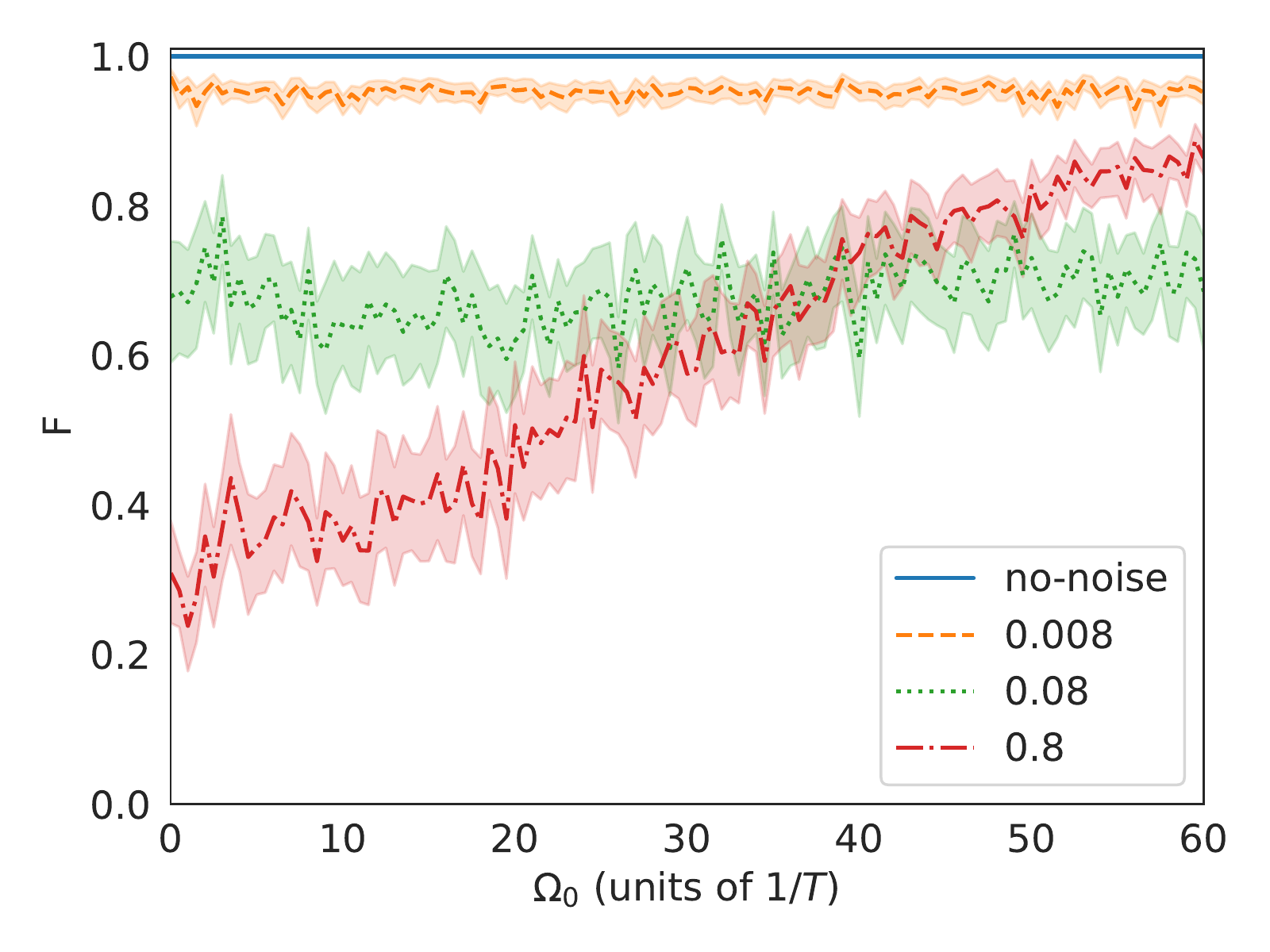}} \\
        \subfigure[$\ $]{
	            \label{fig:gaussian501}
	            \includegraphics[width=.37\linewidth]{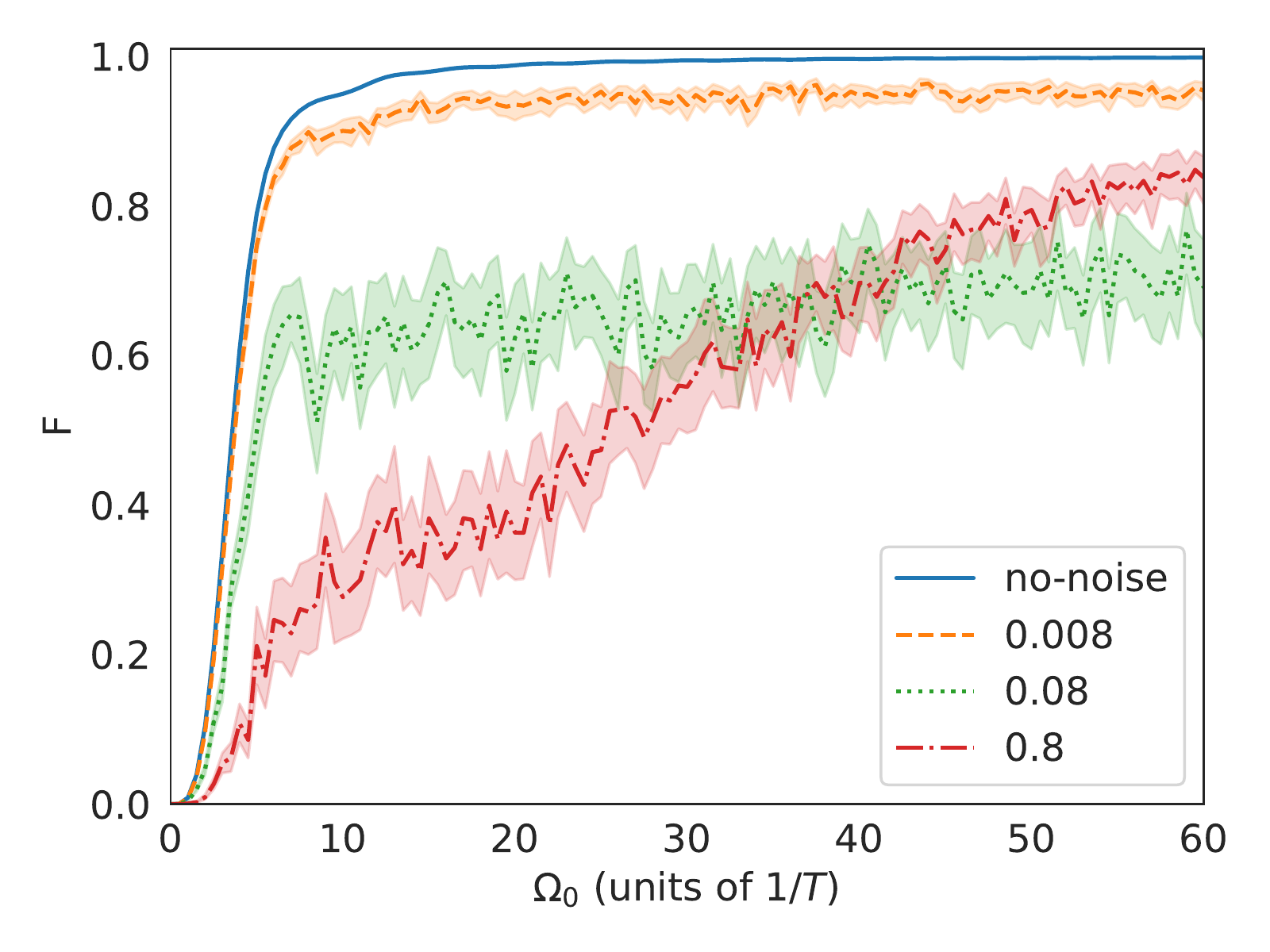}} &
        \subfigure[$\ $]{
	            \label{fig:sa_gaussian501}
	            \includegraphics[width=.37\linewidth]{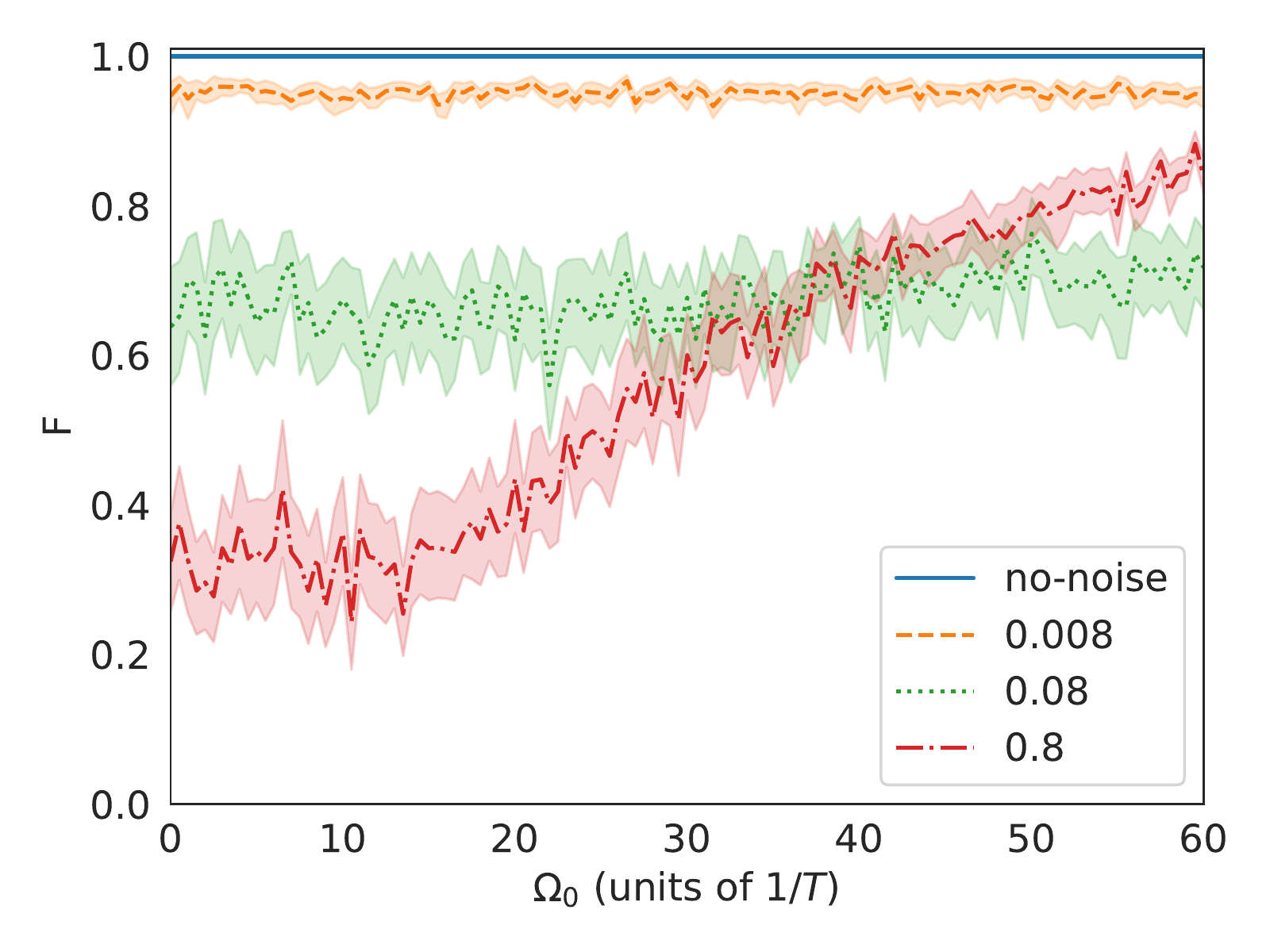}} \\
        \subfigure[$\ $]{
	            \label{fig:gaussian504}
	            \includegraphics[width=.37\linewidth]{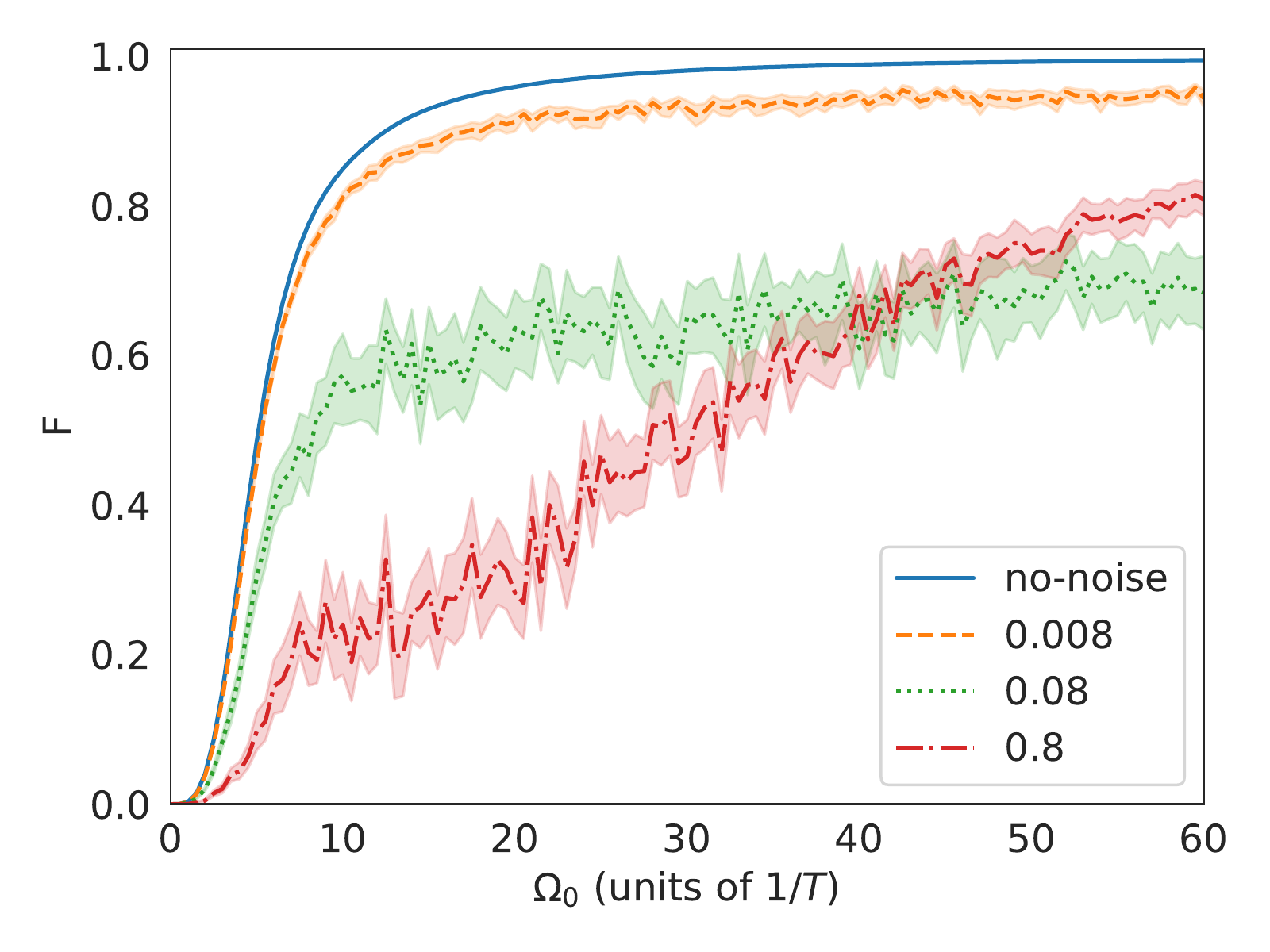}} &
        \subfigure[$\ $]{
	            \label{fig:sa_gaussian504}
	            \includegraphics[width=.37\linewidth]{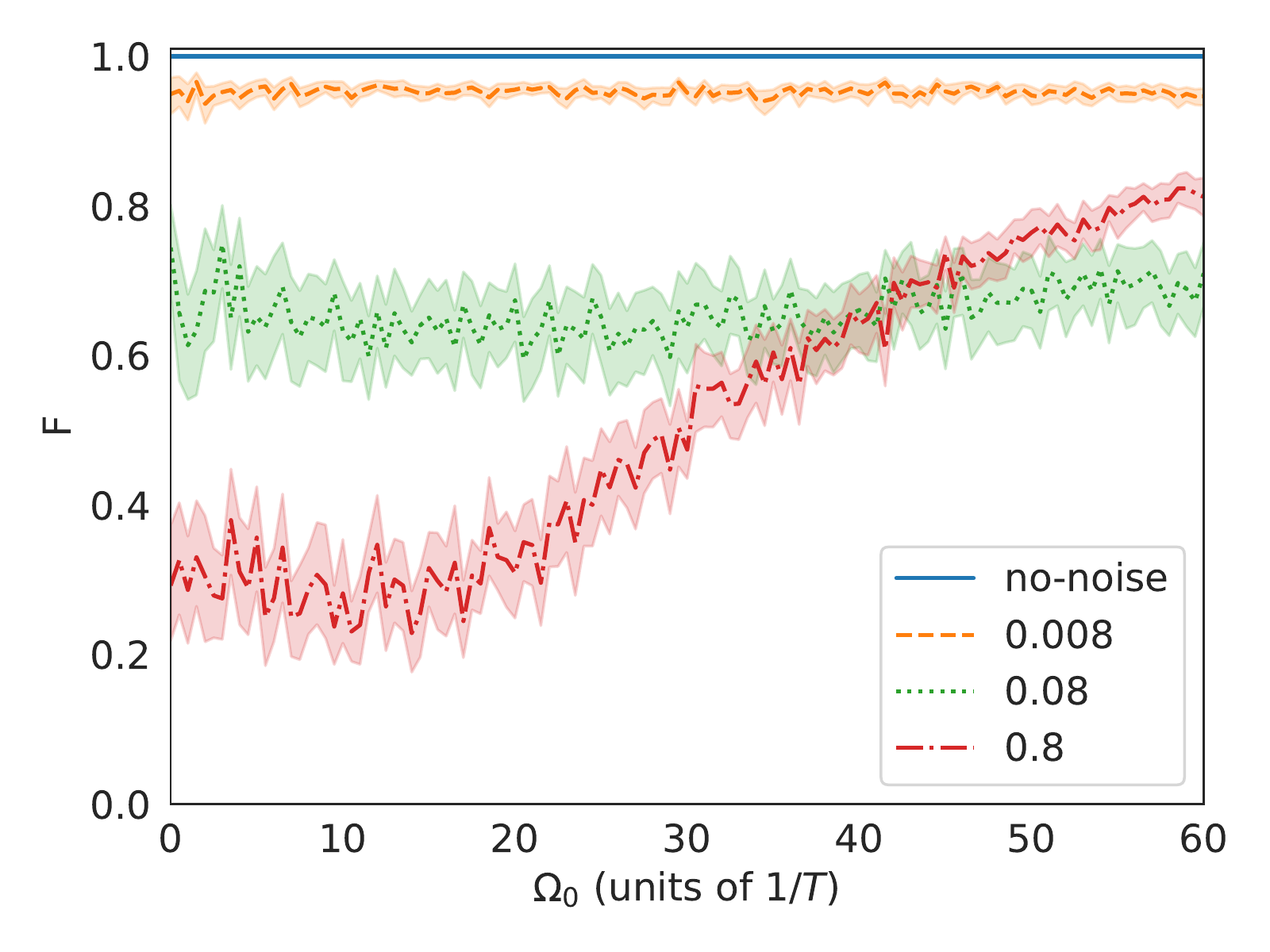}} \\
        \subfigure[$\ $]{
	            \label{fig:gaussian5010}
	            \includegraphics[width=.37\linewidth]{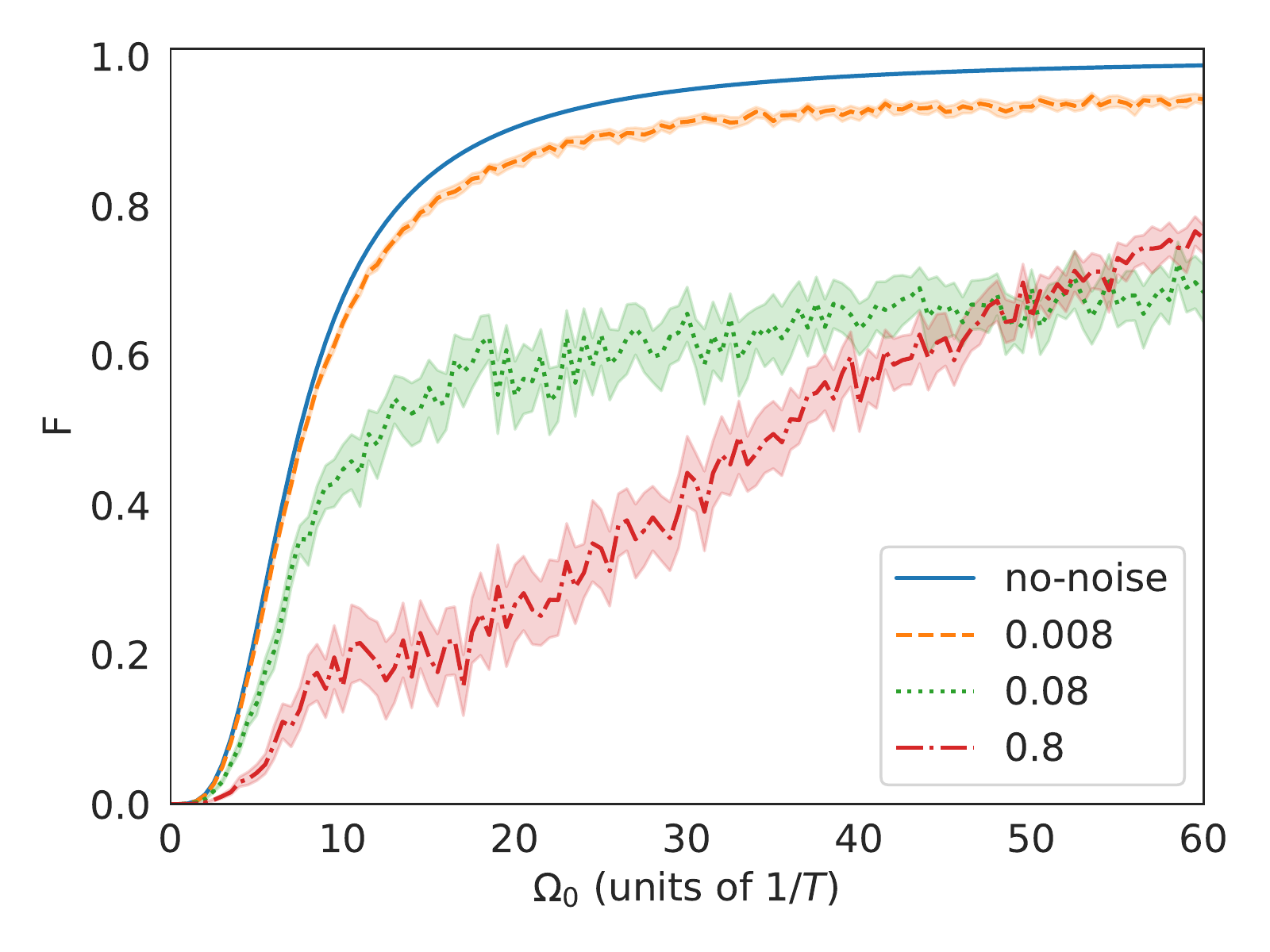}} &
        \subfigure[$\ $]{
	            \label{fig:sa_gaussian5010}
	            \includegraphics[width=.37\linewidth]{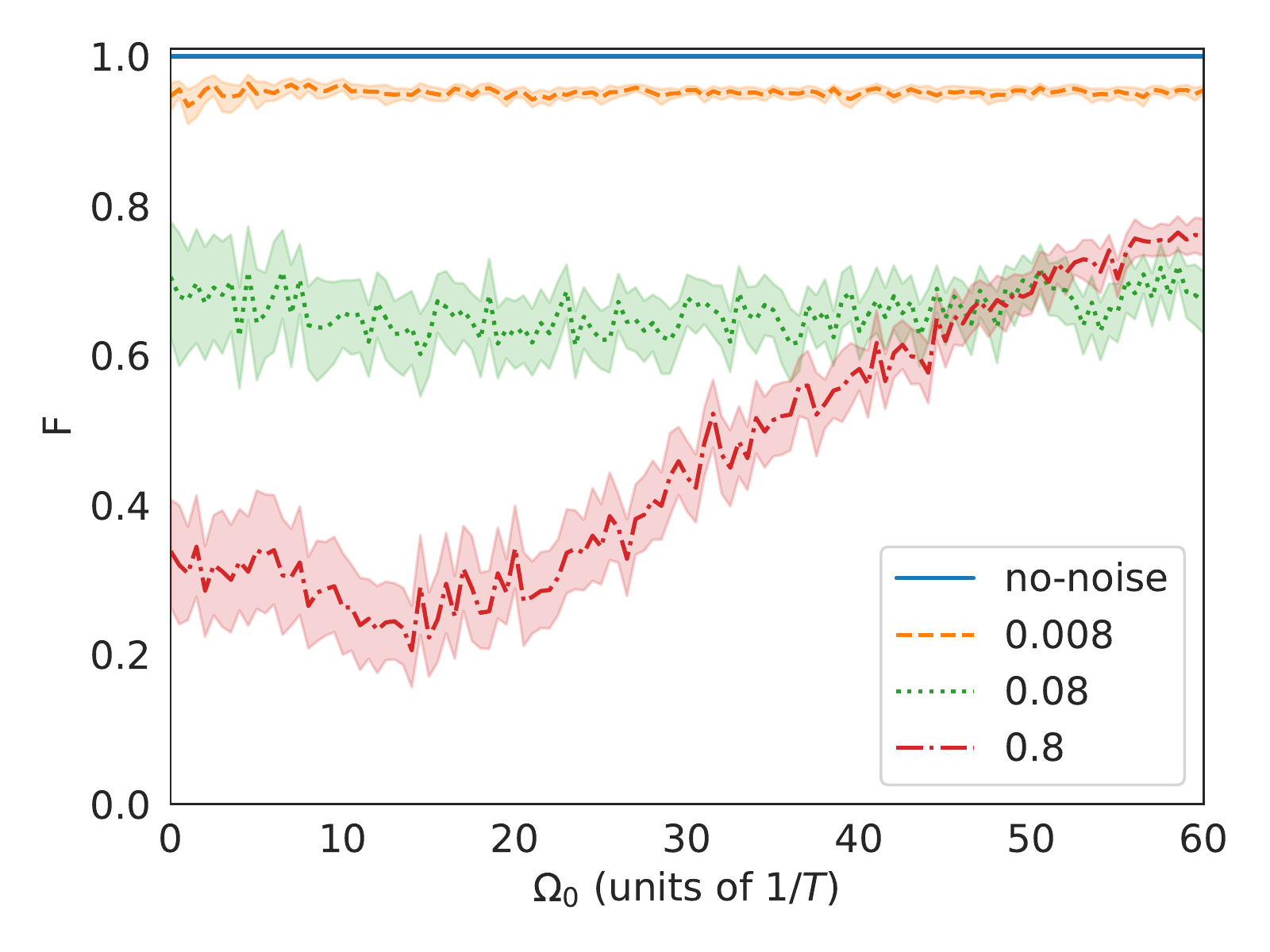}}
		\end{tabular}
\caption{(Color online) Fidelity versus $\Omega_0$ for Gaussian STIRAP and SA-STIRAP with delay $\tau/T=1/2$ between Stokes and pump pulses, for various values of the dephasing noise correlation time $\tau_c/T$, shown in the inset, and dissipation rates $\Gamma$: (a, c, e, g) STIRAP with $\Gamma=0, 1/T, 4/T, 10/T$, respectively, (b, d, f, h) SA-STIRAP with $\Gamma=0, 1/T, 4/T, 10/T$, respectively.}
\label{fig:gaussian50}
\end{figure*}

\begin{figure*}[t]
 \centering
		\begin{tabular}{cc}
     	\subfigure[$\ $]{
	            \label{fig:gaussian750}
	            \includegraphics[width=.37\linewidth]{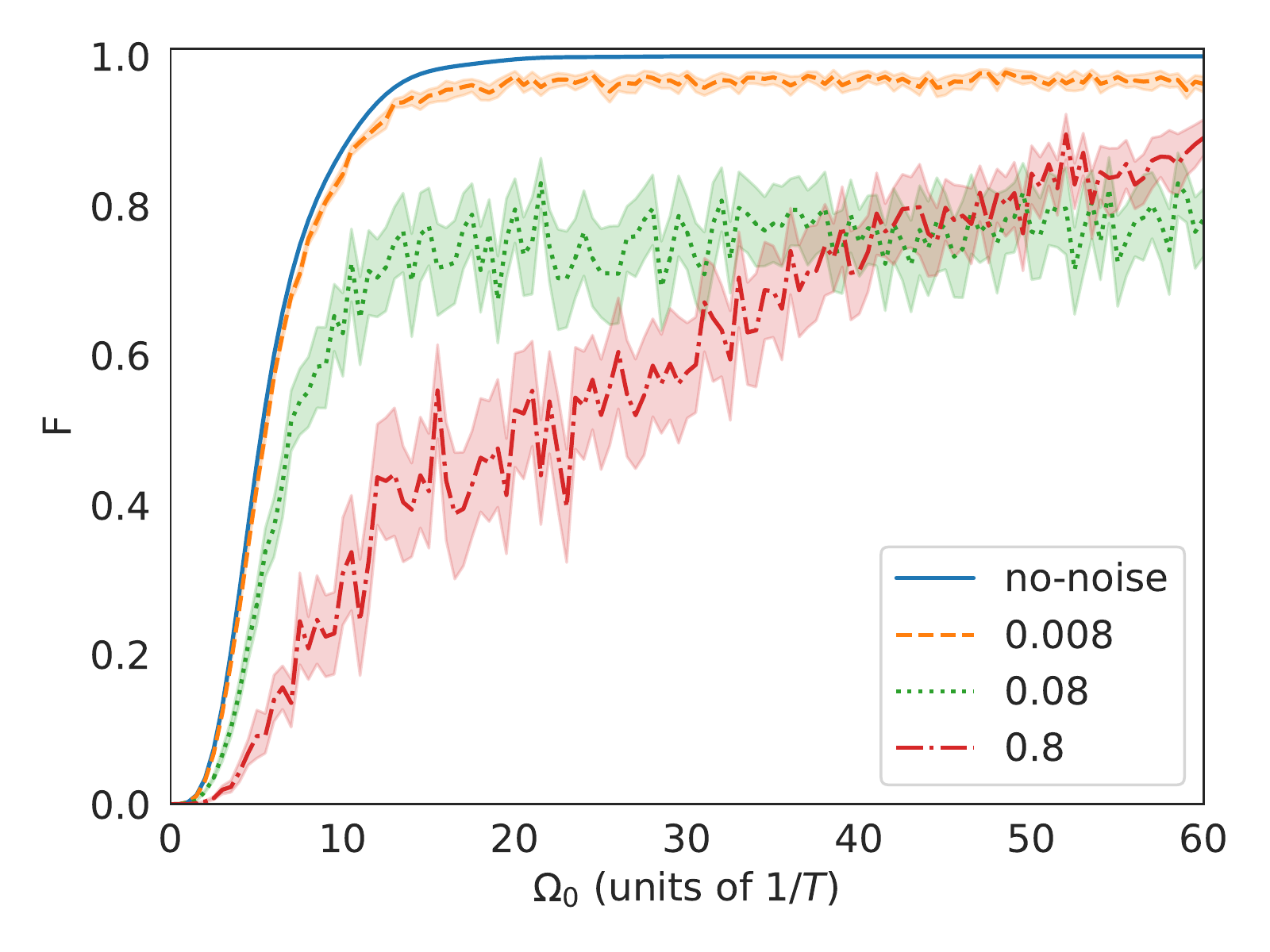}} &
        \subfigure[$\ $]{
	            \label{fig:sa_gaussian750}
	            \includegraphics[width=.37\linewidth]{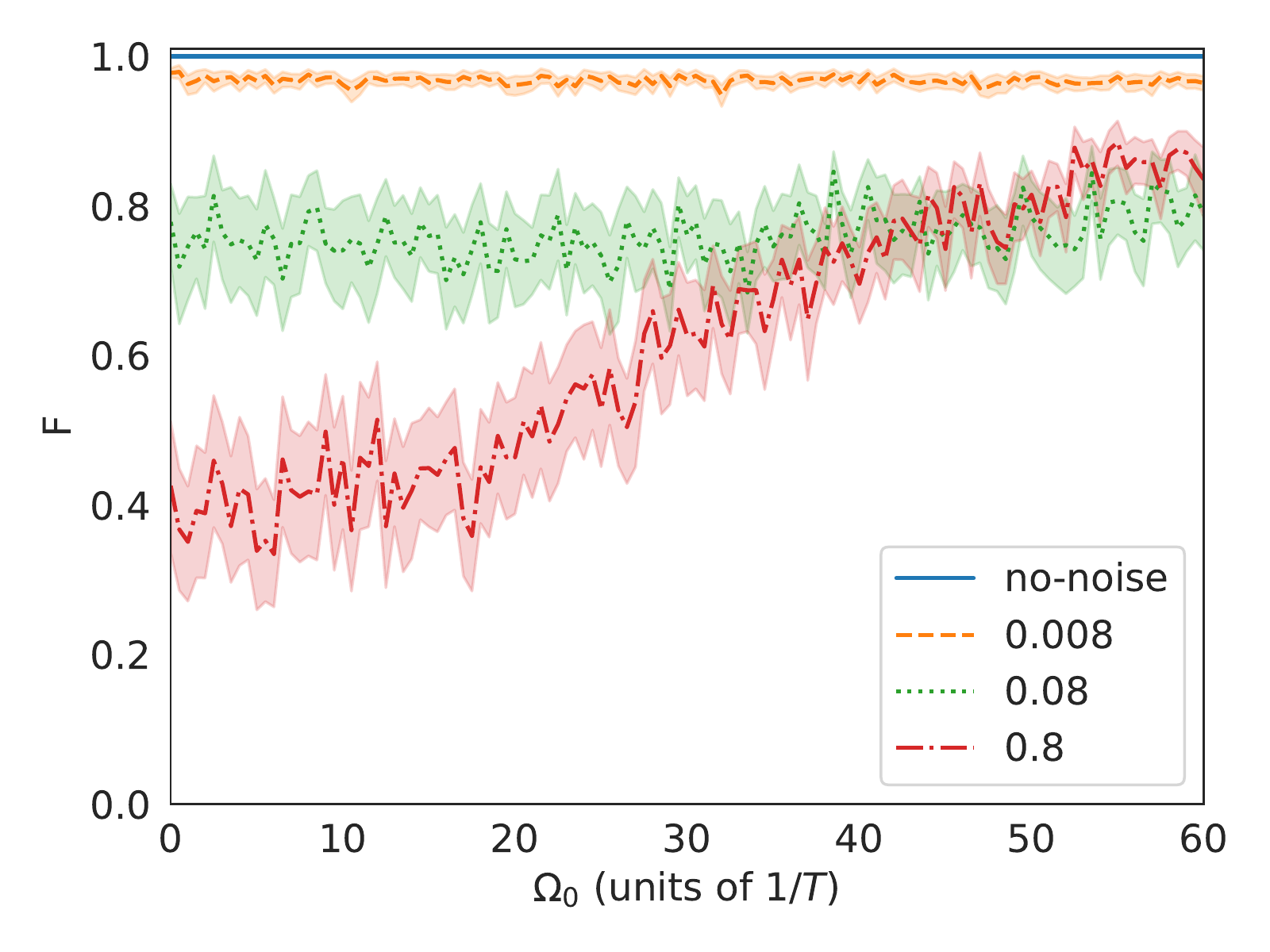}} \\
        \subfigure[$\ $]{
	            \label{fig:gaussian751}
	            \includegraphics[width=.37\linewidth]{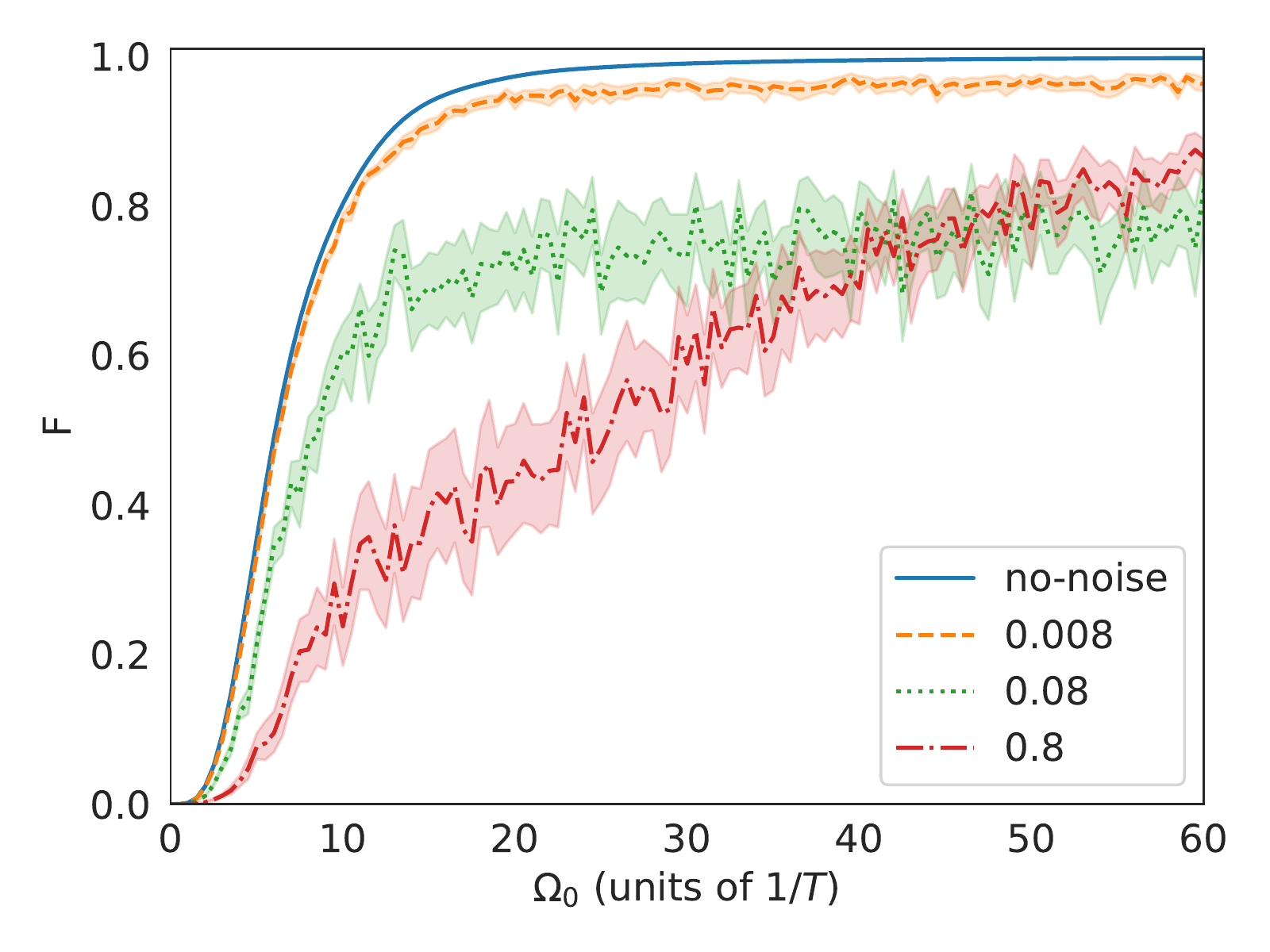}} &
        \subfigure[$\ $]{
	            \label{fig:sa_gaussian751}
	            \includegraphics[width=.37\linewidth]{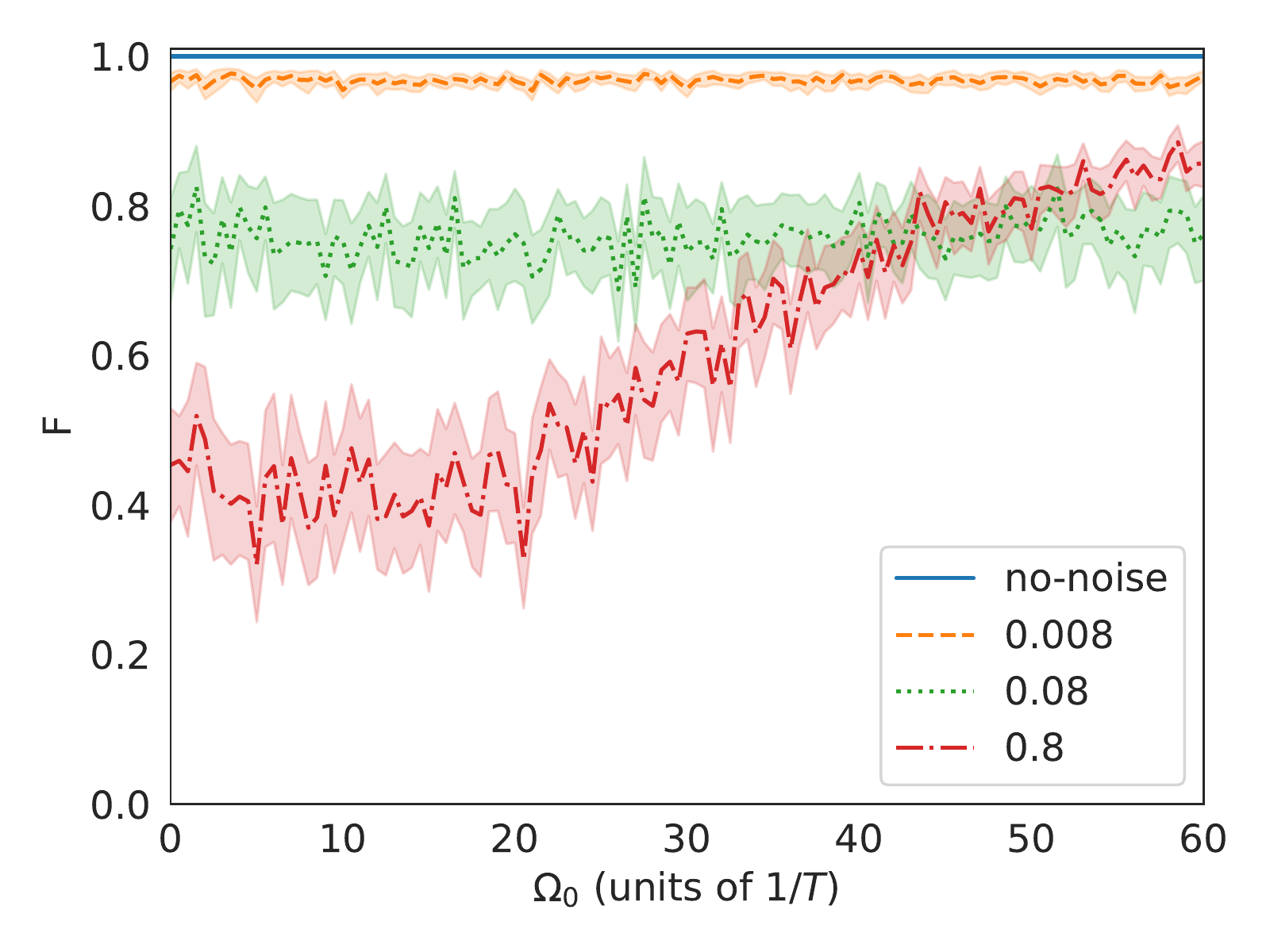}} \\
        \subfigure[$\ $]{
	            \label{fig:gaussian754}
	            \includegraphics[width=.37\linewidth]{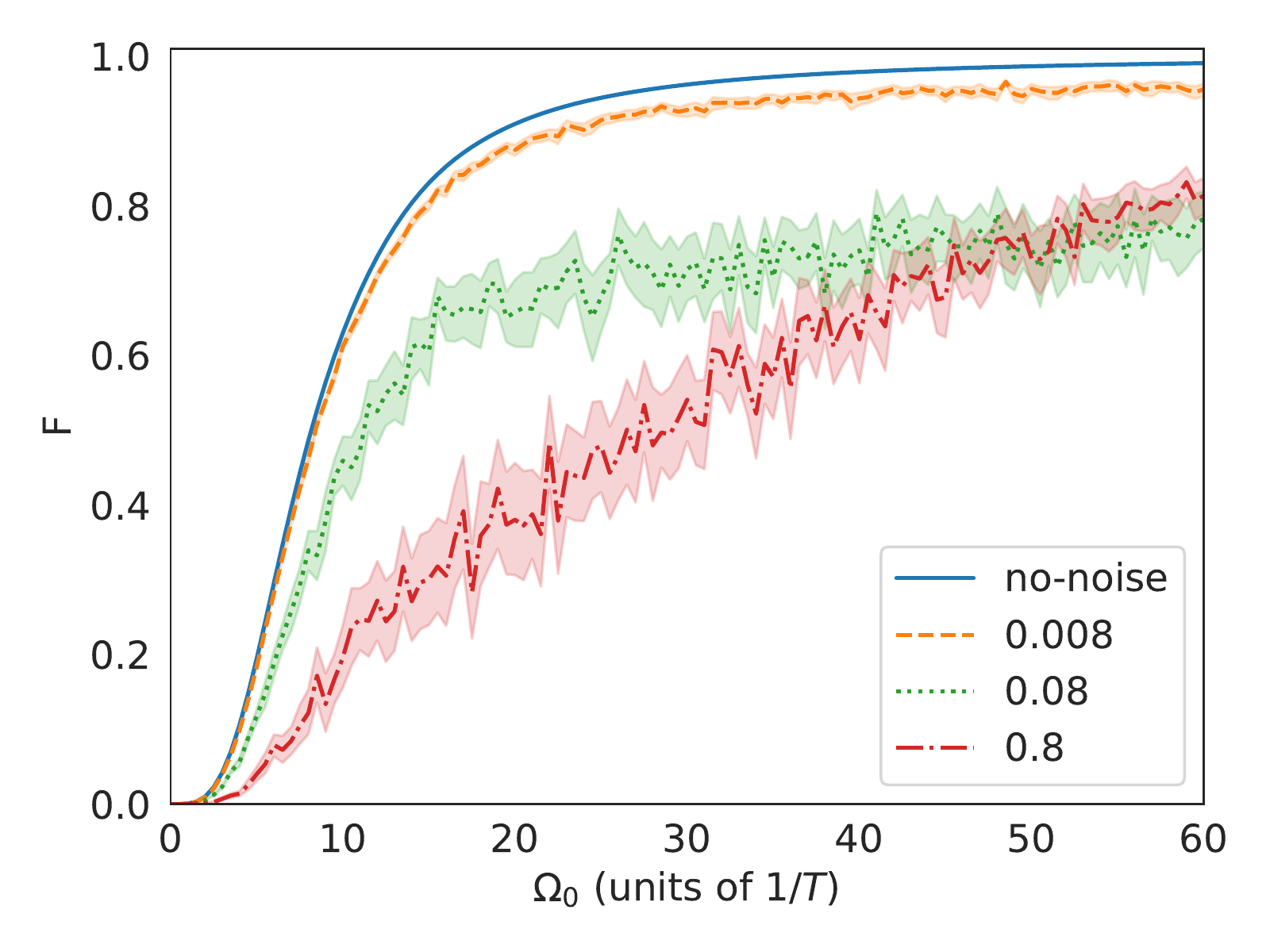}} &
        \subfigure[$\ $]{
	            \label{fig:sa_gaussian754}
	            \includegraphics[width=.37\linewidth]{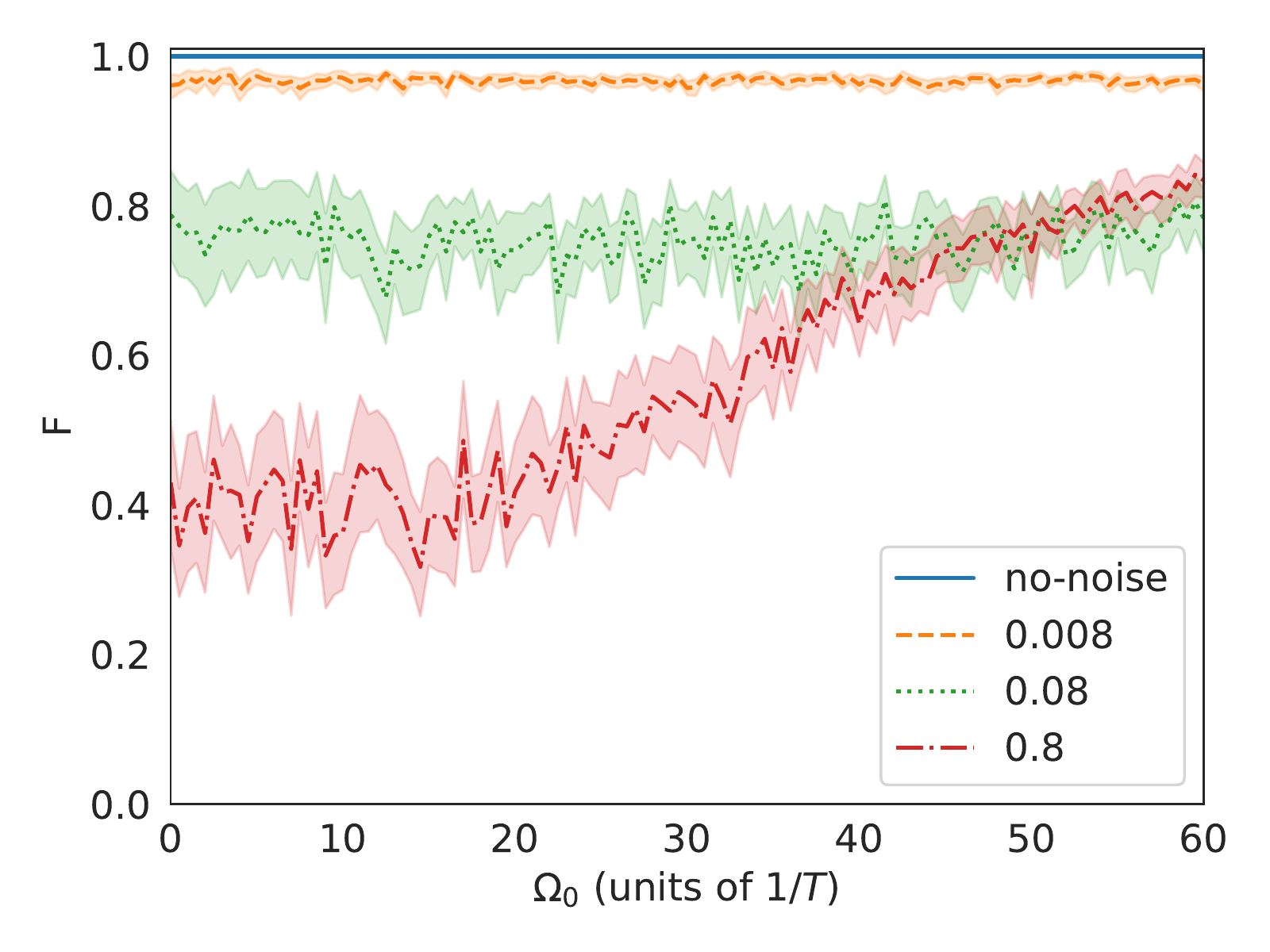}} \\
        \subfigure[$\ $]{
	            \label{fig:gaussian7510}
	            \includegraphics[width=.37\linewidth]{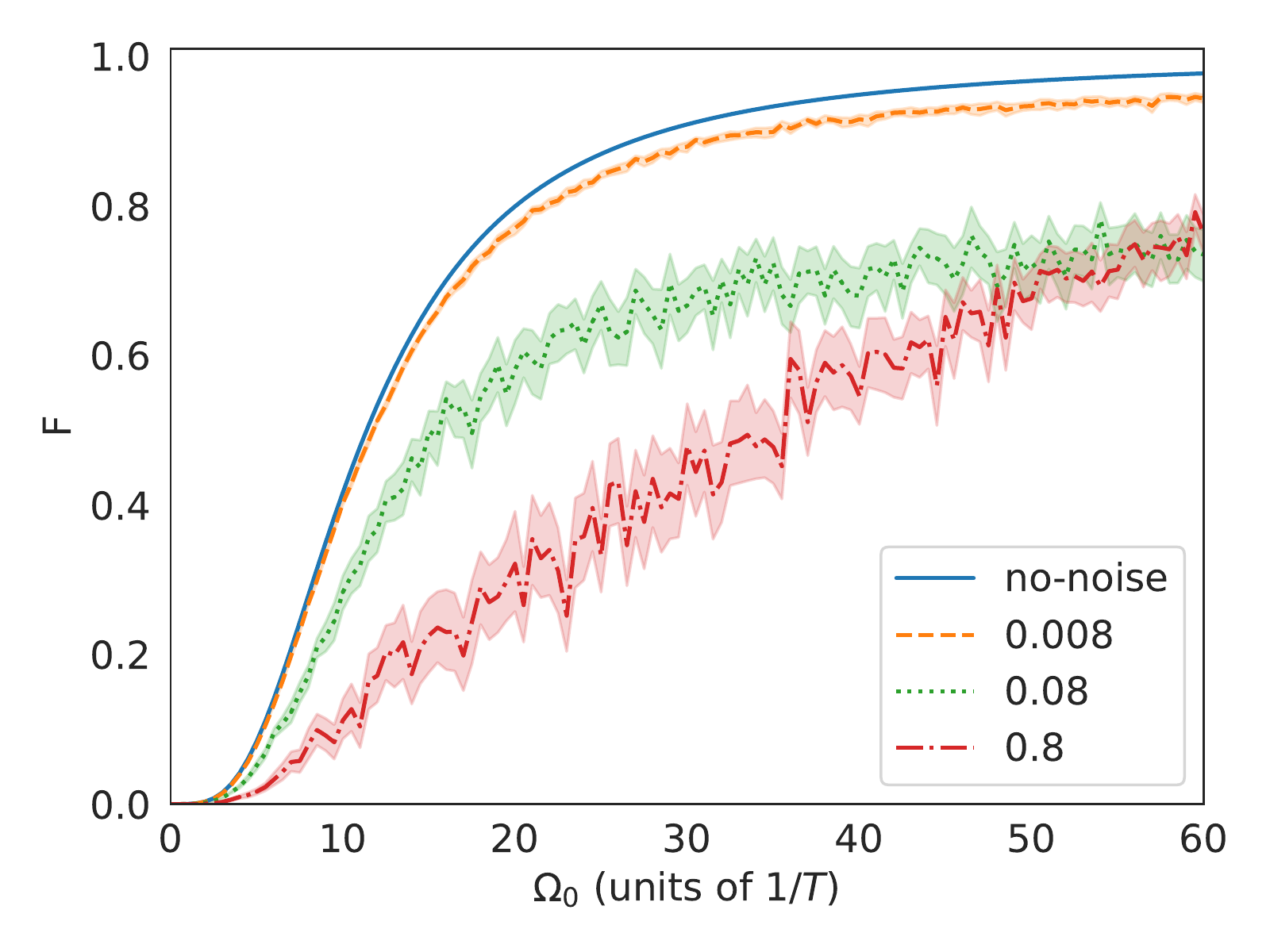}} &
        \subfigure[$\ $]{
	            \label{fig:sa_gaussian7510}
	            \includegraphics[width=.37\linewidth]{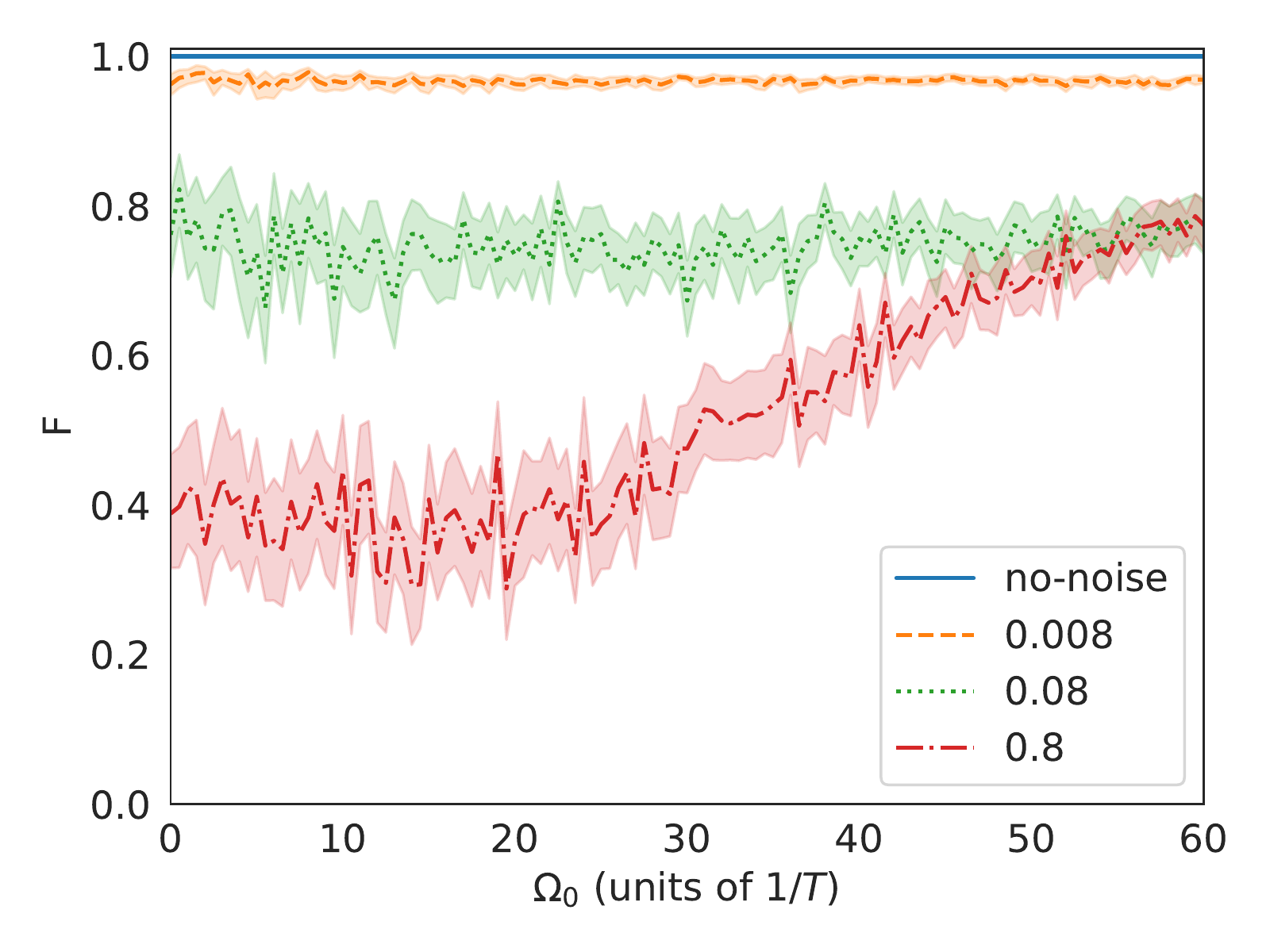}}
		\end{tabular}
\caption{(Color online) Fidelity versus $\Omega_0$ for Gaussian STIRAP and SA-STIRAP with delay $\tau/T=3/4$ between Stokes and pump pulses, for various values of the dephasing noise correlation time $\tau_c/T$, shown in the inset, and dissipation rates $\Gamma$: (a, c, e, g) STIRAP with $\Gamma=0, 1/T, 4/T, 10/T$, respectively, (b, d, f, h) SA-STIRAP with $\Gamma=0, 1/T, 4/T, 10/T$, respectively.}
\label{fig:gaussian75}
\end{figure*}

\begin{figure*}[t]
 \centering
		\begin{tabular}{cc}
     	\subfigure[$\ $]{
	            \label{fig:sin500}
	            \includegraphics[width=.37\linewidth]{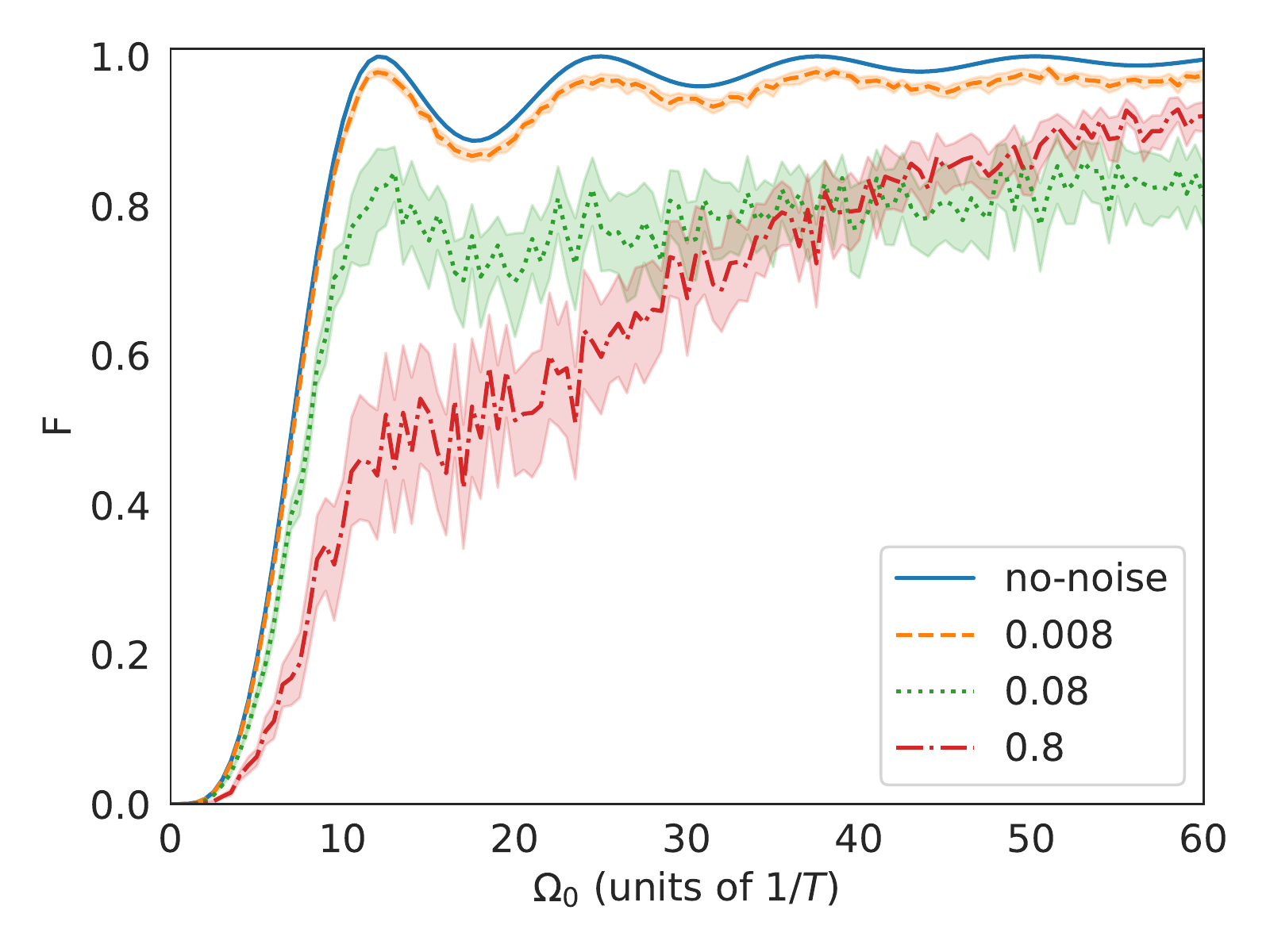}} &
        \subfigure[$\ $]{
	            \label{fig:sa_sin500}
	            \includegraphics[width=.37\linewidth]{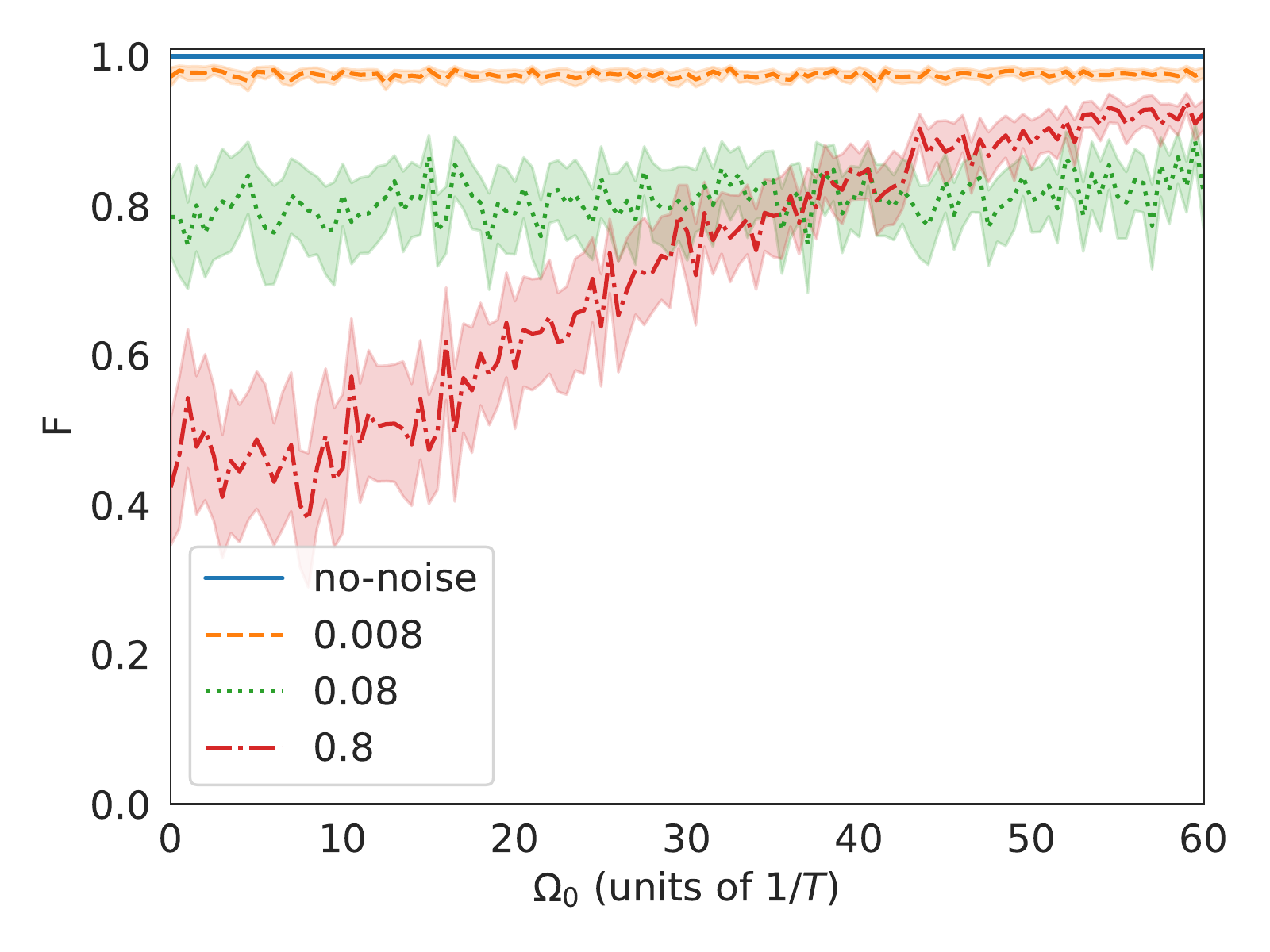}} \\
        \subfigure[$\ $]{
	            \label{fig:sin501}
	            \includegraphics[width=.37\linewidth]{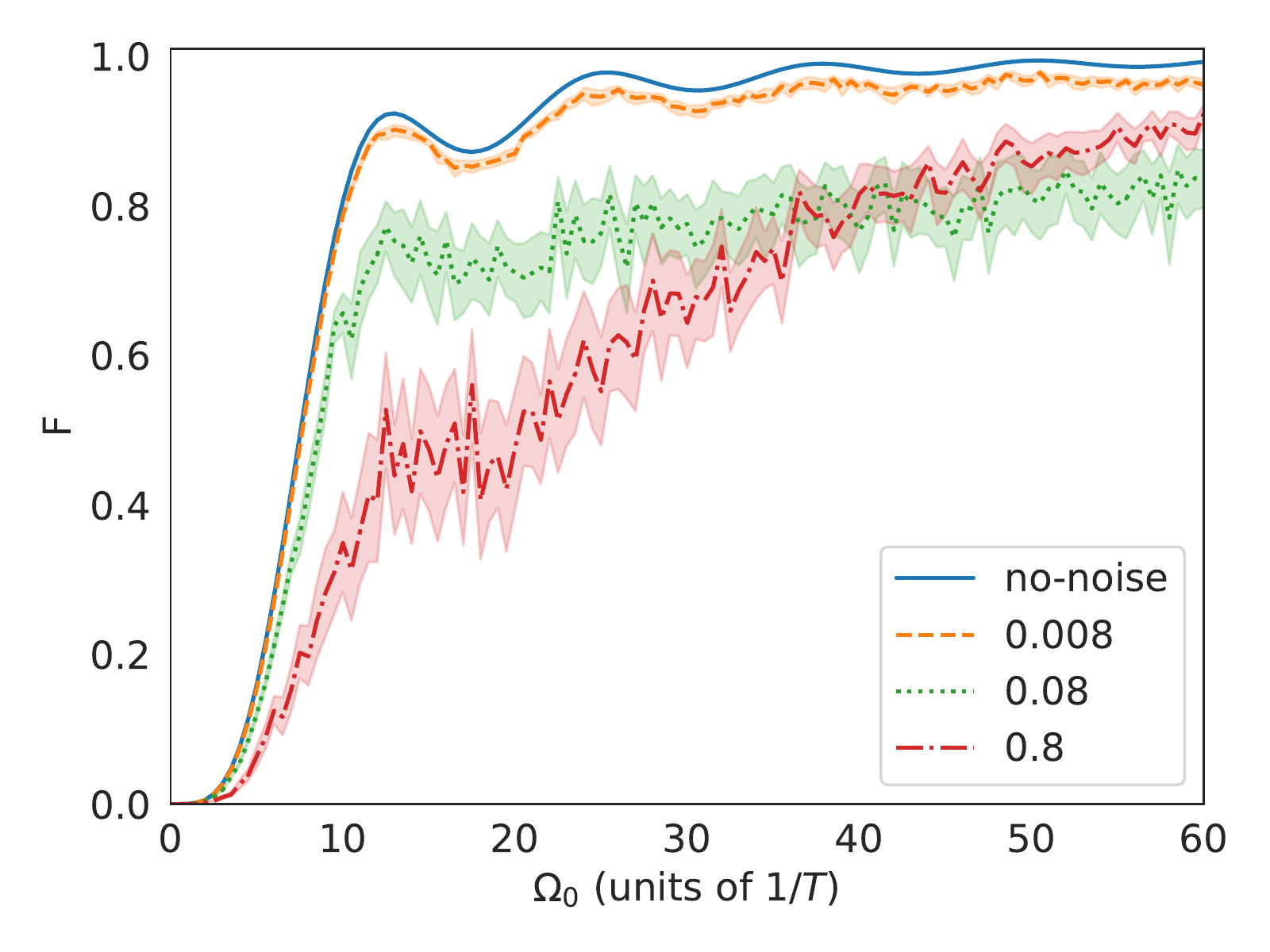}} &
        \subfigure[$\ $]{
	            \label{fig:sa_sin501}
	            \includegraphics[width=.37\linewidth]{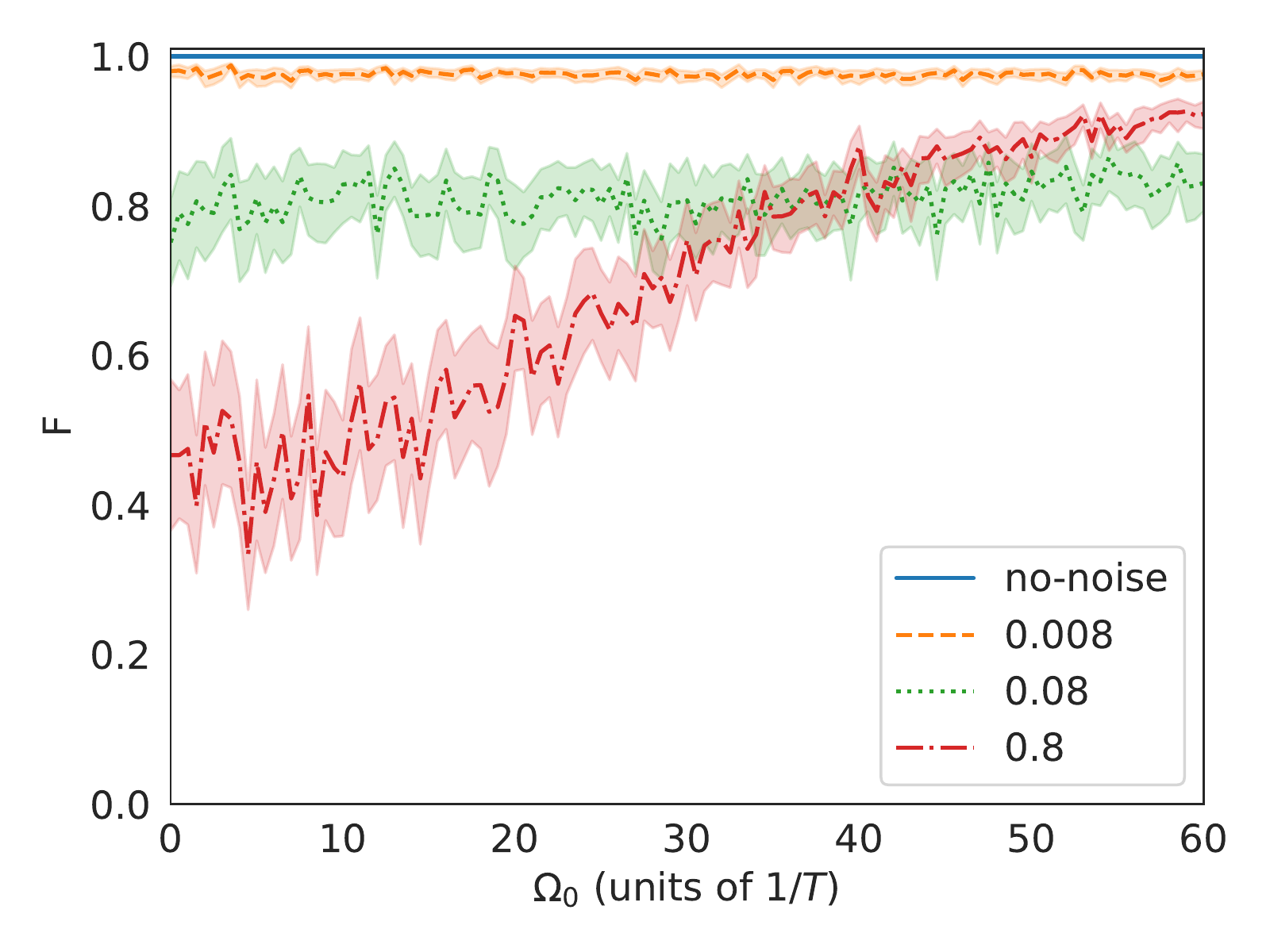}} \\
        \subfigure[$\ $]{
	            \label{fig:sin504}
	            \includegraphics[width=.37\linewidth]{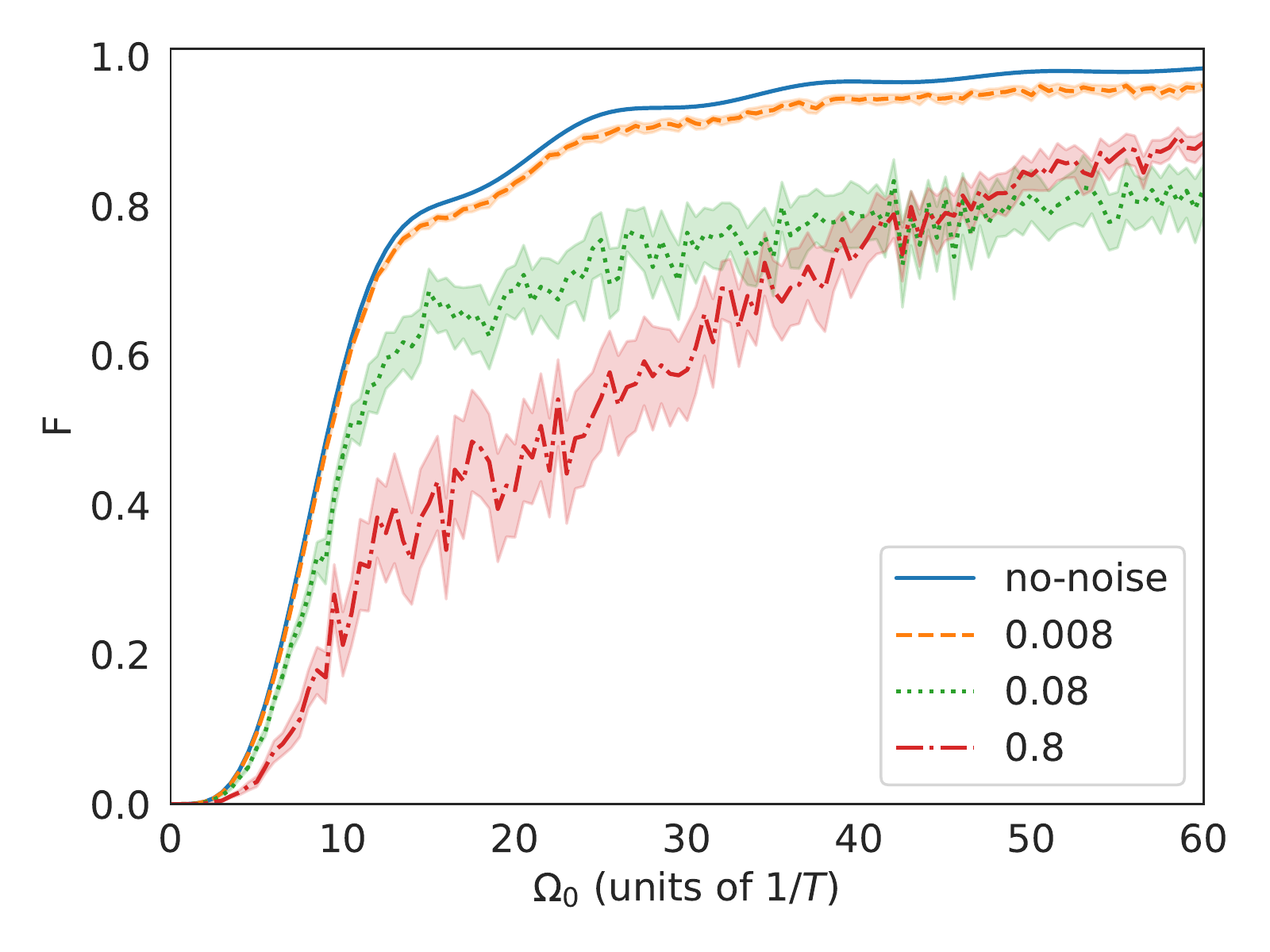}} &
        \subfigure[$\ $]{
	            \label{fig:sa_sin504}
	            \includegraphics[width=.37\linewidth]{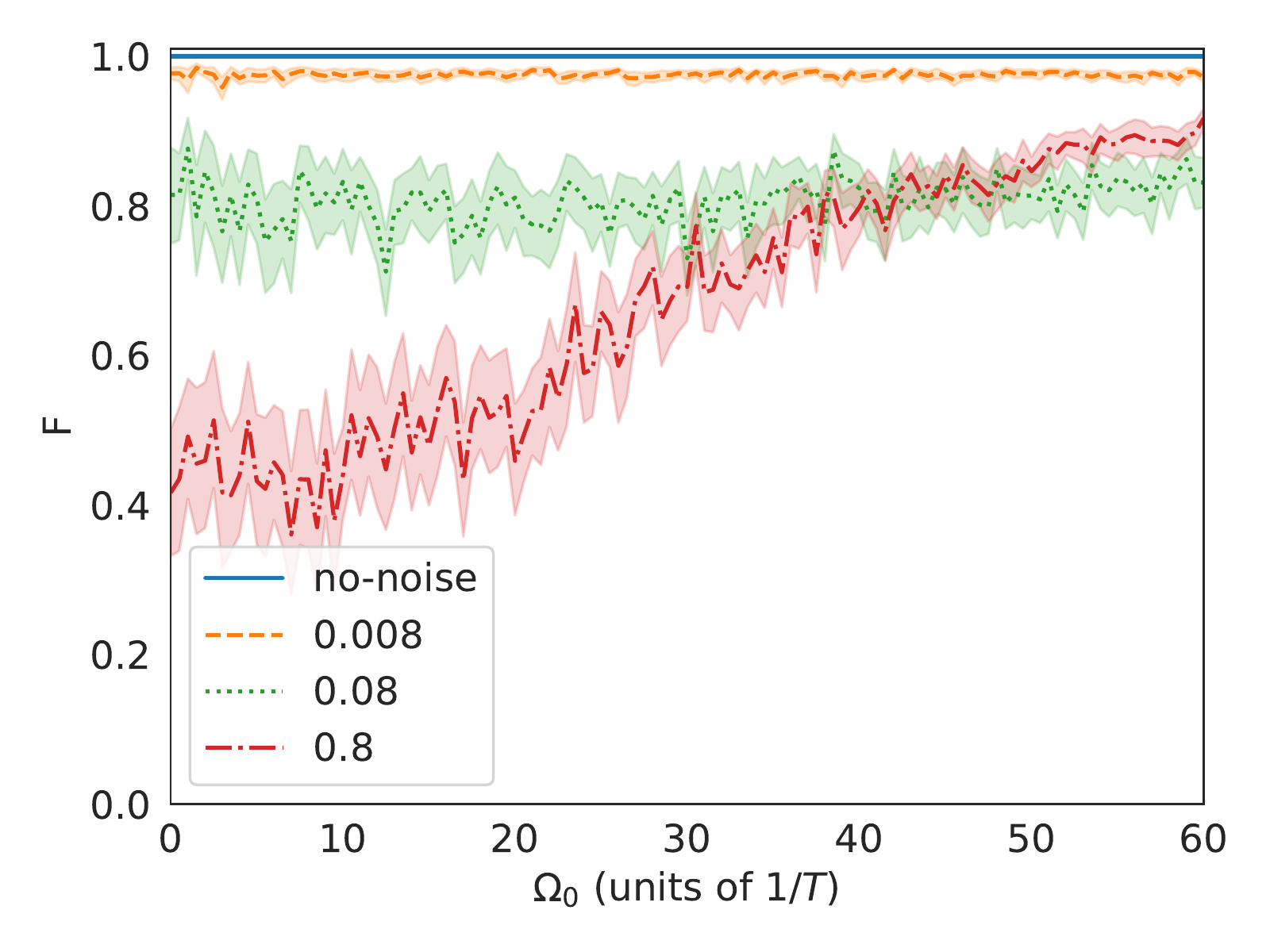}} \\
        \subfigure[$\ $]{
	            \label{fig:sin5010}
	            \includegraphics[width=.37\linewidth]{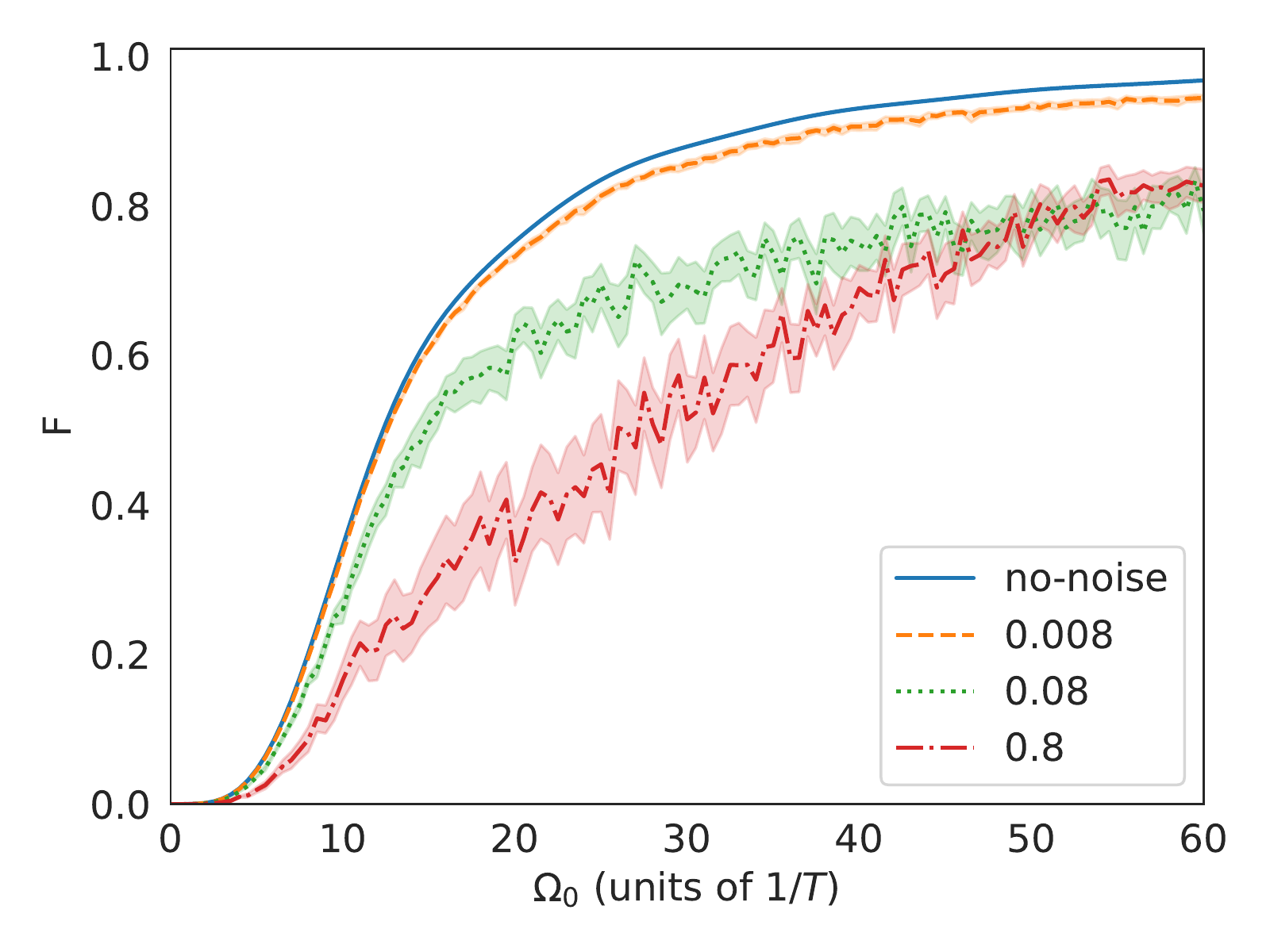}} &
        \subfigure[$\ $]{
	            \label{fig:sa_sin5010}
	            \includegraphics[width=.37\linewidth]{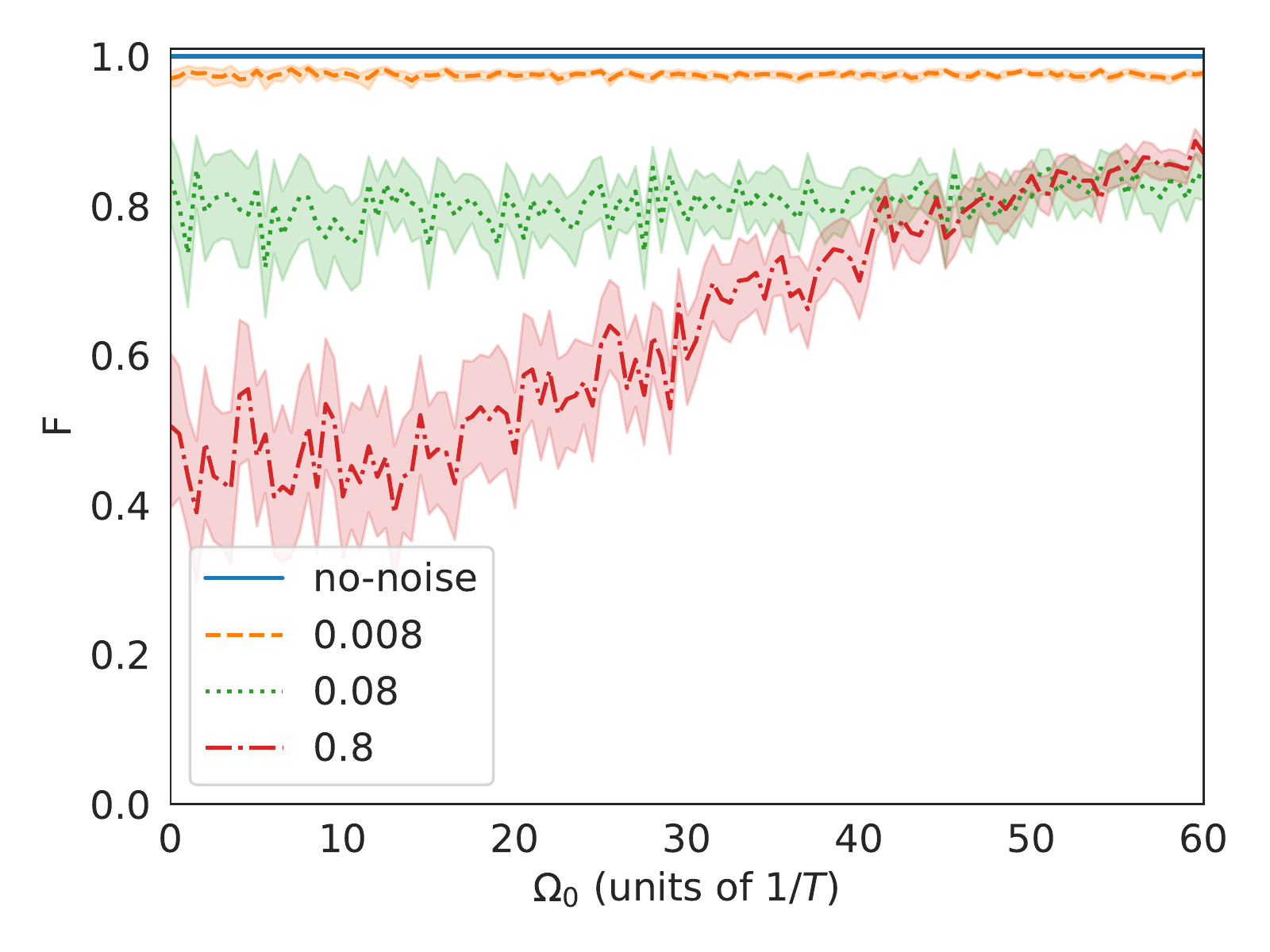}}
		\end{tabular}
\caption{(Color online) Fidelity versus $\Omega_0$ for sin-cos STIRAP and SA-STIRAP, for various values of the dephasing noise correlation time $\tau_c/T$, shown in the inset, and dissipation rates $\Gamma$: (a, c, e, g) STIRAP with $\Gamma=0, 1/T, 4/T, 10/T$, respectively, (b, d, f, h) SA-STIRAP with $\Gamma=0, 1/T, 4/T, 10/T$, respectively.}
\label{fig:sin}
\end{figure*}

\begin{figure}[t]
\centering
\includegraphics[width=0.8\linewidth]{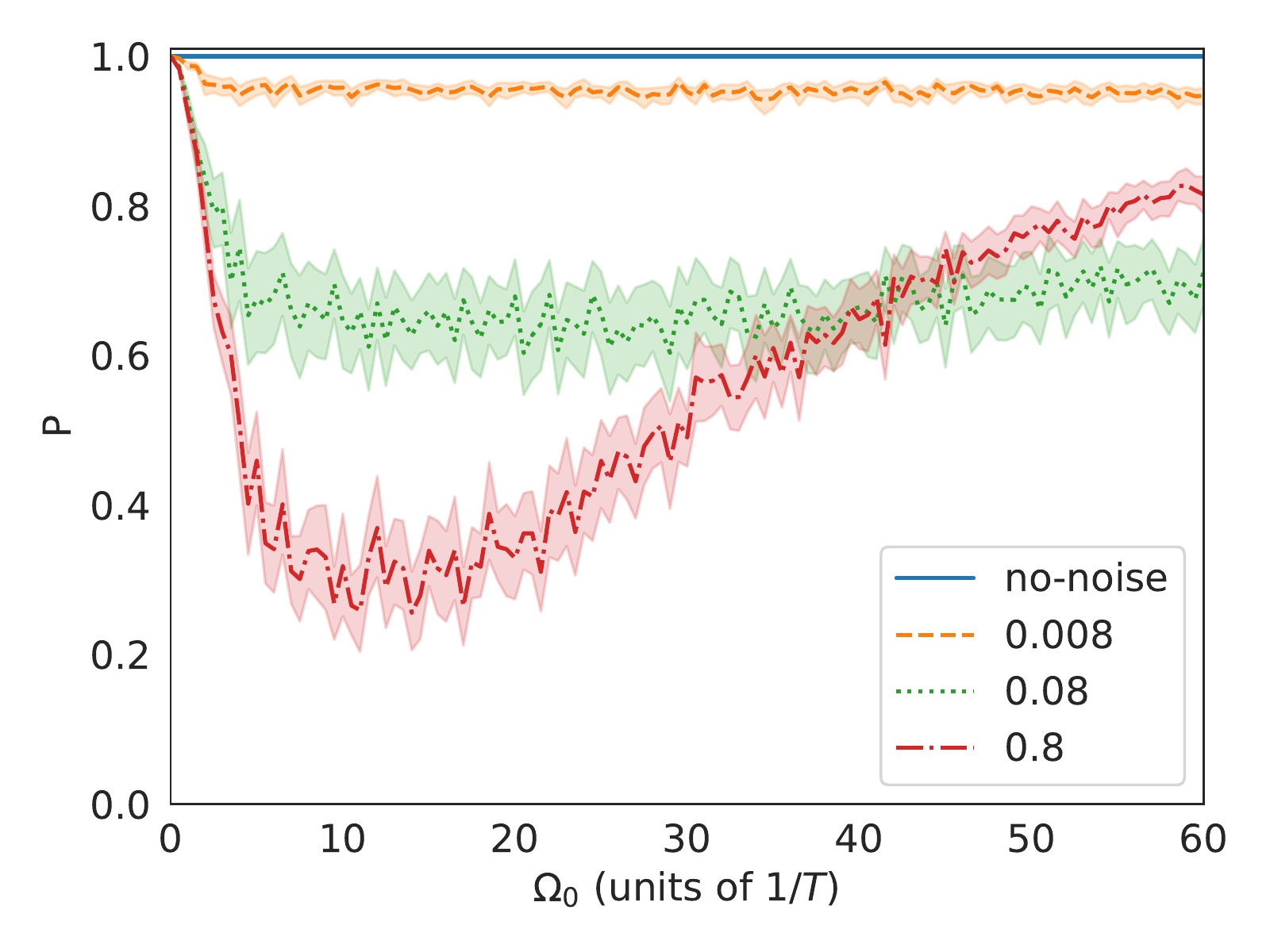}
\caption{Final total population of all three levels versus $\Omega_0$, for Gaussian SA-STIRAP with delay $\tau/T=1/2$ and dissipation rate $\Gamma=4/T$, for different values of the dephasing noise correlation time, shown in the inset.}
\label{fig:sa_gaussian504_P}
\end{figure}

\begin{figure*}[t]
 \centering
		\begin{tabular}{cc}
     	\subfigure[$\ $]{
	            \label{fig:STA_pulses}
	            \includegraphics[width=.45\linewidth]{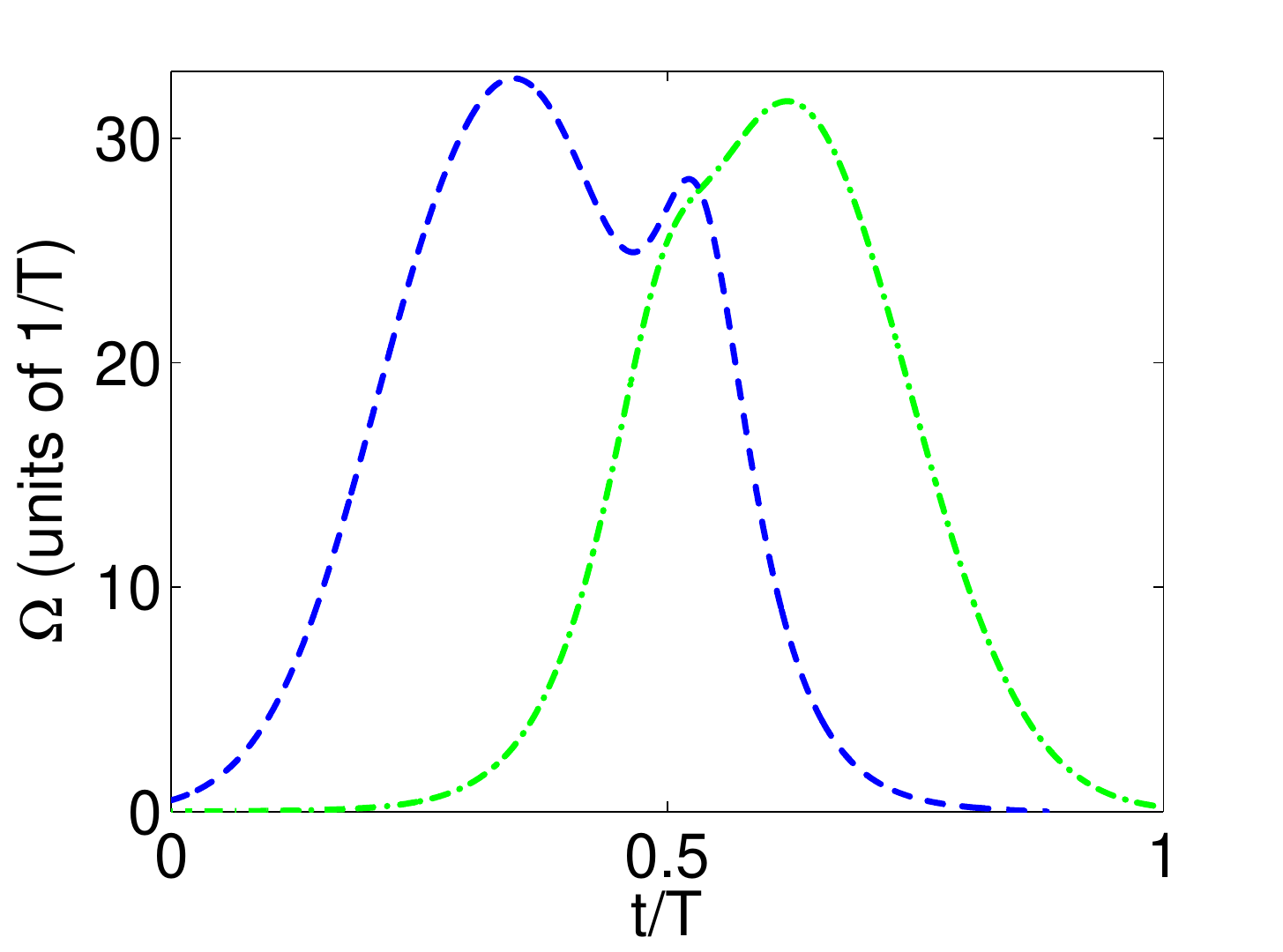}} &
       \subfigure[$\ $]{
	            \label{fig:STA_F}
	            \includegraphics[width=.45\linewidth]{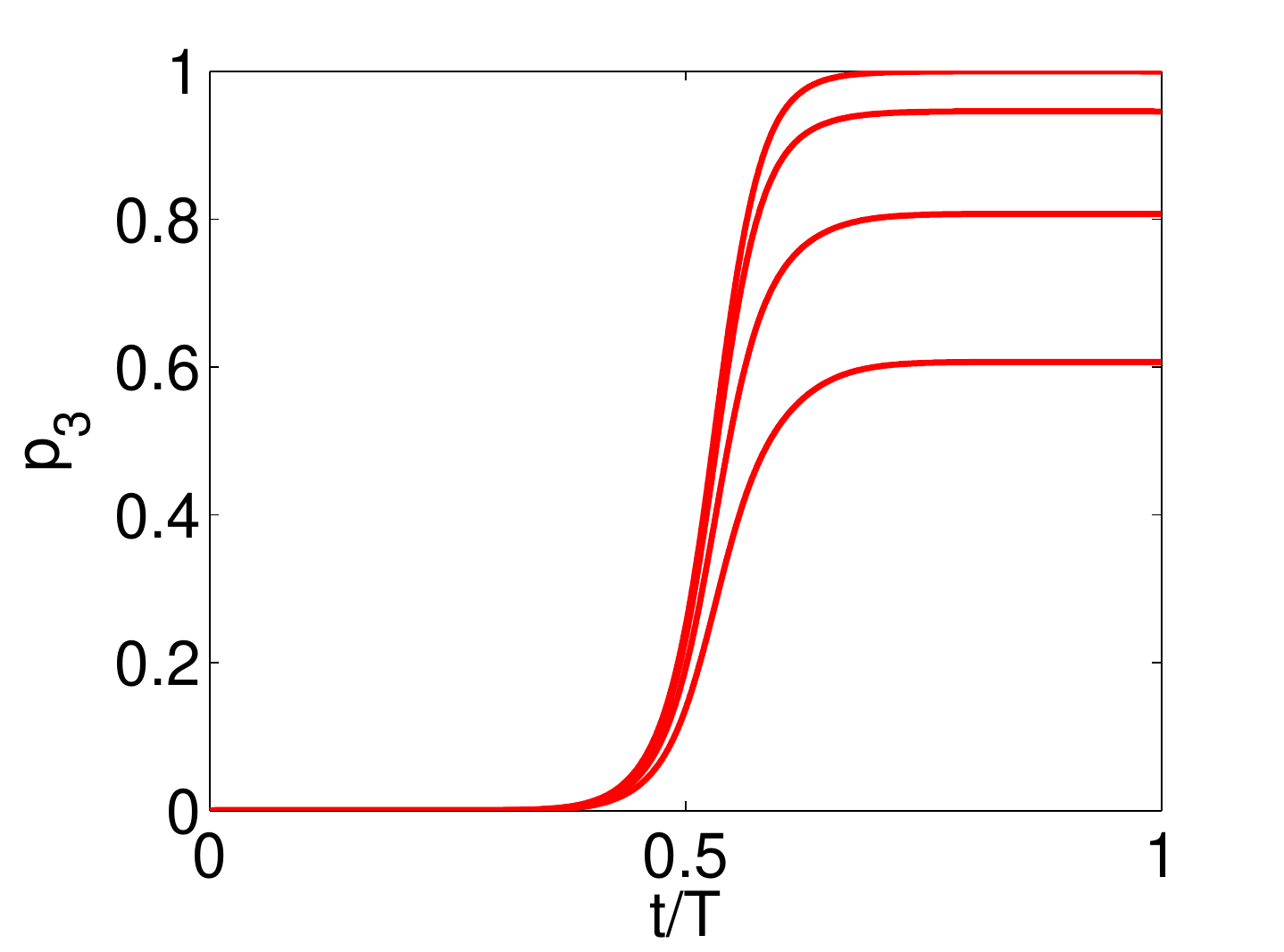}} \\
      \subfigure[$\ $]{
	            \label{fig:SA_ST_pulses}
	            \includegraphics[width=.45\linewidth]{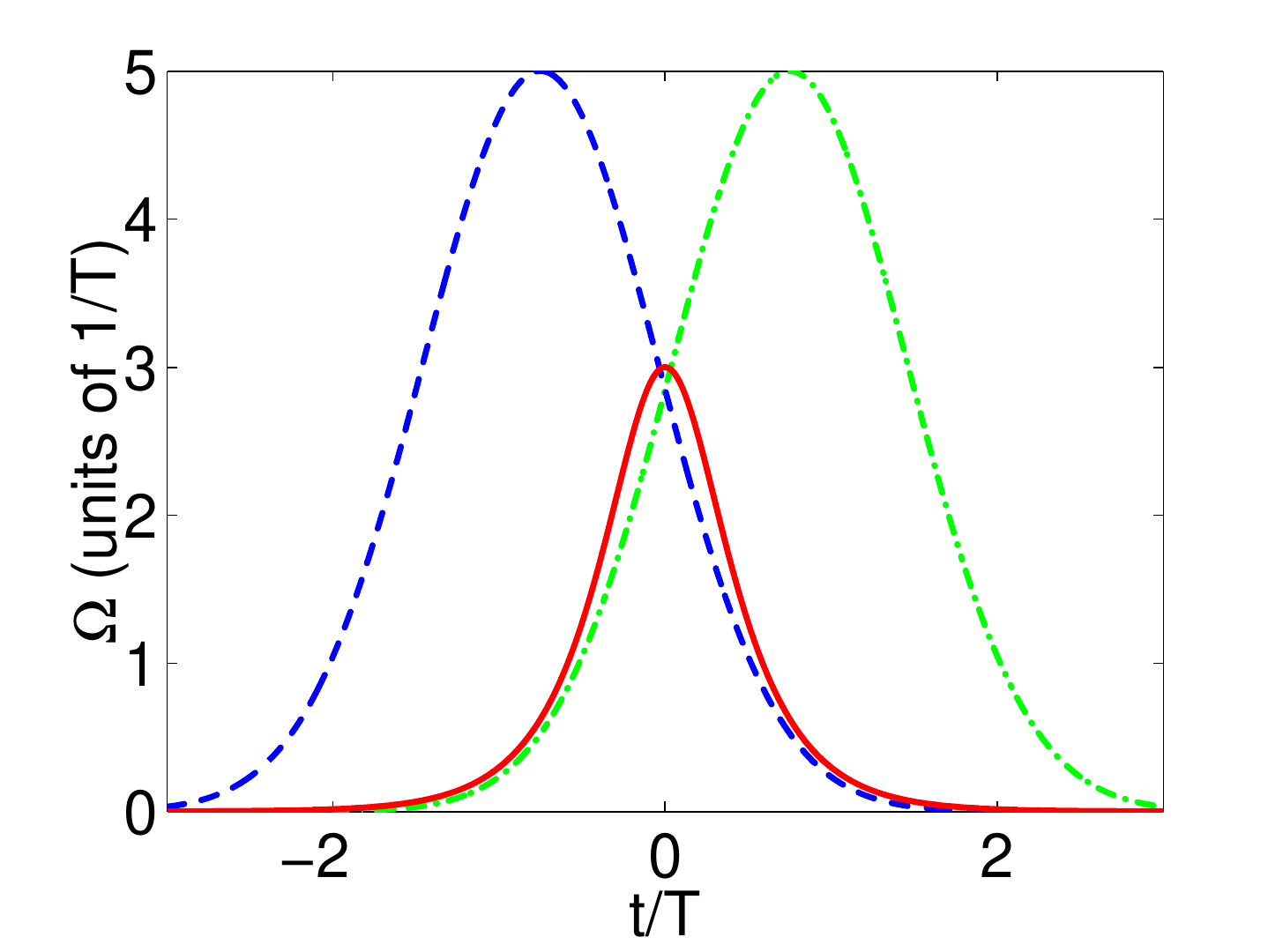}} &
       \subfigure[$\ $]{
	            \label{fig:SA_ST_F}
	            \includegraphics[width=.45\linewidth]{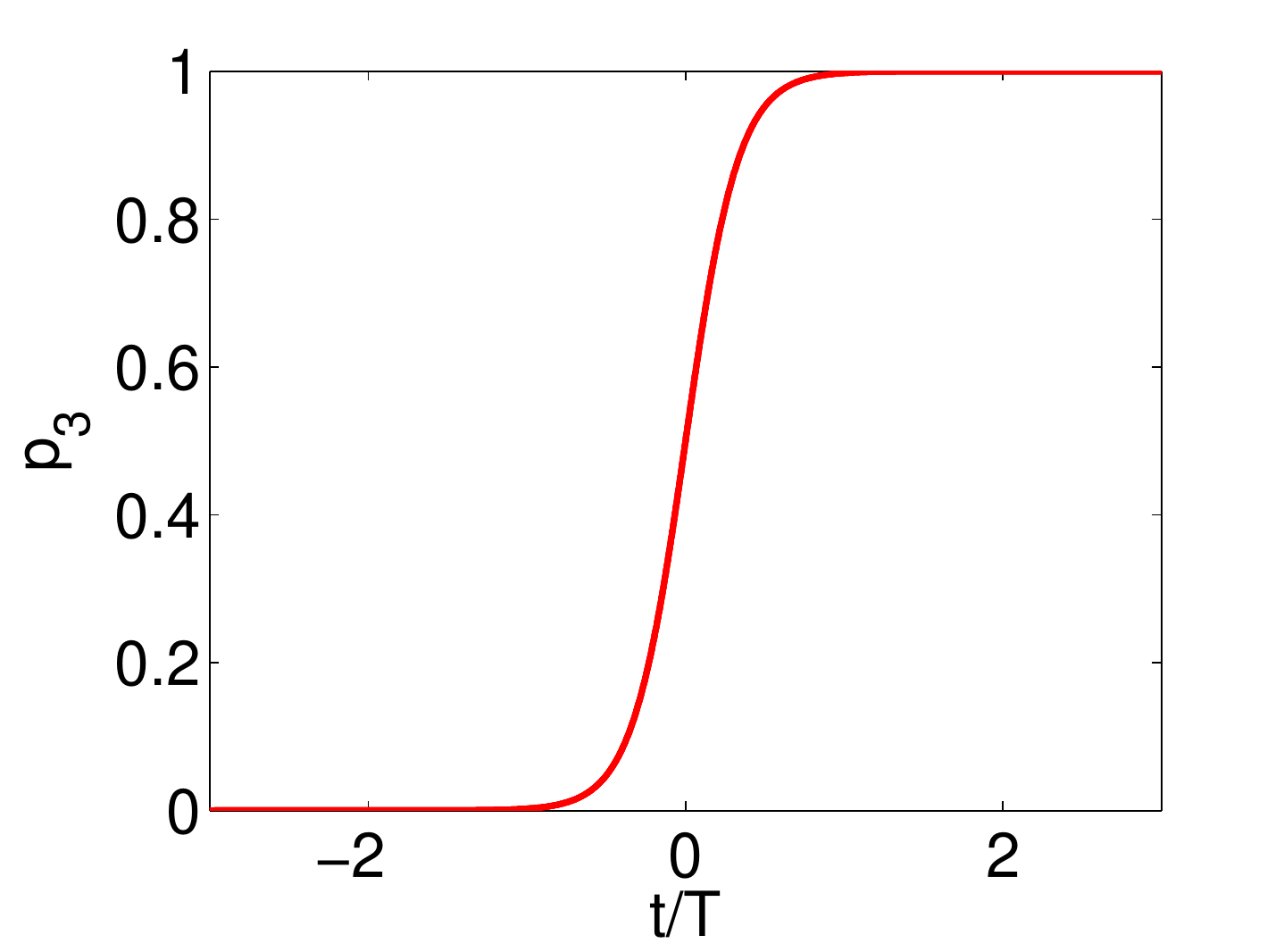}}
		\end{tabular}
\caption{(Color online) (a) Pump (green dashed-dotted line) and Stokes (blue dashed line) pulses for the shortcut method of Ref. \cite{Stefanatos20}. (b) Corresponding time evolution of level $|3\rangle$ population for different values of dissipation rate, $\Gamma=0, 1/T, 4/T, 10/T$ from top to bottom. (c) Pump, Stokes, and counterdiabatic (red solid line) pulses for Gaussian SA-STIRAP with $\Omega_0=5/T$ and $\tau/T=3/4$. (d) Corresponding time evolution of level $|3\rangle$ population for the same dissipation rates as before (the slight dependence can be hardly distinguished and appears as a thicker line).}
\label{fig:SA_ST_vs_STA}
\end{figure*}

\begin{figure*}[t]
 \centering
		\begin{tabular}{cc}
     	\subfigure[$\ $]{
	            \label{fig:changetau_c}
	            \includegraphics[width=.45\linewidth]{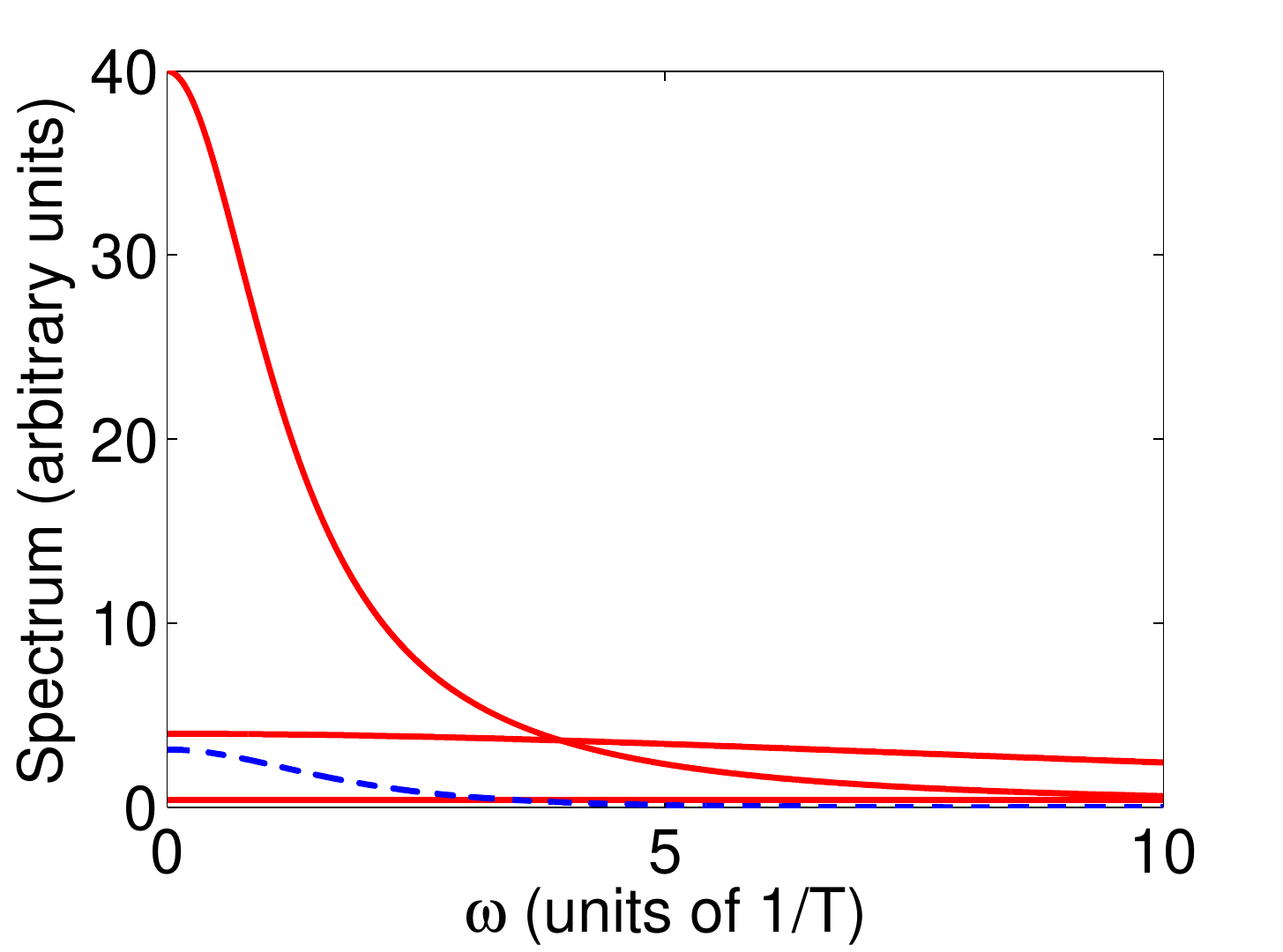}} &
       \subfigure[$\ $]{
	            \label{fig:changetau}
	            \includegraphics[width=.45\linewidth]{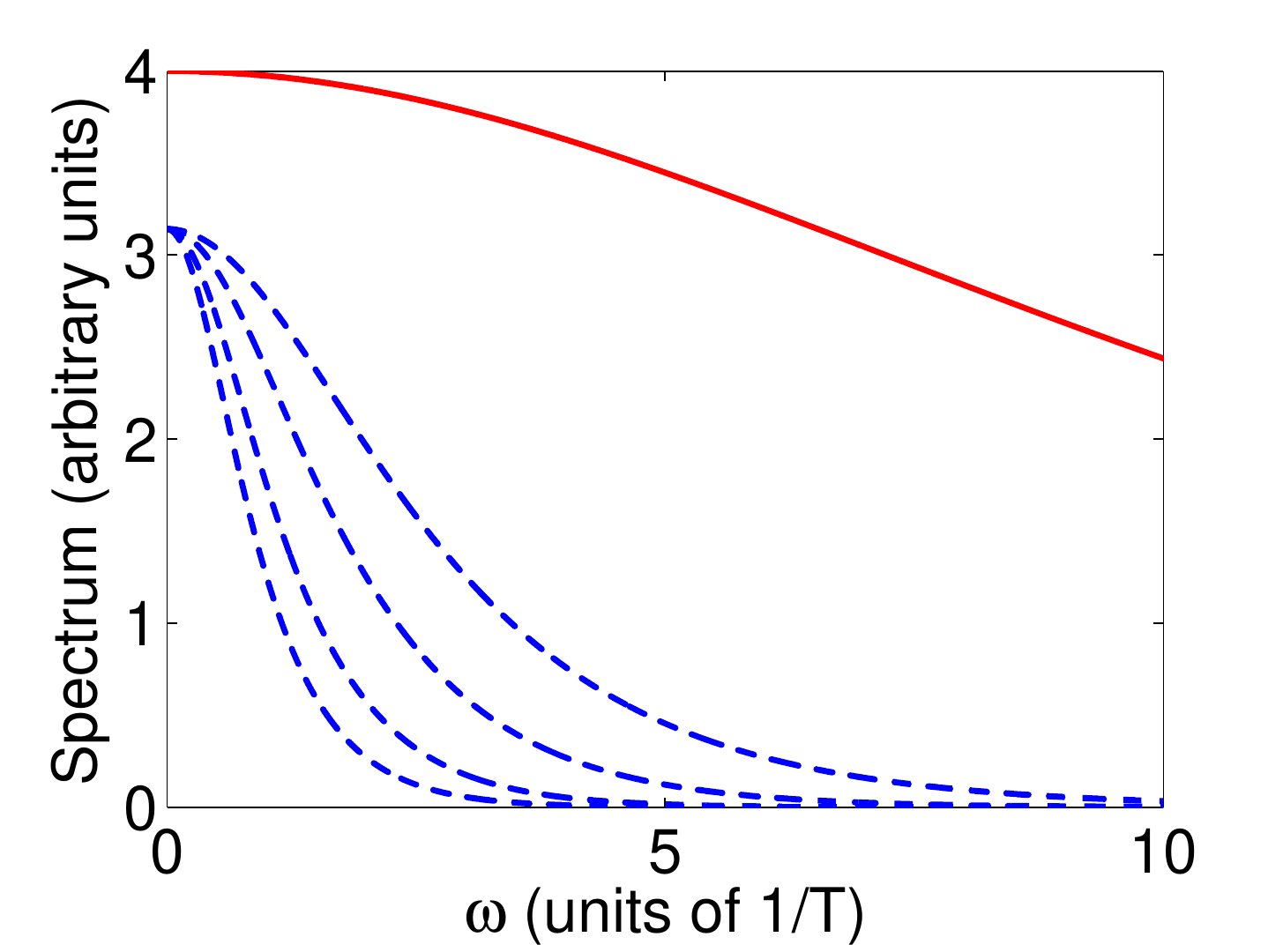}} \\
		\end{tabular}
\caption{(Color online) Red solid lines correspond to the dephasing noise power spectral density for different correlation times $\tau_c$, and blue dashed lines to the Fourier transform of the sech counterdiabatic pulse for various delays $\tau$. (a) For increasing correlation time $\tau_c/T=0.008, 0.08, 0.8$ (red solid lines from bottom to top) and fixed delay $\tau/T=0.5$, the noise power is concentrated in lower frequencies and the counterdiabatic pulse experiences higher noise levels. (b) For increasing delay $\tau/T=1/4, 1/3, 1/2, 3/4$ (blue dashed lines from left to right) and fixed $\tau_c/T=0.08$, the spectrum of the counterdiabatic pulse broadens and the higher frequency components experience lower noise levels.}
\label{fig:spectrum}
\end{figure*}

\begin{figure*}[t]
 \centering
		\begin{tabular}{cc}
     	\subfigure[$\ $]{
	            \label{fig:P2W60}
	            \includegraphics[width=.45\linewidth]{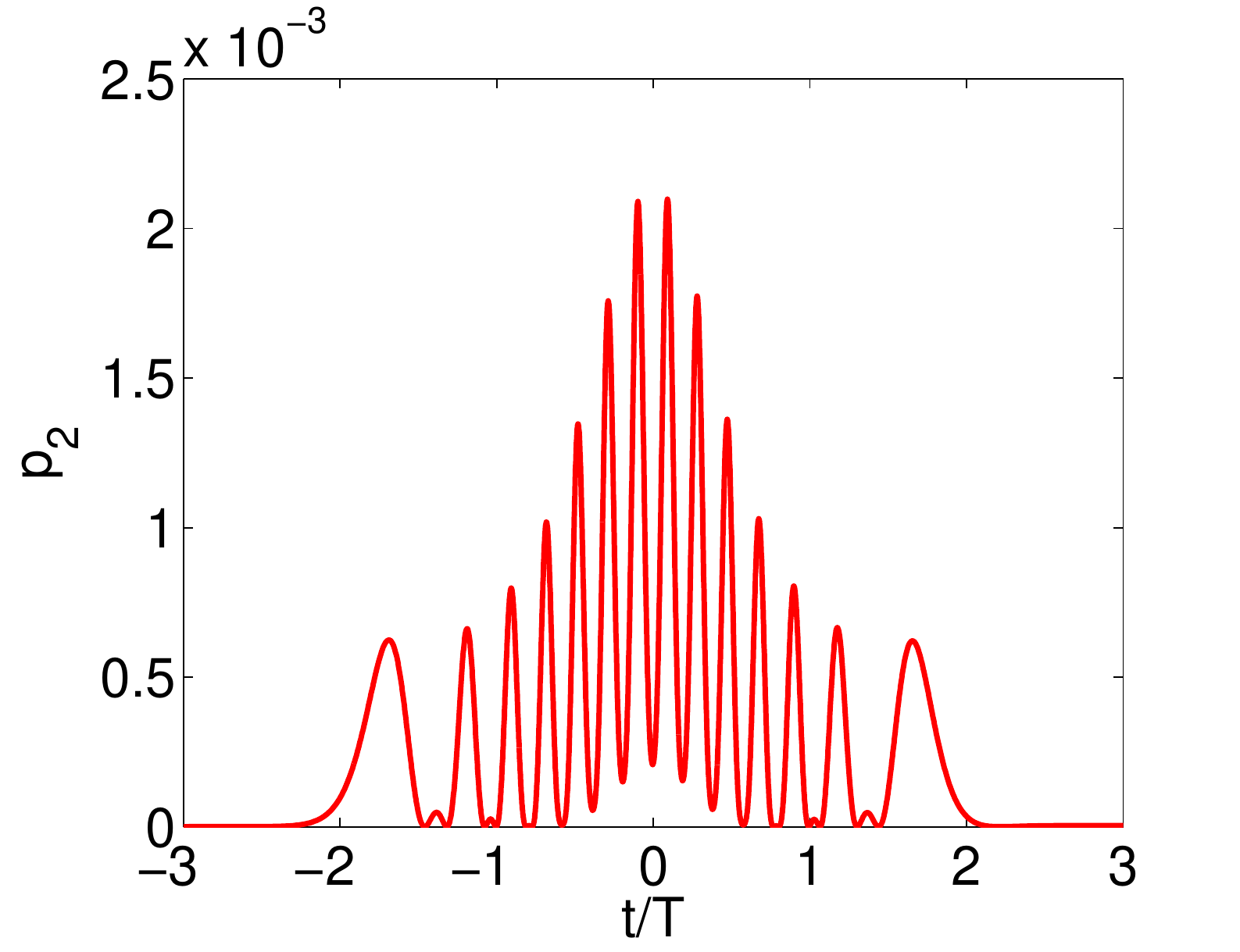}} &
       \subfigure[$\ $]{
	            \label{fig:BlochW60}
	            \includegraphics[width=.45\linewidth]{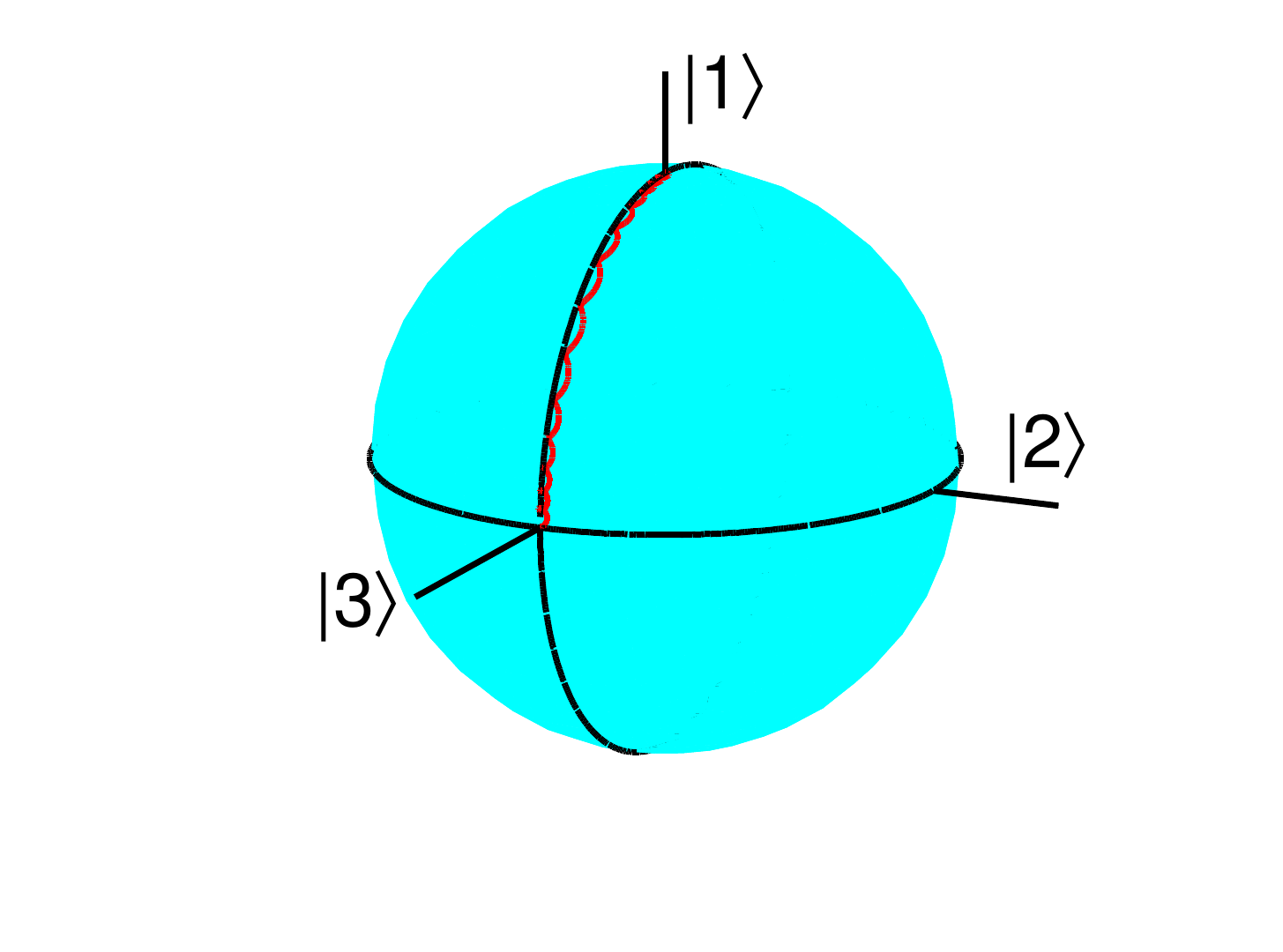}} \\
      \subfigure[$\ $]{
	            \label{fig:P2W15}
	            \includegraphics[width=.45\linewidth]{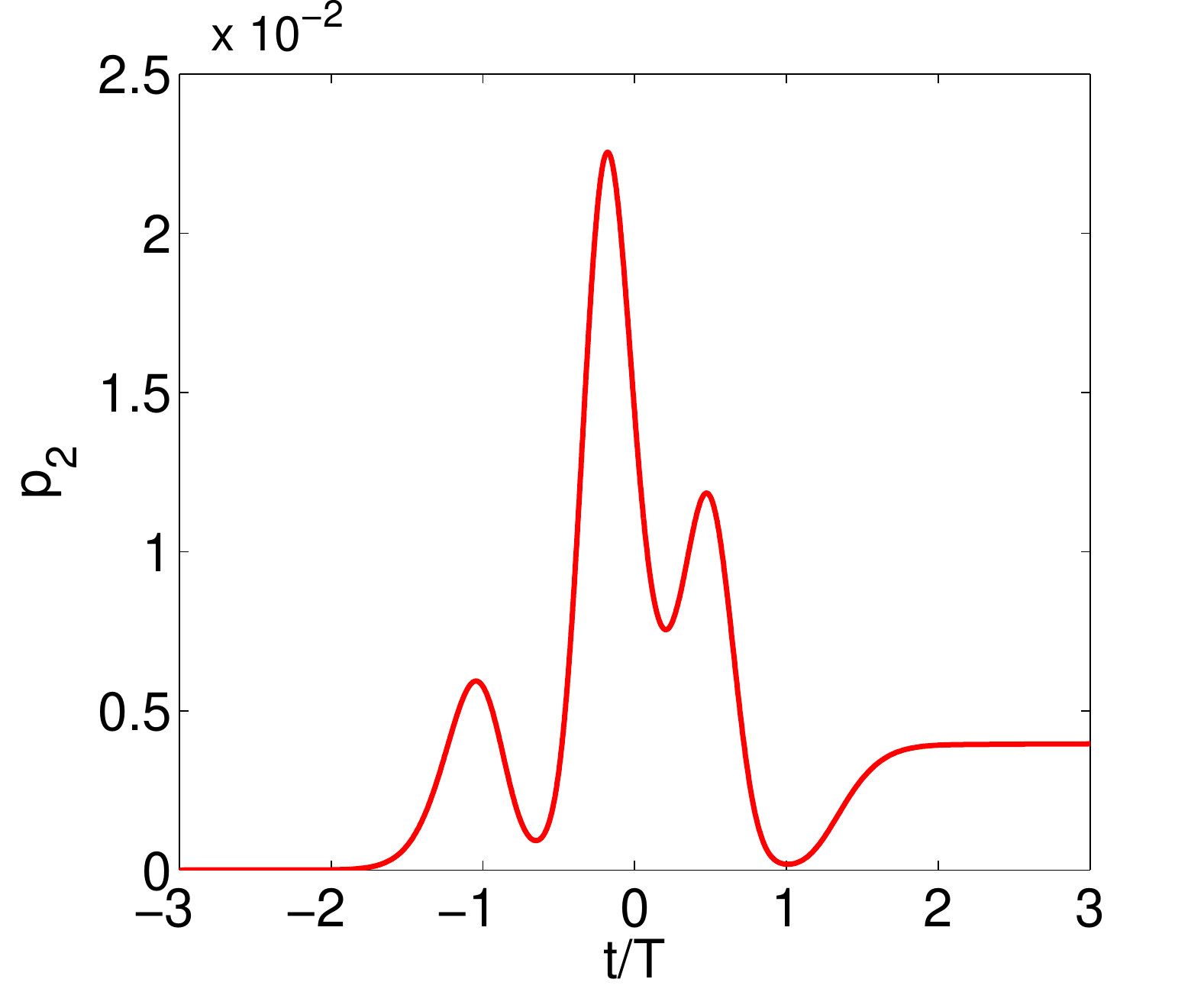}} &
       \subfigure[$\ $]{
	            \label{fig:BlochW15}
	            \includegraphics[width=.45\linewidth]{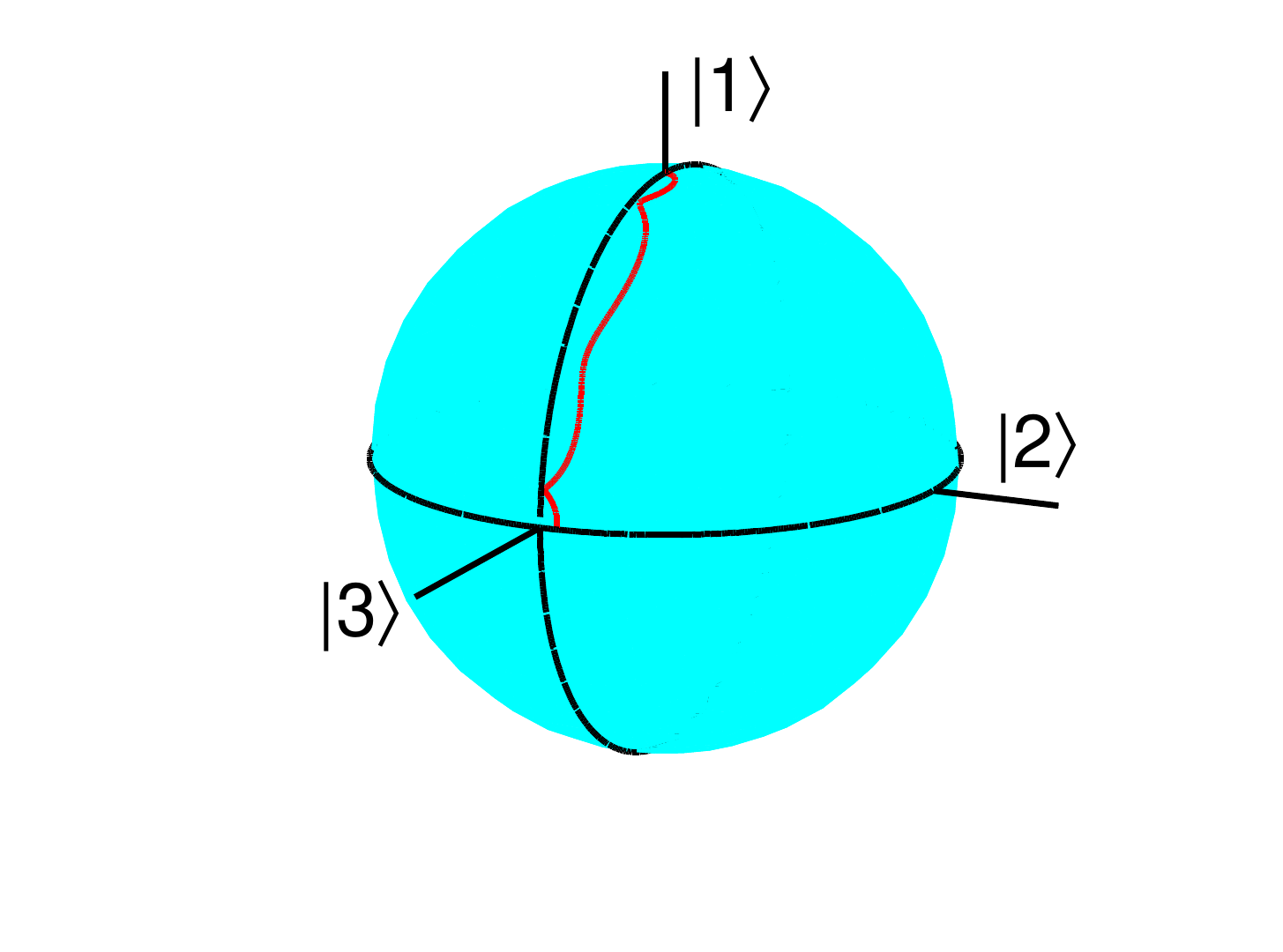}}
		\end{tabular}
\caption{Time evolution of level $|2\rangle$ population and trajectory on the Bloch sphere for Gaussian STIRAP with delay $\tau/T=1/2$, $\Gamma=0$ and no dephasing noise. (a), (b) $\Omega_0=60/T$, (c), (d) $\Omega_0=15/T$.}
\label{fig:large_omega}
\end{figure*}

\section{Conclusion}

\label{sec:conclusion}

We investigated the performance of two SA-STIRAP schemes obtained from Gaussian and sin-cos pulses, in the presence of dissipation and dephasing by exponentially correlated noise. For small amplitudes of Stokes and pump pulses we found that the population transfer is mainly accomplished directly by the counterdiabatic pulse, while for large amplitudes conventional STIRAP dominates. This kind of ``hedging" leads to a remarkable robustness against dissipation in the lossy intermediate level. For small pulse amplitudes and increasing noise correlation time we found a decreasing performance, since the dominant counterdiabatic pulse is affected more, while for large pulse amplitudes, where the STIRAP path dominates, the efficiency is degraded more for intermediate correlation times (compared to the pulse duration). We also studied for the Gaussian SA-STIRAP the effect of delay between pump and Stokes pulses and found that in the presence of noise the performance is improved for increasing delay. Our conclusion is that the Gaussian SA-STIRAP protocol with suitably chosen delay and the sin-cos SA-STIRAP protocol perform quite well even under severe noise conditions. The current work is expected to find application in the implementation of emerging quantum technologies, since STIRAP is a widely used method in this research area.

\appendix*

\section{Notes on numerical method}

\label{sec:appendix}

The simplest type of colored noise is that with exponential correlation function.
It requires only one more parameter than white noise while being a more
realistic noise source in the context of laser noise.
In general, a stochastic differential equation
with an exponentially correlated colored noise term
can be modeled as
\begin{eqnarray}
	\dot{x}_i &= f(x_i) + \epsilon_i \label{eq:noise1} \\
	\dot{\epsilon}_i &= -\frac{1}{\tau_c}\epsilon_i +\frac{1}{\tau_c}g_i\label{eq:noise2}
\end{eqnarray}
where $g_i$ is simple zero-mean Gaussian white noise.


To integrate this stochastic system
an explicit fixed time step method is preferable
to control the stochastic dynamics.
We use an explicit 4th order Runge-Kutta method
to integrate the first, deterministic part of the system
(equation~\ref{eq:noise1})
coupled with
the algorithm by Fox et al.~\cite{Fox88}
to integrate for the noise (Eq.~\ref{eq:noise2}).
A sufficiently small time step must be used to ensure
both that the noise terms are properly integrated
and the deterministic part of the system is accurately
simulated.
In general we have used a time step $\Delta t$ that is at most
$1/10$ the value of correlation time $\tau_c$.

Algorithm~\ref{alg:fox} shows the steps
of the integration of Eq.~\ref{eq:noise2}.
First, an initial noise value $\epsilon_i|_{t=0}$ is generated
and the intensity-related parameter $E$ is calculated for the
integration interval $\Delta t$ and correlation time $\tau_c$.
At each time step of the integration,
a new $\epsilon_i$ value is calculated,
based on the previous value of $\epsilon_i$
and a Gaussian parameter $h$ of mean zero and standard deviation $\sigma$;
then, $\epsilon_i$ are a series of values that
realize a  zero-mean Ornstein-Uhlenbeck process with correlation time $\tau_c$,
standard deviation $\sigma$, obeying the correlation relation
$\langle \epsilon_i(t)\epsilon_i(t+\tau) \rangle = \sigma^2 e^{-\frac{|\tau|}{\tau_c}}$,
as discussed in the previous sections.

\begin{figure*}[t]
 \centering
		\begin{tabular}{c}
        \subfigure[$\ $]{
	            \label{fig:noisefreq}
	            \includegraphics[width=.45\linewidth]{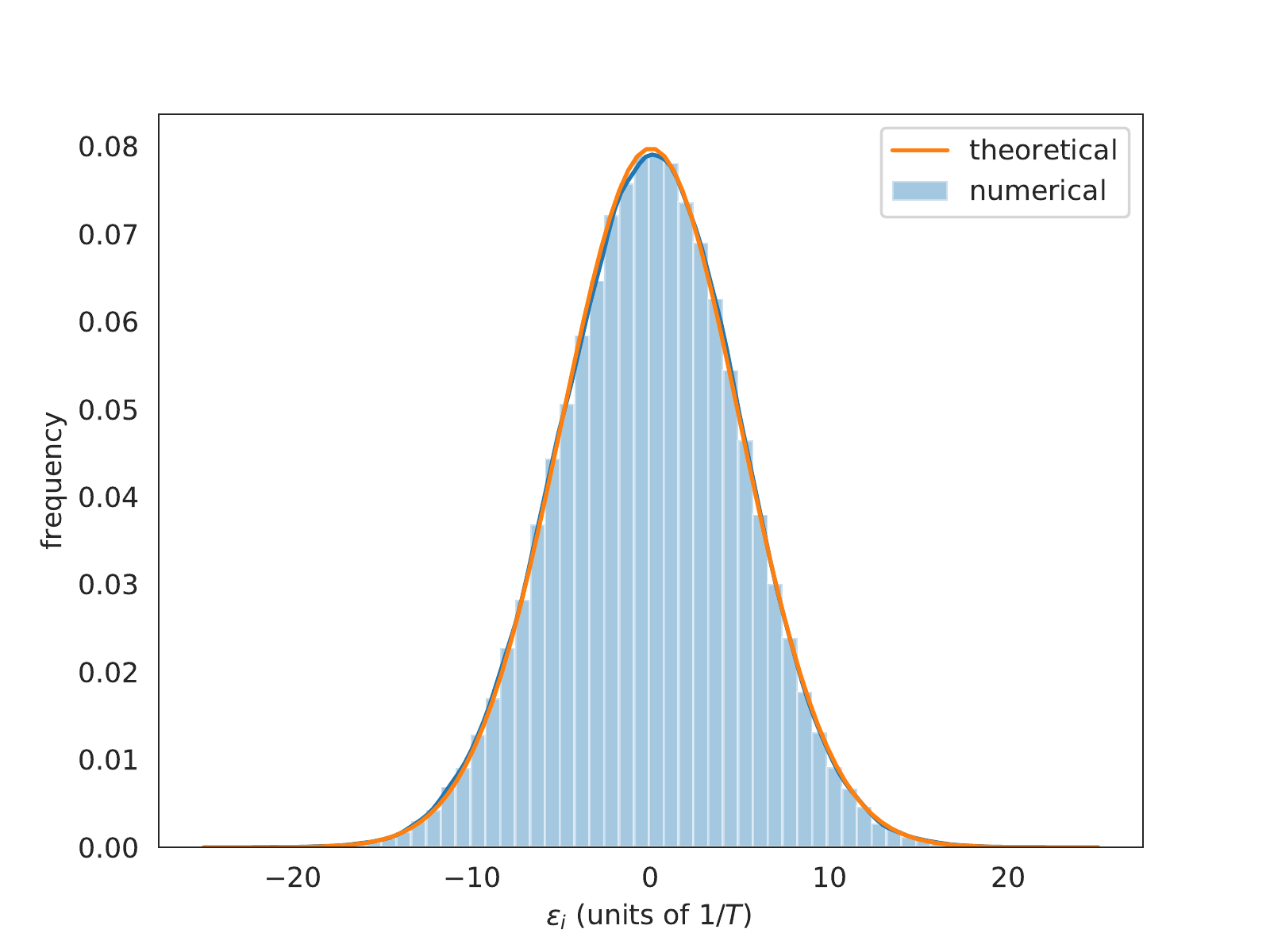}} \\
        \subfigure[$\ $]{
	            \label{fig:noisecorr}
	            \includegraphics[width=.45\linewidth]{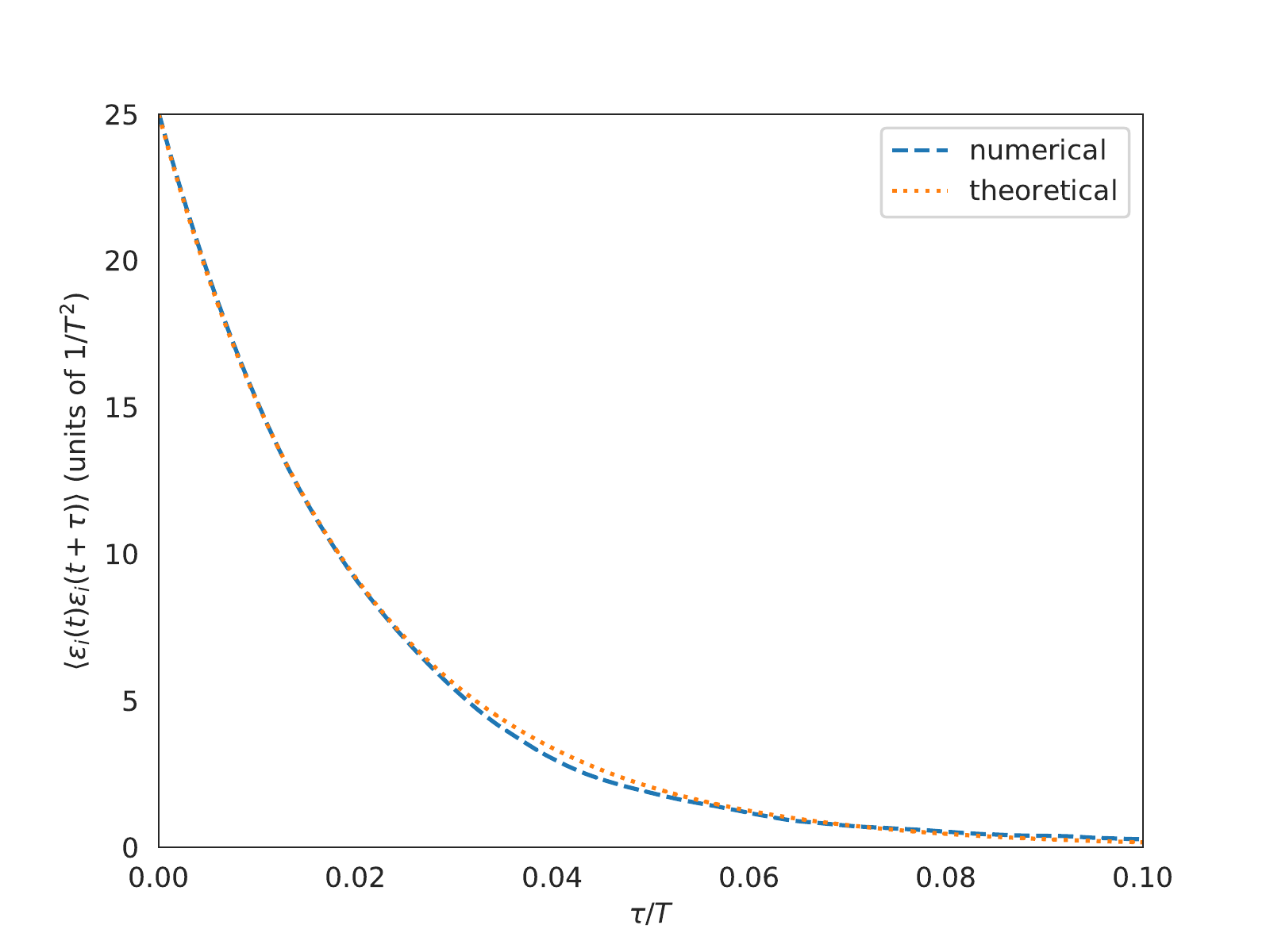}}
		\end{tabular}
\caption{(a) Histogram of noise strength for $\tau_c/T = 0.08, \sigma=5/T$. Theoretical curve
	of $\frac{1}{\sqrt{2\pi}\sigma}e^{-\frac{\epsilon_i^2}{2\sigma^2}}$ and numerical simulation results
averaged over 10 iterations. (b) Numerical results and theoretical curve for $\langle\epsilon_i(t)\epsilon_i(t+\tau)\rangle$ vs $\tau$ for $\tau_c/T = 0.08$ and $\sigma = 5/T$.}
\label{fig:noise}
\end{figure*}

In Figs.~\ref{fig:noisefreq} and~\ref{fig:noisecorr}
we compare the numerical results to the
theoretically desired properties of the noise.
The numerical data are averaged over 10 iterations
of the integration procedure.
Fig.~\ref{fig:noisefreq} shows
the relative frequency of the generated $\epsilon_i$ values.
Fig.~\ref{fig:noisecorr}
shows the mean correlation vs the lag-time $\tau$.
It is obvious that
the numerical results are in excellent agreement
with the theoretically desired properties.

\begin{algorithm}
\SetAlgoLined
\KwResult{An array $[\epsilon_i]$ of noise values for each time step.}
 $n, m = $ random number $\in [0, 1]$\;
 $E = e^{-\frac{\Delta t}{t_c}}$ \;
 $\epsilon|_{t=0} = \sqrt{-2\sigma^2\log{m}}\cos(2\pi n)$\;
 \While{$t < t_{\mathrm{final}}$}{
  $a, b =$ random number $\in [0, 1]$\;
  $h = \sqrt{-2\sigma^2(1 - E^2)\log{a}}\cos(2\pi b)$\;
  $\epsilon|_{t+\Delta t} = \epsilon E + h$\;
  $t \to t + \Delta t$\;
 }
 \caption{Generating exponentially correlated colored noise~\cite{Fox88}}%
 \label{alg:fox}
\end{algorithm}

\end{document}